%% file: apscor.tex
\documentclass{elsarticle-r}

\input{kf-pre}

\biboptions{sort&compress}

\begin{document}

\begin{frontmatter}

\title{Advances of Proof Scores in CafeOBJ}

\author{Kokichi Futatsugi}

\ead{futatsugi@jaist.ac.jp}

\address{Japan Advanced Institute of Science and Technology (JAIST),\\
Nomi, Ishikawa, 923-1292, Japan
}

\date{\today}

\begin{abstract}

Critical flaws continue to exist at the level of
domain, requirement, and/or design specification, and
specification verification (i.e., to check whether
a specification has desirable properties) is still one of
the most important challenges in software/system engineering.
CafeOBJ is an executable algebraic specification
language system and
domain/requirement/design engineers can write
proof scores for improving quality 
of specifications by the specification verification.
This paper describes advances of the proof scores for
the specification verification in CafeOBJ.

\end{abstract}


\end{frontmatter}

\section{Introduction}
\label{sec-intro}

Specification verification is theorem proving where
a specification is a set of axioms and desirable properties are
theorems.
The main goal of a specification verification is not, however,
to prove a property holds for the specification
but to check the quality of the specification against the property;
a specification is supposed to improve via the specification verification.
This is quite different from ordinary theorem proving
where the set of axioms is fixed and not supposed to change.
The improvements of a specification include
the addition of necessary axioms, the deletion of unnecessary axioms,
and the improvement of the specification's module structure.

For achieving the improvements,
specifications and the machinery of the specification verification
are better to be clear and transparent.
That is, the followings are better to be clear and precisely defined,
and hopefully simple and transparent.
(i) Models of a specification (models of a system specified),
(ii) inference (proof, deduction) rules used, and
(iii) rationale for assumptions used
(i.e., why an assumption is justified to be used or not).

CafeOBJ 
\citep{futatsugiN97,rd-kf1998,cafesem02,cafeobj-wp} 
is an executable algebraic specification
language system based on conditional equational logic
and rewriting execution with transparent and precise definition
of the models and the inference/execution rules.  
CafeOBJ's powerful module system makes it possible
to describe the assumption/conclusion relations in
a clear and transparent way.

A fully automatic specification verification often fails to
convey important structures/properties of the systems specified,
and makes it difficult to improve the specification.
One should seek to make balanced optimal use of the respective
abilities of humans and computers, so that computers do the
tedious formal calculations, and humans do the high level
planning.  
Proof scores have been fostered in the OBJ
\citep{futatsugiGJM85,goguen00iobj,goguen06TPA,obj3-wp},
CafeOBJ community,
and intend to meet these
requirements \citep{futatsugi06,futatsugiGO08,futatsugi10}.

A proof score for a specification in CafeOBJ is a piece of code that 
constructs proof modules and does rewritings/reductions on them
with the possibly conditional equations (i.e., axioms) in the
specification and the proof modules 
\citep{futatsugiGo12,futatsugi17}
(see Sections {\bf\ref{sec-bscprs}}, {\bf\ref{sec-adprs}} for more details).
If all the reductions return the expected results,
a proof attempted is achieved.
The formal calculations are encapsulated into
the reductions with the equations, and
the high level planning is encoded into
the proof modules and the reduction commands on them.

Major high level planing includes
lemma, case-split, and induction.
They are supposed to be found by humans and
declared as proof modules by making use of 
CafeOBJ's powerful module system.
Proof modules can also be constructed with fairly flexible
\ttt{open\,${\cdots}$\,close} constructs
(see Section {\bf\ref{sec-bscprs}}), and case-split/induction
can be described in a liberal way.
It involves, however, risks of obscuring the rationale
of the high level planning.
Proof tree calculus (PTcalc) and well-founded induction
(WFI) are incorporated to solve the issue,
by declaring a case-split with exhaustive equations
and an induction via a well-founded relation
\citep{futatsugi17,futatsugi20}.

Transition system is the primary modeling scheme
for software related dynamic systems.
A proof score method making use of
CafeOBJ's builtin search predicates 
is developed for proving leads-to properties of
transition systems \citep{futatsugi15}.
The leads-to properties cover fairly
large part of important properties of
dynamic systems,
and their proof scores can be built with
PTcalc+WFI.

This paper is intended to be a comprehensive and
self-contained account on the recent advances of the proof
scores in CafeOBJ with a nice harmony of theories and cases.
Many papers have been published on CafeOBJ and/or proof scores.
Majority of them, however, are on
theories/methods or examples/case-studies intensively, and
few treat the both in valance and describe how theories
justify cases in CafeOBJ in a concrete fashion.

For achieving the intention, (1) theories are described
so as to justify cases in CafeOBJ as directly as possible,
(2) CafeOBJ's cases are presented to explain
the meaning/intention of theories in concrete ways.
As a result, the paper clarifies the following two things
in a fairly simple and transparent way, because the logics
the CafeOBJ's proof scores based are the simplest among ones
on which other theorem proving systems based.
(a) A minimum cohesive collection of theories that makes 
proof scores work.
(b) Typical cases of proof scores in CafeOBJ
that are justified by the theories.

The following sections are organized as follows.\footnote{%
This paper is a revised version of \citep{Futatsugi22} that is
based on \citep{futatsugi21}.
Appendix {\bf\ref{apps:gpcps}} is totally new. 
Subsection {\it\ref{subsec-trrbtnsp}} is significantly revised and
extended.
Throughout the paper,
some notations are changed to make them consistant with others,
and several parts are revised to improve readability.}
Section {\bf\ref{sec-fth}} describes theories that are
necessary for justifying proof scores.
Section {\bf\ref{sec-bscprs}} explains how to construct
basic proof scores with \ttt{open\,{...}\,close} constructs.
Section {\bf\ref{sec-adprs}} explains how to construct
advanced proof scores with proof tree calculus (PTcalc)
and well-founded induction (WFI).
Section {\bf\ref{sec-prctrsy}} describes theories and methods
for constructing proof scores for transition systems.
Section {\bf\ref{sec-qlkex}} demonstrates the constructions
of the PTcalc+WFI proof scores for 
a simple example of transition system.
Section {\bf\ref{sec-dscs}} discusses related works,
distinctive features of the proof scores,
and concludes the paper.
Appendix {\bf\ref{apps:gpcps}} describes a generic procedure for
constructing proof scores and gives an another illuminating view of
the paper.

\section{Fundamental Theories}
\label{sec-fth}

This section is a preliminary to the following sections,
and is better to be browsed through on first reading.
There are many pointers from the following sections
to related parts in this section, and readers are
encouraged to return to understand necessary
concepts and notations.\\

CafeOBJ supports several kinds of specifications such as
equational specifications, rewriting specifications, and
behavioral specifications.
This paper focuses on equational specifications 
and transition specifications
(a restricted class of rewriting specifications)
on which our proof score method has been mainly developed.
We will proceed with
the following items for CafeOBJ
following the style of 
\citep{rd-kf1998,futatsugiGo12}.

\begin{itemize} \itemsep -1pt

\item a set $\Sign$ of {\bf signatures}.

\item for each signature $\Sg\,{\in}\,\Sign$ a class $\Mod(\Sg)$ 
  of \Sg-{\bf models}.

\item for each signature $\Sg$ a set $\Sen(\varSigma)$ of
\Sg-{\bf sentences}. 

\item for each signature $\varSigma$ a
 {\bf satisfaction relation}
 $\vDash_{\varSigma}$ between \Sg-models and \Sg-sentences,
 i.e., $\vDash_{\varSigma}\subseteq(\Mod(\varSigma)\times\Sen(\varSigma))$.

\end{itemize}

\subsection{Signatures}
\label{subsec-sig}

A signature determines the syntax of terms (or expressions) for
describing specifications (or modules of CafeOBJ).

\subsubsection{Sorts}
\label{sssec-sorts}

A {\bf sort} is a name for entities of the same kind/sort/type,
and interpreted as the set of the entities.
A {\bf subsort} relation can be declared among sorts
and is interpreted as a subset relation.

Let $sl$ = $s_1$\,{$\cdots$}\,$s_n$ be a list of names
$s_i$ separated by blanks.
The CafeOBJ code (i.e., a piece of text in CafeOBJ code)
\ttt{[\,$sl$\,]} declares each $s_i$ ($1 \le i \le n$)
is a sort.
Let $sl_1$ = $s1_{1}$\,{$\cdots$}\,$s1_{m}$ and
$sl_2$ = $s2_{1}$\,{$\cdots$}\,$s2_{n}$ be two lists of names.
The code \ttt{[\,$sl_1$\,<\,$sl_2$\,]} declares that
(i) each of $s1_{i}\,(1 \le i \le m)$ and $s2_{j}\,(1 \le j \le n)$
is a sort, and
(ii) $s1_{i}$ is a subsort of $s2_{j}$ for each pair $(i\,j)$
$(1 \le i \le m, 1 \le j \le n)$
(see {\bf Examples \ref{ex-senreg}, \ref{ex-pnatplus}}).
Let $sl_i\,(1 \le i \le n)$ be $n$ lists of names.
The code \ttt{[\,$sl_1$\,<\,$sl_2$\,<\,$\cdots$\,<\,$sl_n$\,]} 
$(2 \le n)$ is equivalent to 
``\ttt{[\,$sl_1$\,<\,$sl_2$\,]}\,
\ttt{[\,$sl_2$\,<\,$sl_3$\,]}\,$\cdots$%
\ttt{[\,$sl_{n-1}$\,<\,$sl_n$\,]}''.
The code
\ttt{[\,$sl_{11}$\,<\,$\cdots$\,<\,$sl_{1m}${,}%
\,$sl_{21}$\,<\,$\cdots$\,<\,$sl_{2n}$]}%
$(1 \le m, 1 \le n)$
is equivalent to 
``\ttt{[\,$sl_{11}$\,<\,$\cdots$\,<\,$sl_{1m}$\,]}\,%
\ttt{[\,$sl_{21}$\,<\,$\cdots$\,<\,$sl_{2n}$\,]}''.
Note that CafeOBJ code is written in the typewriter font thoughout
this paper.

The subsort declarations define the partially ordered set
(poset) $(S,\le)$ where $S$ is the set of sorts involved
and $\le$ is the minimum partial order relation
including the subsort relations declared.
If $s \le s'$, $s$ is called {\bf subsort} of $s'$ and
$s'$ is called {\bf supersort} of $s$.
Given a poset $(S,\le)$ of sorts, let $\equiv_{\le}$
denote the smallest equivalence relation including the partial order $\le$.
The quotient of $S$ by the equivalence relation $\equiv_{\le}$ is
denoted as $\hat{S} = S\!/\!\!\equiv_{\le}$, and an element of
$\hat{S}$ is called a {\bf connected component}.

\subsubsection{Operators}
\label{sssec-op}

An {\bf operator} (or function) $f$ is declared 
as ``\ttt{op $f$\;{:}\;$w$\;{-\cab}\;$s$ .}'' in CafeOBJ,
where $w \in S^*$ is its {\bf arity} and $s \in S$ is
its {\bf sort} (or {\bf co-arity}) of the operator
(see {\bf Example \ref{ex-pnatplus}}).  
The string $ws$ is called the {\bf rank} of the operator.
A {\bf constant} $f$ is an operator
whose arity is empty (i.e., \ttt{op\;$f$\;{:}\;{-\cab}\;{$s$}\;.}).
Let $F_{ws}$ denote the set of all operators of rank $ws$, then a
whole collection $F$ of operators can be represented as a family of
sets of operators sorted by (or indexed by) ranks as 
$F = \{ F_{ws} \}_{w\in S^*, s\in S}$.  
Note that ``\ttt{op $f$\;{:}\;$w$\;{-\cab}\;$s$ .}''
iff (if and only if) $f \in F_{ws}$.
Operators can be {\bf overloaded}, that is the same name can be used
for two operators of different ranks. In other words, $F_{ws}$ and
$F_{w's'}$ can have a common element for different $ws$ and $w's'$.

\subsubsection{Builtin \mttt{BOOL} and Predicates}
\label{sssec-bibpred}

CafeOBJ has a builtin module \ttt{BOOL} realizing
the Boolean algebra with the sort \ttt{Bool} and 
the following operators:
\ttt{true},
\texttt{false},
\ttt{{not}\us},
\texttt{\us{and}\us},
\texttt{\us{or}\us},
\texttt{\us{xor}\us},
\texttt{\us{implies}\us},
\texttt{\us{iff}\us},
where \ttt{\us} indicates the position of argument.
An operator $f$ with the rank ``$w$\,\texttt{Bool}''
(i.e., with the arity $w$ and the co-arity \texttt{Bool})
is called 
{\bf predicate}, which can be declared as
``\ttt{pred $f$\;{:}\;$w$\;.}'' or
``\ttt{pd $f$\;{:}\;$w$\;.}''.

There is an important builtin equality predicate
``\ttt{pred\,\us{=}\us\,{:}\,{*Cosmos*}\,{*Cos mos*}\,.}''
with equations
``\texttt{eq (CUX:*Cosmos* = CUX) = true .}'' and
``\texttt{eq (true = false) = false .}'',
where \texttt{*Cosmos*} is a builtin sort variable 
that can be interpreted as any sort {\it Srt}.
That is, 
``\texttt{pred \us{=}\us\;{:}\;{\it Srt} {\it Srt}\,.}''
and
``\texttt{eq (CUX:{\it Srt} = CUX) = true .}'' 
is declared for any sort {\it Srt}.

The builtin module \ttt{BOOL} is used 
in a significant way as follows.
\begin{itemize} \itemsep -2pt

\item The condition of an equation
(see {\it \ref{sssec-seneq}}\footnote{%
Numbers in the italic font delimited by ``{\it .}'' 
as {\it 2.3} or {\it 2.3.1} denote
a subsection or a subsubsection of this paper.}%
) is a Boolean term.

\item A property of interest on a specification 
is expressed with the predicates.

\end{itemize}

The semantics (i.e., model) of the builtin \ttt{BOOL} is
the standard propositional calculus and is preserved 
by models of every module (i.e., specification).

\subsubsection{Operator Attributes}
\label{ssec-opattri}

CafeOBJ supports the following three
attributes for a binary operator $f$.
\begin{itemize} \itemsep -2pt

\item {\bf associativity} \{{\tt assoc}\}:
${f(x,f(y,z))}\;{=}\;{f(f(x,y),z)}$

\item {\bf commutativity} \{{\tt comm}\}:
${f(x,y)}\;{=}\;{f(y,x)}$

\item {\bf identity} \{{\tt id:} $e$\}:
  ${f(x,e)}\;{=}\;{f(e,x)}\;{=}\;{x}$ 
  ~~($e$ is an identity constant)

\end{itemize}

If an identity is declared, the identity equations are supposed to be
declared and implemented as rewrite rules.
Associative/commutative declarations are implemented with 
the matching modulo associativity/commutativity and
no equations are incorporated 
(see {\it\ref{ssec-ddrdcex}, \ref{sssec-coex}}).

\subsubsection{Order-Sorted Signatures}
\label{sssec-oss}

An {\bf order-sorted signature} is defined by a tuple $(S,\le,F)$
(see {\it\ref{sssec-sorts}}, {\it\ref{sssec-op}}).

For making construction of symbolic presentations of models
(i.e., term algebras, see {\it\ref{subsubsec-tta}}) possible,
the following {\bf sensibility} condition is the most general
sufficient condition for avoiding ambiguity found until now
\citep{meseguer97mel}.  
An {\bf order-sorted signature} $(S,\le,F)$ is defined to be {\bf sensible}
iff $(w \equiv_{\le} w' \Ra s \equiv_{\le} s')$\footnote{%
Throughout the paper,
$\Ra$ means ``implies'', $\La$ means ``if'', 
$\Lra$ means ``if and only if'',
and $\lnot$ means ``not''.}
for any operator $f \in F_{ws} \cap F_{w's'} $,
where $w \equiv_{\le} w'$ means that 
(i) $w$ and $w'$ are of the same length
and (ii) any element of $w$ is in the same connected component 
with corresponding element of $w'$.  
Note that $[] \equiv_{\le} []$ for the empty arity $[]$.
A sensible signature guarantees a unique interpretation
of each $\Sg(X)$-term ({\it\ref{subsubsec-tta}}).

An order-sorted signature $(S,\le,F)$ is defined to be
{\bf regular}
iff there is a unique least element in the set
\{ $ws \mid f \in F_{ws} \land  w_0 \le w$ \} 
for each $f \in F_{w_1s_1}$ and each $w_0 \le w_1$.
A regular signature guarantees that each term has the unique
minimum parse ({\it\ref{subsubsec-tta}}).

Each order-sorted signature $(S,\le,F)$ is assumed to be
sensible and regular throughout this paper.
By using CafeOBJ's \ttt{check} commands,
sensibility and regularity of a signature
can be checked automatically.

\bex
\label{ex-senreg}
{(i)}\,%
\ttt{{\{}[Bool{\;}Nat]{\;}op{\;}0{\;}:{\;}->{\;}Bool{\;}.%
{\;}op{\;}0{\;}:{\;}-> {\;}Nat{\;}.\}} defines
a non sensible signature,
and \ttt{0} cannot be identified
with any entity of any sort.\\
{(ii)}\,%
\ttt{\{{\,}[Zero{\;}<{\;}Nat{\;}EvenInt]{\;}%
op{\;}2{\;}:{\;}->{\;}Nat{\;}.{\;}%
op{\;}2{\;}:{\;}->{\;}EvenInt{\,}.\}}
defines a sensible but non regular signature,
and \ttt{2} is identified with an entity
that belongs to \ttt{Nat} and \ttt{EvenInt}
but it has no minimum parse.\\
{(iii)}\,%
\ttt{{\{}\,[EvenNat < Nat EvenInt]
op{\;}2{\;}:{\;}->{\;}EvenNat{\;}.{\;}%
op{\;}2{\;}:{\;}->{\;}Nat{\;}. 
op{\;}2{\;}: ->{\;}EvenInt{\;}.\}}
%
%
defines a sensible and regular signature,
and \ttt{2} is identified with the entity that belongs to
\ttt{EvenNat}, \ttt{Nat}, and \ttt{EvenInt}
with the minimum parse \ttt{2:EvenNat}.
\hfill\makebox[0pt][r]{\owari}
\eex

\subsubsection{Constructor-Based Order-Sorted Signatures}
\label{sssec-cboss}

A {\bf constructor-based order-sorted signature} 
$(S,\leq,F,F^c)$
is an order-sorted signature with constructor declarations.
$F^c\,{\subseteq}\,F$ is a distinguished subfamily 
of sets of operators, called {\bf constructors},
such that $(S,\leq,F)$ and ($S$,\,$\leq$, $F^c$) are
order-sorted signatures.
Constructors represent data on which a
specification is constructed, and non-constructors 
represent functions over the data
(see {\bf Example \ref{ex-pnatplus}}).
An operator $f\,{\in}\,F_{ws}^c$ or
$f\,{\in}\,F_{ws'}^c$ with $s'\,{\le}\,s$
is called {\bf a constructor of sort $s$}.
A sort $s\in S$ is called {\bf constrained} if
there exists a constructor of $s$.
Let $S^c$ denote the set of all constrained sorts. 
An element of $S^l\,{\df}\,S-S^c$ is called a {\bf loose} sort.

\subsection{Models}
\label{ssec-model}

A CafeOBJ's signature $\varSigma$ determines the set $\Mod(\varSigma)$
of models called $\varSigma$-{\bf algebras}.  An algebra is a
collection of sorted sets with operators (functions) on the sets.

\subsubsection{$(S,\le,F)$-Algebras}
\label{sssec-sigalg}

Let $(S,\le,F)$ be an order-sorted signature.  
An $(S,\le,F)$-{\bf algebra}
(or an order-sorted algebra of signature $(S,\le,F)$)
$A$ interprets 
{(i)} each sort $s\in S$ as a {\bf carrier set} $A_s$,
{(ii)} each subsort relation $s < s'$ as an inclusion
$A_s \subseteq A_{s'}$, and
{(iii)} each operator $f \in F_{s_1 \dots s_n s}$ as
a function $A_f : A_{s_1} \times \dots \times A_{s_n} \ra A_s$
such that any two functions of the same name return the same value
if applied to the same argument, i.e.,
if $f:w\ra s$ and $f:w'\ra s'$ and $ws \equiv_{\le} w's'$ 
and $\overline{a}\in A_w\cap A_{w'}$ 
then $A_{f:w\ra s}(\overline{a})=A_{f:w'\ra s'}(\overline{a})$.

\subsubsection{ $(S,\le,F)$-Algebra Homomorphisms}

Let $A$, $B$ be $(S,{\le}, F)$-algebras.
An {\bf $(S,\le,F)$-algebra homomorphism}
$h:A \ra B$ is
an $S$-sorted family of functions between the carrier sets
of $A$ and $B$,  $\{ h_s : A_s \ra B_s \}_{s\in S}$, such that  
\begin{itemize} \itemsep -1pt

\item $h_s (A_f (a_1, \dots, a_n)) =
B_f (h_{s_1}(a_1), \dots, h_{s_n}(a_n))$ for all
$f \in F_{s_1 \dots s_n s}$ ($0 \le n$) and $a_i \in A_{s_i}$ for
$i\in \{1,\ldots,n\}$, and

\item if $s \equiv_{\le} s'$ and $a\in A_s\cap A_{s'}$ then
         $h_s(a)=h_{s'}(a)$.

\end{itemize}

\subsubsection{Terms and Term Algebras}
\label{subsubsec-tta}

Let $\vSg = (S,\le,F)$ be an order-sorted signature, and
$X = \{ X_s \}_{s\in S}$ be an $S$-sorted,
mutually disjoint countably infinite, sets of
variables called a {\bf $\vSg$-variable set}.
$\Sg(X)$-{\bf term} is defined inductively as
follows. 
\begin{itemize} \itemsep -1pt

\item Each constant $f\in F_s$ is a \Sg(X)-term of sort $s$.

\item Each variable $x\in X_s$ is a \Sg(X)-term of sort $s$.

\item $t$ is a term of sort $s'$ if $t$ is a term of sort $s$ and
$s < s'$.

\item $f(t_1,\dots,t_n)$ is a term of sort $s$ for each
operator $f\in F_{s_1 \dots s_n s}$ and terms  $t_i$ of sort
$s_i$ for $i\in\{1,2,\ldots,n\}$.

\end{itemize}

The $S$-sorted set of $\varSigma(X)$-terms is denoted as
$T_{\varSigma}(X) \df \{(T_{\varSigma}(X))_s | s \in S\}$. 
For the $S$-sorted empty sets of variables $\{\}$, 
a $\varSigma(\{\})$-term is called $\varSigma$-term
or $\varSigma$-{\bf ground term}.
$T_{\varSigma} \df T_{\varSigma}(\{\})$ denotes the $S$-sorted set of
$\varSigma$-ground terms.
$T_{\varSigma}(X)$ and $T_{\varSigma}$ can be organized as \Sg-algebras
by using the above stated inductive definition of terms.
$T_{\varSigma}$ as a ${\varSigma}$-algebra
is the {\bf initial algebra} of \Sg-algebras
as stated as follows.
\bfact \citep{goguenM92OSA,goguen06TPA}
For any \Sg-algebras $A$ there exists a unique
\Sg-algebra homomorphism $T_\varSigma\,{\ra}\,A$.
\hfill\makebox[0pt][r]{\owari}
\efact

The {\bf least sort} \ttt{ls}($t$) of a term $t\in T_{\varSigma}(X)$
is defined to be the least sort $s$ such that $t \in (T_{\varSigma}(X))_s$.
A regular signature \Sg\, ({\it\ref{sssec-oss}})
guarantees that each term $t\in T_{\varSigma}(X)$
has the unique minimum parse with
the least sort (\citep{goguenM92OSA,goguen06TPA}).

\subsubsection{Valuations and Term Interpretation}
\label{subsubsec-vti}

A {\bf valuation} (substitution, instantiation) assigns values
to variables, that is, instantiates each variable with a value
in a given model.  Let $\Sg$ be $(S,\le,F)$-signature.  Given
a \Sg-model $A$ and an $S$-sorted set $X$ of variables, a
valuation $\theta : X \ra A$ is an $S$-sorted family of maps
$\{ \theta_s : X_s \ra A_s \}_{s\in S}$.
Each \Sg(X)-term $t\,{\in}\,T_{\vSg}(X)$ can be {\bf interpreted as a value
$\theta(t)$} in the model $A$ as follows.
That is, $\theta$ can extend to 
$\theta : T_{\varSigma}(X) \rightarrow A$.

\begin{itemize} \itemsep -1pt

\item  $\theta(t) = A_f$ if $t$ is a constant $f$.

\item  $\theta(t) = \theta(x)$ if $t$ is a variable $x$.

\item  $\theta(t) = A_f (\theta(t_1),\dots,\theta(t_n))$ if $t$ is of the form
$f(t_1,\dots,t_n)$ for some $f\in F_{s_1 \dots s_n s}$
and terms $t_i\,{\in}\,(T_{\varSigma}(X))_{s_i}$ ($1 \le i \le n$).

\end{itemize}

\noindent
Note that $\theta$ can be seen as a ${\varSigma(X)}$-algebra
homomorphism $\theta : T_{\varSigma(X)} \rightarrow A$
by interpreting ${\varSigma(X)}$ as a signature
with fresh constants $X$ (i.e., $\vSg{\cup}X$).

\subsubsection{ $(S,\leq,F,F^c)$-Algebras}
\label{sssec-cbosa}

An $(S,\leq,F,F^c)$-{\bf algebra}
$A$ is an $(S,\leq,F)$-algebra 
with the carrier sets for the constrained sorts (\ref{sssec-cboss})
consisting of interpretations of terms formed with constructors and elements of
loose sorts (\ref{sssec-cboss}).  
That is, the following holds for $\varSigma^c = (S,\leq,F^c)$,
the set $S^l$ of the loose sorts,
the set $S^c$ of the constrained sorts.
There exists an $S^l$-sorted set of variables 
$Y$ = $\{Y_s\}_{s \in S^l}$ and a valuation
$\theta : Y \rightarrow A$ 
= $\{\theta_s:Y_s\rightarrow A_s\}_{s \in S^l}$
such that 
$\theta : T_{\varSigma^c}(Y) \rightarrow A$ is 
surjective 
for each constrained sort $s {\in} S^c$
(i.e., 
$\theta_s\,{:}\,(T_{\varSigma^c}(Y))_s {\rightarrow} A_s$ 
is a surjection for each $s\in S^c$).

\bex
\label{ex-seqsg}
The CafeOBJ code\\ 
\hspace*{2em}
\ttt{\ob[Elt\,{<}\,Seq] op\,{\us \us}\,{:}\,Seq Seq\,{->}\,Seq \ob
  constr assoc\cb .\cb}\\ 
defines the constructor-based order-sorted signature
(i.e., $(S,\leq,F,F^c)$)\\
\hspace*{2em}\ttt{SEQsg} = (\{\ttt{Elt},\ttt{Seq}\},\{(\ttt{Elt},\ttt{Seq})\},%
\{\{\ttt{\us \us}\}\},\{\{\ttt{\us \us}\}\}).\\
Then \ttt{NATsq} (sequences of natural numbers)
defined by\\
\hspace*{2em}$\ttt{NATsq}_{\ttt{Elt}} = \{1,2,\cdots \}$,\\
\hspace*{2em}$\ttt{NATsq}_{\ttt{Seq}} =
\{n_1 n_2 \cdots n_k \mid
   (1 \le k),\,
   n_i\,{\in}\,\ttt{NATsq}_{\ttt{Elt}} (1 \le i \le k) \}$\\
is a \ttt{SEQsg}-algebra.  
It can be seen as follows.
Let $Y = \{Y_{\ttt{Elt}}\}$, $Y_{\ttt{Elt}} = \{y_1,$  $y_2,$ $\cdots\}$ then\\
\hspace*{2em}$(T_{\ttt{SEQsg}^c}(Y))_{\ttt{Seq}} =
\{z_1 z_2 \cdots z_k \mid
   (1 \le k),\,
   z_i\,{\in}\,Y_{\ttt{Elt}} (1 \le i \le k) \}$\\
and
``$\theta_{\ttt{Seq}}:(T_{\ttt{SQEsg}^c}(Y))_{\ttt{Seq}}\rightarrow 
\ttt{NATsq}_{\ttt{Seq}}$''
is a surjection by taking
``$\theta_{\ttt{Elt}}\,{:}\,Y_{\ttt{Elt}}\rightarrow
\ttt{NATsq}_{\ttt{Elt}}$'' 
with $\theta_{\ttt{Elt}}(y_i) = i$ ($1 \le i$).
\hfill\makebox[0pt][r]{\owari}
\eex

\subsection{Sentences and Specifications}
\label{ssec-sntspc}

\subsubsection{$\varSigma$-Equations}
\label{sssec-seneq}

A $\vSg$-equation is a main sentence of CafeOBJ
and is declared as follows;
\ttt{cq} stands for conditional equation
and \ttt{ceq} can be used instead.\\
\hspace*{2em}\ttt{cq $l(X)$ = $r(X)$ {\rm if} $c(X)$ .}\\
\noindent
$X$ is a finite $\vSg$-variable set ({\it\ref{subsubsec-tta}}),
$c(X) {\in} (T_{\varSigma}(X))_{\mathtt{Bool}}$ and
$l(X),r(X) {\in} (T_{\varSigma}(X))_s$
for some sort $s{\in}S$.
If $c(X)$ is \ttt{true} (a builtin constant of the sort \ttt{Bool})
the equation is unconditional and written as 
``\ttt{eq $l(X)$ = $r(X)$ .}''.

Labels $lb_1\,\cdots\,lb_n$ can be put to an equation by declaring 
them just after the keyword \ttt{eq} or \ttt{cq}
as ``\ttt{eq[$lb_1\,\cdots\,lb_n$]:\;$l(X)$\,{=}\,$r(X)$\,{.}}''.
A label is the builtin attribute \ttt{:nonexec} 
or an ordinary label
(see {\bf Example \ref{ex-nonexec}}).

A Boolean term $c(X)$ is called conjunctive if
it can be expressed using 
the builtin equality predicate \ttt{\us{=}\us} and
the builtin Boolean predicate \ttt{\us{and}\us} ({\it\ref{sssec-bibpred}})
as ``($l_1(X)$ \ttt{=} $r_1(X)$) \ttt{and} $\cdots$ \texttt{and}
($l_n(X)$ \ttt{=} $r_n(X)$)'' ($1 \le n$) 
and each of $l_i(X)$ and $r_i(X)$ ($1 \le i \le n$)
(i) is the Boolean constant \ttt{true} ({\it\ref{sssec-bibpred}}), or
(ii) does not contain the builtin Boolean predicates ({\it\ref{sssec-bibpred}})
or the builtin equality predicate \ttt{\us{=}\us}.
An equation is called {\bf conjunctive} if its condition is \ttt{true} or
conjunctive.

An equation is called {\bf sort decreasing} iff
\ttt{ls}($\theta(r(X))$) $\le$ \ttt{ls}($\theta(l(X))$)
(see {\it\ref{subsubsec-tta}} for \ttt{ls}($t$))
for any substitution ({\it\ref{subsubsec-vti}}) 
$\theta : X \ra T_{\vSg}(X)$ such that
$\theta(x) = x'$ ($x' \in X$) and
\ttt{ls}($x'$) $\le$ \ttt{ls}($x$).

A set of equations $E$ is conjunctive or sort decreasing if
each element of $E$ is so.

The conjunctiveness and the sort decreasingness can be checked automatically,
but not implemented in the current CafeOBJ system.

\subsubsection{Equational Specifications}
\label{sssec-eqspec}

An {\bf equational specification} $\mathit{SP}$ is defined to be a
pair of signature $\varSigma$ and a set of \Sg-equations $E$, and
denoted as $\mathit{SP} = (\Sg,E)$.

\subsubsection{Structured Specifications}
\label{sssec-strspec}

A specification $\mbox{\it SP}$ can be defined 
by importing other already defined specification $\mbox{\it SP}_a$
and declaring newly added signature $\varSigma$
and equations $E$,
and denoted as $\mbox{\it SP} = (\mbox{\it SP}_{a};\varSigma,E)$.
If $\mbox{\it SP}_a = (\varSigma_a,E_a)$
then $\mbox{\it SP} = (\varSigma_a{\cup}\varSigma,E_a{\cup}{E})$.

It is a casual definition of structured specifications.
More formal treatments can be found in
\citep{rd-kf1998,futatsugiGo12}.

\subsection{Satisfaction}

\subsubsection{Equation Satisfaction $A\,{\vDash_{\varSigma}}\,e$\;}
\label{subsubsec-algst}

A $\vSg$-equation
$e =$
``\ttt{cq $l(X)$ = $r(X)$ if $c(X)$ .}''
is {\bf satisfied} by 
a \Sg-algebra $A$, in symbols
$A\;{\vDash_{\varSigma}}\;e$, %
iff
$\theta(l(X)) = \theta(r(X))$ whenever $\theta(c(x)) = \mttt{true}$ 
for any valuation $\theta:X \ra A$.  
That is, an equation is satisfied by an algebra iff any possible
way to assign values to variables evaluates both sides of the equation
to the same value whenever the condition is evaluated to \ttt{true}.
${\varSigma}$ of ${\vDash_{\varSigma}}$ can be dropped if it is clear
from the context.
The following facts are implied from the fact that the
builtin \ttt{BOOL} realizes the propositional calculus with equality.

\bfact
\label{fact-algsatis}
Let \mttt{\us{=}\us} and \mttt{\us{implies}\us} be
Boolean operators of the CafeOBJ's builtin module \mttt{BOOL},
and $(l(X)\,\mttt{=}\,r(X))$,
$(c(X)\,\mttt{implies}\,(l(X)\,\mttt{=}\,r(X)))$
$\in (T_\vSg(X))_{\mathtt{Bool}}$.\\
\hspace*{1.3em}%
(i) $A\,{\vDash_{\varSigma}}$\,``\mttt{eq $l(X)$ = $r(X)$ .}'' $\Lra$
$A\,{\vDash_{\vSg}}$\,%
``\mttt{eq}\,$(l(X)\,\mttt{=}\,r(X))$ \mttt{= true .}''\\
%
\hspace*{1em}%
(ii) %
$A\,{\vDash_{\varSigma}}$\,``\mttt{cq $l(X)$ = $r(X)$ if $c(X)$ .}''
$\Lra$\\
\hspace*{2.7em}$A\,{\vDash_{\vSg}}$\,%
``$\mttt{eq}\,(c(X)\,\mttt{implies}\,(l(X)\,\mttt{=}\,r(X)))$
\mttt{=\;true\,.}''
\hfill\makebox[0pt][r]{\owari}
\efact

\subsubsection{$\Mod(\mathit{SP})$ and Loose/Tight Denotations}
\label{sssec-modsp-loti}

The set {\bf $\Mod(\mathit{SP})$} 
of models of a specification $\mathit{SP}\,{=}\,(\Sg,E)$
is defined as
$\Mod(\mathit{SP}) \df
\{\,{A}\,{\in}\,\Mod(\varSigma)\,{|}\,%
\forall\,{e}\,{\in}\,{E}\,(A\,{\vDash_{\varSigma}}\,{e})\,\}$.
An element of $\Mod(\mathit{SP})$ is called 
an {\bf $\mathit{SP}$-algebra} (or {\bf $\mathit{SP}$-model}).

A specification $\mathit{SP}$
in CafeOBJ (i.e., a module)
can have {\bf loose} or {\bf tight} denotation
that are indicated by two keywords
\ttt{mod*} or \ttt{mod!} respectively
(see CafeOBJ codes for the module \ttt{TRIV} and \ttt{LIST} 
in {\it \ref{ssec-spcsys}}-\ttt{02:-07:}\footnote{%
\label{fn-coderef}
Refers to the CafeOBJ code with line numbers from \ttt{02:} to \ttt{07:}
in the Subsection {\it \ref{ssec-spcsys}}.%
}).
The loose denotation intends to denote
$\Mod(\mathit{SP})$, 
and the tight denotation intends to denote 
the initial or free algebra of $\Mod(\mathit{SP})$
assuming its existence.
That is, tight denotation intends to specify a specific data
type, and loose denotation intends to specify a class of data
types with common constraints.

\subsubsection{$\mathit{SP} \vDash p(Y)$ and Theorem of Constants}
\label{sssec-sppx-thct}

Let $SP = (\vSg,E)$ be a specification,
and $Y$ be a $\vSg$-variable set ({\it\ref{subsubsec-tta}}).
A Boolean term 
$p(Y) {\in} (T_{\varSigma}(Y))_{\mathtt{Bool}}$,
which is supposed to express a property
of interest for a specification $\mathit{SP}$,
is defined to be {\bf satisfied} by
$\mathit{SP}$ (in symbols $\mathit{SP} \vDash p(Y)$)
iff 
$\forall {A}\,{\in}\,\Mod(\mathit{SP})$%
($A\;{\vDash_{\varSigma}}$
``\ttt{eq{\;}$p(Y)${\;}{=}{\;}true{\;}.}''). 

Let $\vSg\,{\cup}\,Y$ be a signature
obtained by adding $Y$ to $\vSg$
as fresh constants, 
and $p^Y{\in}\,(T_{\varSigma{\cup}Y})_{\mathtt{Bool}}$ 
be the term obtained from $p(Y)$ by considering each $y\,{\in}\,Y$ 
as a fresh constant.
Let $A$ be a $\varSigma$-algebra,
$\theta : Y \ra A$ be a valuation,
then there is a one to one correspondence between
a pair $($A$,\theta)$ and
a $(\varSigma\,{\cup}\,Y)$-algebra $A'$.
Hence the following Fact 
(Theorem 3.3.11 of \citep{goguen06TPA})
which will be used to obtain
{\bf Facts \ref{fact-stind}, \ref{fact-wfishe}}
and play an important role in proof scores.

\bfact {\it (Theorem of Constants)}
\label{fact-thcst}
\hspace*{1em}
$(\varSigma,E)\,{\vDash}\,p(Y)\;{\Lra}\;
(\varSigma\,{\cup}\,Y,E)\,{\vDash}\,p^Y$%
\hfill\makebox[0pt][r]{\owari}
\efact

The following facts are implied from the definition of
$\vDash$ and {\bf Fact \ref{fact-thcst}}.

\bfact 
\label{fact-msatp}
Let $p,q{\in}(T_{\vSg})_{\mathtt{Bool}}$,
$p'(Y),q'(Y){\in}(T_{\varSigma}(Y))_{\mathtt{Bool}}$,
and 
$p''(Y){\in}(T_{\vSg}(Y))_{\mathtt{Bool}}$
be the term gotten 
by substituting non overlapping strict
subterms of $p$ with
variables in $Y$.\\
\hspace*{4.3em}(i)\:\:$(\varSigma,E)\,{\vDash}\,%
(p'(Y)\,\mttt{and}\,q'(Y))\;{\Lra}\;\\
\hspace*{5.8em}(\varSigma,E)\;{\vDash}\;p'(Y)\:{\land}\:
(\varSigma,E)\;{\vDash}\;q'(Y)$\\
\hspace*{4.1em}(ii)~$(\varSigma,E)\,{\vDash}\,(p'(Y)\,\mttt{implies}\,q'(Y))\;{\Ra}\\
\hspace*{5.8em}(\varSigma,E{\cup}%
\{\mttt{eq}\:{p'(Y)}\:\;\mttt{=}\:\;\mttt{true\:.}\})\;{\vDash}\;q'(Y)$\\
\hspace*{3.9em}(iii)~$(\varSigma,E)\,{\vDash}\,(p\,\mttt{implies}\,q)\;{\Lra}\\
\hspace*{5.9em}(\varSigma,E{\cup}%
\{\mttt{eq}\:{p}\:\;\mttt{=}\:\;\mttt{true\:.}\})\;{\vDash}\;q{\:\:}{\La}\\
\hspace*{5.9em}(\varSigma,E{\cup}\{\mttt{eq}\:p''(Y)\:\;\mttt{=}\:\;%
  \mttt{true\:.}\})\;{\vDash}\;q$\\
\hspace*{3.9em}(iv)\;%
$((\varSigma,E)\,{\vDash}\,p'(Y)\,\Ra\,
  (\varSigma,E{\cup}E')\,{\vDash}\,q'(Y))\,\La\,\\
\hspace*{5.7em}(\varSigma,E{\cup}E'{\cup}%
\{\mttt{eq}\:{p'(Y)}\:\;\mttt{=}\:\;\mttt{true\:.}\})\;{\vDash}\;q'(Y)$
\hfill\makebox[0pt][r]{\owari}
\efact


\subsection{Quotient Algebra $T_{\varSigma,E}$ and Initial Algebras}
\label{subsec-initalg}

A {\bf congruence} $\equiv$ on an $(S,\le,F)$-algebra $A$ is 
an $S$-sorted equivalence on $A$
(i.e., $\{\,\equiv_s \subseteq A_s{\times}A_s \mid  s\in S\,\}$) such that 
(i)
if $a_i \equiv_{s_i} a'_i$ ($i\in \{1,\ldots,n\}$) then
$A_{f}(a_1,\dots,a_n) \equiv_s A_{f}(a'_1,\dots,a'_n)$
for each $f\,{\in}\,{F_{s_1 \dots s_n s}}$, and
(ii) if $s \le s'$ ($s, s' \in S$) and $a, a' \in A_s$ then
$a \equiv_s a'$ iff $a \equiv_{s'} a'$.

For an equational specification $SP = (\vSg,E)$,
$\varSigma = (S, {\le}, F)$,
a {\bf quotient algebra} $T_{\varSigma,E}$ 
is constructed as follows:
\begin{itemize} \itemsep -1pt

\item for each $s\,{\in}\,S$ let $(T_{\varSigma,E})_s$ be the set of
equivalence classes 
of \Sg-terms in $(T_{\varSigma})_s$ under the congruence
${\equiv_s}\!\!\!^E$ defined as
($t\,{\equiv_s}\!\!\!^E\,t'$ 
iff $(\varSigma,E)\,{\vDash}\,{{t}\,\mttt{=}\,{t'}}$).
That is, $(T_{\varSigma,E})_s =
\{\,[t]_{{\equiv_s}\!\!\!^E}\,\mid\,t\,{\in}\,(T_{\varSigma})_s\,\}$.

\item each operator $f\,{\in}\,F_{s_1 \dots s_n s}$ is
interpreted as\\
\hspace*{2em}\((T_{\varSigma,E})_f ([t_{1}]_{{\equiv_{s_1}}\!\!\!\!\!^E},
\dots,[t_{n}]_{{\equiv_{s_n}}\!\!\!\!\!^E})
= [f(t_1,\dots,t_n)]_{{\equiv_s}\!\!\!^E}\)\\
for all $t_i\,{\in}\,(T_{\varSigma})_{s_i}$
($i\in \{1,\ldots,n\}$) by using the congruence property
of $\equiv^E$ on $T_{\varSigma}$. 

\end{itemize}

\subsubsection{The Initial Algebra of $\Mod((S,\le,F),E)$}
\label{subsubsec-iaosa}

\bfact \citep{goguenM92OSA,goguen06TPA} 
Let $\mathit{SP}\,{=}\,(\vSg,E)$, $\vSg = (S,\le,F)$ with no constructors.
If $E$ is conjunctive, 
for any algebra $A\,{\in}\,\Mod(\mathit{SP})$
there exists a unique $\varSigma$-algebra homomorphism
$T_{\varSigma,E} \ra A$.
\hfill\makebox[0pt][r]{\owari}
\efact

\subsubsection{The Initial Algebra of $\Mod((S,\le,F,F^c),E)$}
\label{sssec-iacbosa}

Let $SP = ((S,\le,F,F^c),E)$ be
a constructor-based order-sorted specification and
$S^c$ be the set of constrained sorts and
$S^l$ be the set of loose sorts ({\it\ref{sssec-cboss}}), and
$F^{S^c} \df \{f:w\ra s \,|\, f\in F, s\in S^c \}$, and 
$\varSigma^{S^c} \df (S,\leq,F^{S^c})$, and
$\varSigma^c \df {(S,\leq,F^c)}$, and
$Y$ be any $S^l$ sorted set of variables.
A specification $\mathit{SP}$ is defined to be {\bf sufficiently complete}
iff for any term $t\,{\in}\,T_{\varSigma^{S^c}}(Y)$ 
there exists a term $t'\,{\in}\,T_{\varSigma^c}(Y)$
such that $\mathit{SP}\,{\vDash}\,{t}\,\mttt{=}\,{t'}$.

\bfact \citep{gainaF15,futatsugiGo12}
Let $\mathit{SP}\,{=}\,(\vSg,E)$, $\vSg = (S,\le,F,F^c)$
be a constructor-based order-sorted specification.
If the specification $\mathit{SP}$ is sufficiently complete
and $E$ is conjunctive,
for any algebra $A\,{\in}\,\Mod(\mathit{SP})$
there exists a unique $\varSigma$-algebra homomorphism
$T_{\varSigma,E} \ra A$.
\hfill\makebox[0pt][r]{\owari}
\efact

\subsection{Deduction, Reduction, CafeOBJ Execution}
\label{ssec-ddrdcex}

Let $\mathit{SP} = (\Sg,E)$, $\Sg {=} (S,\le,F)$
be an equational specification.
If there are operators in $F$ that have associative and/or
commutative attributes then let $\mathit{AC}$ denote the set of
equations that define associativity or commutativity (AC) 
of the operators ({\it\ref{ssec-opattri}}).
Let $=_{\!\mtnit{AC}}$ denote the smallest $\Sg$-congruence 
({\it\ref{subsec-initalg}}) on the $\Sg$-term algebra 
$T_{\varSigma}$ ({\it\ref{subsubsec-tta}}) 
that includes the equivalence relation defined by
the equations in $\mathit{AC}$.
$t =_{\!\mtnit{AC}} t'$ can be decided with the AC matching algorithm.
If there are no AC operators in $F$, 
$\mathit{AC}$ is empty and $=_{\!\mtnit{AC}}$ is
the usual equality $=$ on $T_{\varSigma}$.

Let 
\sm{$\Box_s$} be a special fresh constant of a sort $s{\in}S$,
and let $t_c[\sm{\Box_s}]$ denote a ground term composed of
operators in $F{\cup}\{\sm{\Box_s}\}$
that includes one \sm{$\Box_s$}.
The term $t_c[\sm{\Box_s}]$ is called
an $F{\cup}\{\sm{\Box_s}\}$-term or a context,
and $t_c[t]$ denotes the $\varSigma$-term 
obtained by substituting \sm{$\Box_s$} 
with a term $t \in (T_\vSg)_s$.
Let $p \in (T_\vSg)_{\mathtt{Bool}}$
(i.e., a Boolean ground $\vSg$-term).

\subsubsection{Equational Deduction}
\label{sssec-eqded}

An equational specification 
$\mathit{SP}\,{=}\,(\vSg,E\,{\cup}\mathit{AC})$ defines 
(i) {\bf one-step deduction relation}
${\dedu{E}{AC}}\,{\subseteq}\,T_\vSg\,{\times}\,T_\vSg$ and
(ii) its reflexive and transitive closure $\dedus{E}{AC}$
called {\bf deduction relation}.
Inference (proof, deduction) rules for $\dedu{E}{AC}$ and $\dedus{E}{AC}$
are given as follows for $t_l, t_r, t_m\,{\in}\,T_\vSg$.
\begin{enumerate}[(a)]

\item $(t_l = t_r) \Ra t_l {\dedus{E}{AC}}\, t_r$.
That is, $t_l\,{\dedus{E}{AC}}\,t_r$ is deduced if $(t_l = t_r)$.

\item $t_l\,{\dedu{E}{AC}}\,t_m \land t_m\,{\dedus{E}{AC}}\,t_r \Ra 
t_l\,{\dedus{E}{AC}}\,t_r$.
That is, 
$t_l\,{\dedus{E}{AC}}\,t_r$ is deduced 
if $t_l\,{\dedu{E}{AC}}\,t_m$ and $t_m\,{\dedus{E}{AC}}\,t_r$
are deduced for some $t_m$.
The following (c1), (c2) are interpreted in a similar way.

\item For an equation
``\ttt{cq{\,}$l(X)${\,}={\,}$r(X)${\,}if{\,}$c(X)${\,}.}''\,%
${\in}\,E\,{\cup}\mathit{AC}$,
a valuation $\theta:X \ra T_\vSg$,
and a context $t_c[\sm{\Box_s}]$,

\begin{enumerate}[(c1)]

\item $\theta(c(X))\,{\dedus{E}{AC}}\,\mttt{true} \Ra
t_c[\theta(l(X))]\,{\dedu{E}{AC}}\,t_c[\theta(r(X))]$, 

\item $\theta(c(X))\,{\dedus{E}{AC}}\,\mttt{true}$
$\Ra$ $t_c[\theta(r(X))]\,{\dedu{E}{AC}}\,t_c[\theta(l(X))]$.

\end{enumerate}  

\end{enumerate}  

Let $\mathit{SP}\,{\tse}\,p$ denote
$p\,{\dedus{E}{AC}}\,\mttt{true}$.

\bfact\label{fact-edemdl} \citep{goguen06TPA}%
\hspace*{2em} (i) \hspace*{1em}
$\mathit{SP} {\tse} p\;{\Ra}\;\mathit{SP}\,{\vDash}\,p$\\
\hspace*{7.1em} (ii) \hspace*{1em}
If $E$ is conjunctive, 
$\mathit{SP} {\tse} p\;{\Lra}\;\mathit{SP}\,{\vDash}\,p$\,.
\hfill\makebox[0pt][r]{\owari}
\efact

Equational deduction can be done also on the equivalence
classes defined by the congruence relation $=_{\!\mtnit{AC}}$,
and an equational specification 
$\mathit{SP}\,{=}\,(\vSg,E)$ defines 
{\bf one-step deduction relation modulo AC} 
${\ded{E/AC}}\,{\subseteq}\,T_\vSg{\times}T_\vSg$
and
{\bf deduction relation modulo AC}
$\dedstr{E/AC}\,{\subseteq}\,T_\vSg{\times}T_\vSg$.
Inference rules for $\ded{E/AC}$ and
$\dedstr{E/AC}$ are given as follows for $t_l, t_r, t_m\,{\in}\,T_\vSg$.
\begin{enumerate}[(a)]

\item $(t_l =_{\!\mtnit{AC}} t_r) \Ra t_l {\dedstr{E/AC}}\, t_r$.

\item $t_l\,{\ded{E/AC}}\,t_m \land t_m\,{\dedstr{E/AC}}\,t_r \Ra 
t_l\,{\dedstr{E/AC}}\,t_r$.

\item For an equation
``\ttt{cq{\,}$l(X)${\,}={\,}$r(X)${\,}if{\,}$c(X)${\,}.}''\,${\in}\,E$,
a valuation $\theta:X \ra T_\vSg$,
and a context $t_c[\sm{\Box_s}]$,
if $t_l =_{\!\mtnit{AC}} t_c[\theta(l(X))]$ $\land$
$t_r =_{\!\mtnit{AC}} t_c[\theta(r(X))]$, 

\begin{enumerate}[(c1)]

\item $\theta(c(X))\,{\dedstr{E/AC}}\,\mttt{true} \Ra t_l\;{\ded{E/AC}}\;t_r$, 

\item $\theta(c(X))\,{\dedstr{E/AC}}\,\mttt{true} \Ra t_r\;{\ded{E/AC}}\;t_l$.

\end{enumerate}  

\end{enumerate}  

The following is proved via a similar argument as
Lemma 4 of \citep{Meseguer17}.\\
\hspace*{2em}{\bf (E/AC)}\hspace*{4em} $\forall\,t_l, t_r{\in}\,T_\vSg$\,%
($t_l\,{\dedus{E}{AC}}\,t_r$ $\Lra$ $t_l\,{\dedstr{E/AC}}\,t_r$)

\subsubsection{Rewriting Reduction}
\label{sssec-rwrd}

For a term $t\,{\in}\,T_\vSg(X)$ let var($t$)\,$\,{\subseteq}\,X$
denote the set of variables in $t$.
An equation
``\ttt{cq[$lbs$]:\,$l(X)${\;}={\;}$r(X)${\;}if{\;}$c(X)${\;}.}''\,%
is called a {\bf rewrite rule}
if (var($r(X)$)\,{$\cup$} var($c(X)$)\;{$\subseteq$}\;var($l(X)$))
and \ttt{:nonexec} is not in $lbs$.
If $lbs$ is empty \ttt{[$lbs$]:} can be omitted.
If each equation $e\,{\in}\,E$ is a rewrite rule,
$\mathit{SP}$ is called a {\bf reduction specification}
(or {\bf term rewriting system (TRS)}) and defines
{\bf one-step reduction relation modulo AC}
${\rdc{E/AC}}$ ${\subseteq}\,T_\vSg{\times}T_\vSg$
and
{\bf reduction relation modulo AC}
${\rdcstr{E/AC}}$ ${\subseteq}\,T_\vSg{\times}T_\vSg$.
Inference rules for $\rdc{E/AC}$ and ${\rdcstr{E/AC}}$ are the same as
for $\ded{E/AC}$ and $\dedstr{E/AC}$ in {\it\ref{sssec-eqded}},
by replacing $\ded{E/AC}$ with $\rdc{E/AC}$ and $\dedstr{E/AC}$ with
$\rdcstr{E/AC}$,
and delete the rule (c2)
because rewriting is only from left to right direction.

Let 
$\mathit{SP}\,{\tsr}\,p$ denote
$p\,{\rdcstr{E/AC}}\,\mttt{true}$,
{\it\ref{sssec-eqded}}-{\bf (E/AC)} implies the following.

\bfact\label{fact-rwrede} 
\hspace*{6em}$\mathit{SP} {\tsr} p\;{\Ra}\;\mathit{SP} {\tse} p$%
\hfill\makebox[0pt][r]{\owari}
\efact

\bex \label{ex-nonexec}

An equation with the \texttt{:nonexec} attribute,
which is not used for the application of the rule (c1),
can be declared as\\
\hspace*{2em}\ttt{cq[trans :nonexec]: X = Z if (X = Y and Y = Z) .}\\
where \ttt{X,Y,Z} are variables of some sort.
This equation is logically valid because of the transitivity of equality,
but is not a rewrite rule because
[var(\ttt{Z})\,{$\cup$}\,var(\ttt{(X = Y and Y = Z)})
$\subseteq$ var(\ttt{X})] does not hold.
This equation can become an executable rewrite rule
by instantiating its variables with ground terms,
and deleting the \texttt{:nonexec} attribute,
as the equation 
{\it \ref{sssec-prswfisch}}-\ttt{11:-13:}
is instantiated with the \ttt{:init} command
declared at 
{\it \ref{sssec-prswfisch}}-\ttt{21:-22:}
and applied at {\it \ref{sssec-prswfisch}}-\ttt{23:}.
\hfill\makebox[0pt][r]{\owari}
\eex

\subsubsection{CafeOBJ Execution}
\label{sssec-coex}

Let $\mathit{SP}\,{=}\,(\vSg,E)$ be
a reduction specification (or TRS), then
\textit{SP} defines
{\bf one-step weak reduction relation modulo AC}
${\rdc{E,AC}}$ ${\subseteq}\,T_\vSg{\times}T_\vSg$
and
{\bf weak reduction relation modulo AC}
${\rdcstr{E,AC}}$ ${\subseteq}\,T_\vSg{\times}T_\vSg$.
Inference rules for $\rdc{E,AC}$ and $\rdcstr{E,AC}$ are given as follows
for $t_l, t_r, t_m, t_p\,{\in}\,T_\vSg$.
\begin{enumerate}

\item[(a)] $(t_l =_{\!\mtnit{AC}} t_r) \Ra t_l {\rdcstr{E,AC}}\, t_r$.

\item[(b)] $t_l\,{\rdc{E,AC}}\,t_m \land t_m\,{\rdcstr{E,AC}}\,t_r \Ra 
t_l\,{\rdcstr{E,AC}}\,t_r$.

\item[(c1)] For a rewrite rule
``\ttt{cq{\,}$l(X)${\,}={\,}$r(X)${\,}if{\,}$c(X)${\,}.}''\,${\in}\,E$,
a valuation $\theta:X \ra T_\vSg$,
and a context $t_c[\sm{\Box_s}]$,
if $t_p =_{\!\mtnit{AC}} \theta(l(X))$,\\
\hspace*{4em}$\theta(c(X))\,{\rdcstr{E,AC}}\,\mttt{true} \Ra
t_c[t_p]\;{\rdc{E,AC}}\;t_c[\theta(r(X))]$.

\end{enumerate}  

The check with $=_{\!\mtnit{AC}}$ in (c1) via AC-matching
is restricted to ``$t_p\,{=_{\!\mtnit{AC}}}\,\theta(l(X))$'', and
$\rdc{E,AC}, \rdcstr{E,AC}$ can be implemented much more
efficiently than $\rdc{E/AC}, \rdcstr{E/AC}$ of {\it \ref{sssec-rwrd}}.

(a) and (b) tell that $t_0\,{\rdcstr{E,AC}}\,t_n$ iff 
there exists a sequence of one-step weak reductions
$t_0\,{\rdc{E,AC}}\,t_1\,\cdots\,t_{n-1}\,{\rdc{E,AC}}\,t_n$
or $t_0\,{=_{\!\mtnit{AC}}}\,t_n$.

(c1) tells the way to compute $t_{r}$ from $t_l$ for
establishing a one-step weak reduction $t_l\,{\rdc{E,AC}}$ $t_{r}$
if a context $t_c[\sm{\Box_s}]$ and a rewrite rule
``\ttt{cq{\,}$l(X)${\,}={\,}$r(X)${\,}if $c(X)${\,}.}''${\in}E$
are chosen.
If $b = \theta(c(X))$ then (c1) declares that
$b\,{\rdcstr{E,AC}}\,\mttt{true}$ is a prerequisite
for establishing $t_l\,{\rdc{E,AC}}$ $t_{r}$.
Let $\{b\}$ denote the necessary computation
for checking $b\,{\rdcstr{E,AC}}\,\mttt{true}$,
then the computation step for establishing $t_l\,{\rdc{E,AC}}\,t_{r}$
is written as $t_l\{b\}t_{r}$ 
and called a {\bf one-step conditional reduction} ({\bf oc-red} for
short). 

For a Boolean ground term $b_0\,{\in}\,(T_\vSg)_{\ttt{Bool}}$,
let $[b_0]$ denote a sequence of one-step weak reductions
from $b_0$ to \ttt{true}.
That is, 
let $[b_0]$ =
``$b_0\,{\rdc{E,AC}}\,b_1\cdots\,b_n\,{\rdc{E,AC}}$ $\ttt{true}$''
($0 \le n$)
if $\lnot(b_0 = \ttt{true})$, 
and $[b_0]$ = \ttt{true} if $b_0$ = \ttt{true}.
%
%
$\{b_0\}$ is defined as follows based on $[b_0]$,
where ${b_{i(i+1)}}$ ($0 \le i \le n$) is the condition (i.e., $\theta(c(X))$)
of the one-step weak reduction 
$b_i\,{\rdc{E,AC}}\,b_{i+1}$.\\
\hspace*{1em}$\{b_0\}$ = 
$b_0\,\{b_{01}\}\,b_1\cdots\,b_i\,\{b_{i(i+1)}\}\,b_{i+1}\cdots
b_n\,\{b_{n(n+1)}\}\,\ttt{true}$\\
\hspace*{3.2em}if $[b_0]$ =
$b_0\,{\rdc{E,AC}}\,b_1\cdots\,b_i\,{\rdc{E,AC}}\,b_{i+1}\cdots$ 
$b_n\,{\rdc{E,AC}}\,\ttt{true}$, and\\
\hspace*{1em}\{$b_0$\} = $\{\ttt{true}\}$ if [$b_0$] = \ttt{true}.\\
\{\ttt{\us}\} applies recursively and
can be nested infinitely as
shown in {\bf Example \ref{ex-trstrmt}} below,
and 
$\{b_0\}$ might be undefined because
(i) it continues infinitely or
(ii) it stops in the middle without
getting to \ttt{true}.
The above definition implies that 
$t_0\{b_0\}t_1{\cdots}t_i\{b_i\}t_{i+1}{\cdots}t_n\{b_n\}t_{n+1}$
($0 \le n$)
is a {\bf sequence of oc-reds} iff
each $b_i$ of $\{b_i\}$ ($0 \le i \le n$) is a sequence of oc-reds
or \ttt{true}.

A term $t {\in} T_{\vSg}$ is called 
{\bf ${\rdc{E,AC}}$-reduced modulo AC},
or simply just {\bf reduced},
if there is no term $t' {\in} T_{\vSg}$ such that
$t\,{\rdc{E,AC}}\,t'$.
For a ground term $t_0 {\in} T_{\vSg}$,
a CafeOBJ's reduction command
``\ttt{reduce{\,}in\,$\mathit{SP}$:{\,}$t_0${\,}.}''\!\!
computes a sequence of oc-reds
$t_0\{b_0\}t_1{\cdots}t_i\{b_i\}t_{i+1}{\cdots}$
from left to right as much as possible
by choosing a context and a rewrite rule
for each oc-red using a pre-defined algorithm,
and, if the sequence terminates,
returns {\bf the reduced term $t_0^{\mathtt{red}}$} that ends
the sequence.
Note that $t_0^{\mathtt{red}}$ is uniquely determined
if the reduction terminates,
and \ttt{\us}$^{\mathtt{red}}$ is a partial function.

\bex
\label{ex-pnatplus}

The following CafeOBJ code 
declares, with the keyword \ttt{mod!},
a module (i.e., a
specification) \ttt{PNAT+} 
(Peano NATural numbers with \verb!_+_!).
\ttt{PNAT+} declares
between \ttt{\{} and \ttt{\}}
(i) three sorts \ttt{Zero}, \ttt{NzNat}, \ttt{Nat} with subsort relations, 
(ii) two constructor operators \ttt{0}, \ttt{s\us} and
one non-constructor operator \ttt{\us+\us},
(iii) two variables \ttt{X}, \ttt{Y} after the keywords
\ttt{vars},
and (iv) two equations that reduce a term with non-constructors
to a term only with constructors\footnote{%
All the CafeOBJ modules and proof scores explained in
this paper are posted at the following web page:\\
\hspace*{2em} \url{https://cafeobj.org/~futatsugi/misc/apsco-220907/}\\
and reader can execute them.
The module \ttt{PNAT+} is in the file \ttt{peano-nat-spc.cafe} on the Web.
}%
.
\begin{cosmall}
\begin{verbatim}
      mod! PNAT+ {
      [Zero NzNat < Nat]
      op 0 : -> Zero {constr} .
      op s_ : Nat -> NzNat {constr} .
      op _+_ : Nat Nat -> Nat .
      vars X Y : Nat .
      cq X + Y = Y if (X = 0) .
      eq (s X) + Y = s(X + Y) . }
\end{verbatim}
\end{cosmall}
The CafeOBJ command:
``\ttt{red {in} PNAT+\,{:}\,{s 0} {+} 0 .}''
returns \ttt{s 0}
after computing the following sequence of oc-reds.\\
\hspace*{3em}
\verb|(s 0) + 0 {true} s(0 + 0) {0 = 0 {true} true} s 0|
\hfill\makebox[0pt][r]{\owari}
\eex

\subsubsection{Completeness of CafeOBJ Execution}
\label{sssec-compspec}

Since ${\rdcstr{E,AC}} \subseteq {\rdcstr{E/AC}}$, 
it could happen that 
$(t\;{\rdcstr{E/AC}}\;t')$ but $\lnot(t\;{\rdcstr{E,AC}}\;t')$.
CafeOBJ's implementation of ${\rdc{E,AC}}, {\rdcstr{E,AC}}$ is
complete enough such that\\
\hspace*{1em} {\bf (E,AC)}
\hspace*{2em}$\forall t, u\,{\in}\,T_{\vSg} (t\;{\rdc{E/AC}}\;u \;\Ra\;
\exists u'\,{\in}\,T_{\vSg} (t\;{\rdc{E,AC}}\;u' \land
u =_{\!\mtnit{AC}} u'))$ .\\
This is realized by {\bf extending} (or {\bf completing}) each equation $e \in E$
involving operators with AC attributes%
\footnote{%
\citep{Meseguer17} provides the most advanced study of 
conditional order-sorted rewriting modulo axioms including
important issues like AC-extended equations 
and ${\rdcstr{E/AC}}$ vs. ${\rdcstr{E,AC}}$.
}.

A TRS $\mathit{SP} = (\Sg,E)$
is defined to be
\begin{enumerate}[(i)]

\item {\bf operationally terminating} \citep{LucasMM05}\footnote{%
\citep{LucasMM05} defines
operational termination based on applications of
inference rules but does not introduce oc-red.
}%
iff
  there is no infinite sequence of oc-reds or infinitely 
  nested oc-reds no matter what
  context and rewrite rule are chosen for each oc-red,

\item {\bf terminating} iff ${\rdc{E,AC}}$ is well-founded,

\item {\bf confluent} iff\\
\hspace*{3.5em}$\forall\,t_1, t_2, t_3 {\in} T_\vSg
((t_1\,{\rdcstr{E,AC}}\,t_2\: \land\: t_1\,{\rdcstr{E,AC}}\,t_3) \;{\Ra}\\
\hspace*{10em}\exists t_4\,{\in}\,T_\vSg
  (t_2\,{\rdcstr{E,AC}}\,t_4\: \land\: t_3\,{\rdcstr{E,AC}}\,t_4))$,

\item {\bf sufficiently complete} iff $t\,{\rdcstr{E,AC}}\,t'$
for $t$ and $t'$ in {\it\ref{sssec-iacbosa}}
and $Y$ being considered to be the set of fresh constants.

\end{enumerate}

All the above four properties (i), (ii), (iii), (iv) are undecidable,
but usable sufficient conditions for guaranteeing them are known.

\bex
\label{ex-trstrmt}
If a TRS has only one equation
``\ttt{cq b = true if b .}'' for a fresh Boolean constant \ttt{b},
the system is terminating because 
(i) the builtin module BOOL without the equation is terminating, and
(ii) ``\ttt{b} $\rdc{E,AC}$ $t$'' does not hold for any $t \in T_{\vSg}$.
The TRS is, however, not operationally terminating because
there is an infinitely nested oc-reds
 ``\ttt{b} \{\ttt{b} \{\ttt{b} $\cdots$''.
\hfill\makebox[0pt][r]{\owari}
\eex

Let $\mathit{SP}\,\tsc\,{p}$ denote
that ``\ttt{reduce\,{in}\,$\mathit{SP}$\,{:}\,$p$\,{.}}''
returns \ttt{true}, i.e., $p^{\mathtt{red}} = \ttt{true}$.
The following is obtained via
{\bf Facts \ref{fact-edemdl}, \ref{fact-rwrede}}.

\bfact\label{fact-crdmdl}
\hspace*{0.1em}
{\bf (PR1)}\hspace*{4em}$\mathit{SP}\,{\tsc}\,p\;{\Ra}\;\mathit{SP}\,{\mdl}\,p$%
\hfill\makebox[0pt][r]{\owari}
\efact

Note that the above Facts hold even if $\mathit{SP}$ has no
initial model, or is not operationally terminating,
terminating, confluent, or sufficiently complete.

If a reduction specification (TRS) $\mathit{SP}\,{=}\,(\vSg,E)$
is
(1) conjunctive, i.e., $E$ is conjunctive ({\it\ref{sssec-seneq}}),
(2) sort decreasing, i.e., $E$ is sort decreasing ({\it\ref{sssec-seneq}}),
(3) operationally terminating, 
(4) confluent, and
(5) sufficiently complete
then $\mathit{SP}$ is called {\bf red-complete}.

\bfact\label{fact-cansys}\citep{Meseguer17}
For a red-complete reduction specification
$\mathit{SP}\,{=}\,(\vSg,E)$,\\
\hspace*{6em}
$\forall t, u\,{\in}\,T_{\vSg}
((t\,{\dedus{E}{AC}}\,u) \Lra 
 (t\,{\dedstr{E/AC}}\,u) \Lra (t^{\mathtt{red}} =_{\!\mtnit{AC}} u^{\mathtt{red}}))$.
\hfill\makebox[0pt][r]{\owari}
\efact

\section{Basic Proof Scores}
\label{sec-bscprs}

Basic techniques for constructing specifications and
proof scores are presented, and justified by the Facts shown
in Section {\bf \ref{sec-fth}}.

\subsection{Specifying Systems}
\label{ssec-spcsys}

The following CafeOBJ code \ttt{01:-12:}\footnote{%
Refers to the CafeOBJ code with line numbers from 
\ttt{01:} to \ttt{12:} in this subsection.
See also the code reference notation explained in
the footnote \ref{fn-coderef}.
}%
specifies a generic list data structure.
Numbers in 1st to 3rd columns like \ttt{01:},
\ttt{10:} are for explanation, and CafeOBJ code
starts from the 7th column.
CafeOBJ code starting \ttt{--} or
\ttt{-->} followed by a blank,
like in \ttt{01:} or \ttt{08:},
is a comment until the end of a line.
A comment with ``\ttt{-- }'' or
``\ttt{--> }'' can start at any point of a line.

\vspace*{0.2em}
\noindent
@\footnote{%
A footnote on @ shows in which files the following
CafeOBJ code exists.
\ttt{01:-17:} is in the file
\ttt{list-append.cafe}
on the Web:\\
\hspace*{2em}
\url{https://cafeobj.org/~futatsugi/misc/apsco-220907/}%
\,.}
\vspace*{-1em}
\begin{cosmall}
\begin{verbatim}
01:   -- generic collection of objects
02:   -- mod* TRIV {[Elt]} -- TRIV is a builtin module

03:   --> generic list
04:   mod! LIST (X :: TRIV) {
05:   [NnList < List]
06:   op nil : -> List {constr} .
07:   op _|_ : Elt List -> NnList {constr} . }

08:   --> generic list with enhanced _=_
09:   mod! LIST= (X :: TRIV) { 
10:   pr(LIST(X))
11:   eq (nil = E:Elt | L:List) = false .
12:   eq (E1:Elt | L1:List = E2:Elt | L2:List) = (E1 = E1)and(L1 = L2).}
\end{verbatim}
\end{cosmall}

The keyword \ttt{mod*} (\ttt{02:})
declares,
with loose denotation ({{\it\ref{sssec-modsp-loti}}),
a module with name \ttt{TRIV}
and its body \ttt{\ob[Elt]\cb} that 
just declares the sort ({\it\ref{sssec-sorts}}) \ttt{Elt}
with no constraints.
The keyword \ttt{mod!} (\ttt{04:}) declares,
with tight denotation ({\it\ref{sssec-modsp-loti}}),
a module with name \ttt{LIST},
its parameter list \ttt{(...)} (\ttt{04:}),
and its body \verb#{...}# (\ttt{04:-07:}).
The parameter list \ttt{(X\;{::}\;TRIV)}
declares that this module has
a parameter module \ttt{X},
and that can be replaced by a module 
that satisfies the module \ttt{TRIV},
i.e., by any module with at least one sort.
The body declares,
(i) two sorts \ttt{NnList} and \ttt{List}
with the former being a subsort ({\it\ref{sssec-sorts}})
of the latter (\ttt{05:}),
(ii) constructor operator ({\it \ref{sssec-cboss}})
\ttt{nil} with 
rank ({\it\ref{sssec-cboss}}) \ttt{List} (\ttt{06:}) and 
constructor operator \verb!_|_! with rank
``\ttt{Elt List NnList}'' (\ttt{07:}).

The module \ttt{LIST=} declares two equations
(\ttt{11:-12:}) for enhancing the builtin
equality \ttt{\us{=}\us} on the sort \ttt{List}.
\ttt{E:Elt}, \ttt{L1:List}, etc. are on-line
variable declarations each of them
is effective until the end of each equation.
\ttt{pr(LIST(X))} (\ttt{10:}) declares that the module
\ttt{(LIST(X))} is imported in {\bf pr}otecting mode,
i.e., with no changes on its models.

Because the module \ttt{LIST} has the tight denotation
and its model is the initial algebra of
$\Mod(\mttt{LIST})$ ({\it\ref{sssec-modsp-loti}}),
the equality of the initial model does not
change with the addition of the two equations.
It is sometimes necessary to add equations
to refine the definition of the equality
on the initial algebra, for only one equation:
``\ttt{eq (L:List = L) = true .}'' is
builtin originally.

The following module \ttt{APPEND} (\ttt{14:-17:}) specifies
the append operator \verb!_#_! of rank ``\ttt{List List List}'' via
the two equations (\ttt{16:-17:}).

\begin{cosmall}
\begin{verbatim}
13:   --> append operator _#_ on List 
14:   mod! APPEND (X :: TRIV, Y :: LIST(X)) {
15:   op _#_ : List List -> List .
16:   eq nil # L2:List = L2 .
17:   eq (E:Elt | L1:List) # L2:List = E | (L1 # L2) . }
\end{verbatim}
\end{cosmall}

The parameter declaration \ttt{(X\,{::}\,TRIV,\,Y\,{::}\,LIST(X))}
(\ttt{01:}) includes the 2nd parameter ``\ttt{Y\,{::}\,LIST(X)}''
and also indicates that the module \ttt{LIST(X)} is imported
in protecting mode.
The 2nd parameter is prepared for 
instantiating it with a more elaborated module \ttt{LIST=(X)}.
That is, we need\\
\hspace*{2em}\ttt{APPEND(X,LIST=(X))}
instead of \ttt{APPEND(X,LIST(X))}\\
in some occasion 
(see {\it\ref{ssec-spcprin}}-\ttt{10:}).

\subsection{Specifying Properties of Interest}
\label{ssec-spcprin}

Assume that we want to construct a proof score
for proving that the operator \ttt{\us{\shp}\us} of
the module \ttt{APPEND} is associative.
For expressing the proof goal,
the following module \ttt{APPENDassoc} (\ttt{01:}) defines
the predicate \ttt{appendAssoc} and 
the module \ttt{APPENDassoc@} (\ttt{06:}) 
introducing the three fresh constants
\ttt{l1@,l2@,l3@} of the sort \ttt{List}
(\ttt{08:}).

\vspace*{0.2em}
\noindent
@\footnote{%
\ttt{01:-10:} is in the file
\ttt{list-append.cafe}
on the Web.}
\vspace*{-1em}
\begin{cosmall}
\begin{verbatim}
01:   mod APPENDassoc (X :: TRIV,Y :: LIST(X)) {
02:   ex(APPEND(X,Y))
03:   pred appendAssoc : List List List .
04:   eq appendAssoc(L1:List,L2:List,L3:List) =
05:      ((L1 # L2) # L3 = L1 # (L2 # L3)) .     }

06:   mod APPENDassoc@(X :: TRIV,Y :: LIST(X)){
07:   ex(APPENDassoc(X,Y))
08:   ops l1@ l2@ l3@ : -> List . }

09:   mod APPENDassoc@= (X :: TRIV) {
10:   pr(APPENDassoc@(X,LIST=(X)))  }
\end{verbatim}
\end{cosmall}

The keyword \ttt{mod} (\ttt{01:,06:,09:}) indicates
no specific intention of the loose or tight denotation
of the module, and used for proof modules in proof scores.
The declaration \ttt{ex(...)} (\ttt{02:,07:})
indicates the importation of a module in {\bf ex}tending mode.
That is, no change except the addition of
new operators like \ttt{l1@,l2@,l3@},\,\ttt{appendAssoc}.
The module \ttt{APPENDassoc@=} (\ttt{09:}) is using 
the module \ttt{LIST=} instead of \ttt{LIST}
by putting \ttt{LIST=} to the 2nd parameter of
\ttt{APPENDassoc@} (\ttt{10:}), then of \ttt{APPENDassoc} (\ttt{07:}),
and then of \ttt{APPEND} (\ttt{02:}).

The associativity of the operator \ttt{\us\shp\us} is
formalized as\\
\hspace*{2em}$\mttt{APPENDassoc@=} \vDash \mttt{appendAssoc(l1@,l2@,l3@)}$\\
and by {\bf Fact \ref{fact-crdmdl}}\\
\hspace*{2em}$\mttt{APPENDassoc@=} \tsc \mttt{appendAssoc(l1@,l2@,l3@)}$\\
is sufficient for the proof.
However, the CafeOBJ reduction:\\
\hspace*{2em}\ttt{red\;in\;APPENDassoc@=\;{:}\;appendAssoc(l1@,l2@,l3@)\;{.}}\\
does not return \ttt{true}, where \ttt{red} is the abbreviation of
\ttt{reduce}.

\subsection{Case-Split and Induction}
\label{ssec-csi}

Let $M\,{=}\,(\vSg,E)$
($\vSg = (S,\leq,F,F^c)$ ({\it\ref{sssec-cboss}}))
be a module (i.e., a specification) in CafeOBJ.
If $\mathit{M} \tsc p$ does not hold,
the universal strategy for proving $\mathit{M} \mdl p$
is making use of a lemma
(including for discharging a contradictory case),
a case-split, and/or an induction.
Specification $M$ may need some improvements as well.

\subsubsection{Case-Split with Exhaustive Equations}
\label{sssec-cswee}

Let $Y_i = \{Y_{i_s}\}_{s{\in}S}$ be an $S$-sorted set of
fresh constants and $e_i$ be
a ($\vSg {\cup} Y_i$)-equation ({\it\ref{sssec-seneq}}).
That is, $e_i$ may contain fresh constants in 
$Y_i$ (i.e., in $Y_{i_s}$ for some $s{\in}S$)
that are not in $\vSg$.
An $M$-model $A {\in} \Mod(M)$ ({\it\ref{sssec-modsp-loti}})
is defined to satisfy ($\vSg {\cup} Y_i$)-equation $e_i$
(in symbols $A\,{\vDash_{\vSg  {\cup} Y_i}}\,e_i$)
iff $A\,{\vDash_{\vSg}}\,e_i$ ({\it\ref{subsubsec-algst}})
holds for some interpretation
of every element in $Y_i$ in $A$
(i.e., when for each $s{\in}S$ every constant in $Y_{i_s}$
is interpreted as some element in $A_s$).
Equations $e_1,\cdots\!,e_n$ ($1 \le n$) 
are defined to be {\bf exhaustive} for $M$ iff
$\forall A {\in} \Mod(M) 
(\exists i{\in}\{1,\cdots\!,n\} (A\,{\vDash_{\vSg \cup Y_i}}\,e_i))$.

Let $e_1,\cdots\!,e_n$ ($1 \le n$) be exhaustive equations
and $M_{+e_i}\,{=}\,(\vSg {\cup} Y_i, E {\cup} \{e_i\})$
($1 \le i \le n$),
then each $M$-model $A {\in} \Mod(M)$ is an
$M_{+e_i}$-model $A {\in} \Mod(M_{+e_i})$ for
some $i {\in} \{1,\cdots\!,n\}$ by
interpreting $Y_i$ appropriately in $A$,
and $A\,{\vDash_{\vSg}}\,p$ if ${M_{+e_i}}{\!\mdl}{p}$.
Hence the following proof rule on which 
every case-split in CafeOBJ can be based.

\bfact \label{fact-csee}
({\it Case-Split with Exhaustive Equations})\\
\hspace*{2em}%
{\bf (PR2)}
\hspace*{2em}%
$(M_{+e_1}{\!\mdl}{p}\land{M_{+e_2}}{\!\mdl}{p}\land
\cdots\land{M_{+e_n}}{\!\mdl}{p})\Rightarrow{M}{\!\mdl}{p}$
\hfill\makebox[0pt][r]{\owari}
\efact

Because the sort \ttt{List} of the module \ttt{LIST}
({\it\ref{ssec-spcsys}}-\ttt{04:-07:})
is constrained ({\it\ref{sssec-cboss}})
with the two
constructors \ttt{nil} and \ttt{\us{|}\us},
two equations
``\ttt{eq[e1]:{\;}l1@ = nil .}'',
``\ttt{eq[e2]:{\;}l1@ = e\dl\,{|}\,l1\dl .}''
with fresh constants \ttt{e\dl}, \ttt{l1\dl}
are exhaustive for the module \ttt{APPENDassoc@=}
(see {\it\ref{sssec-cbosa}} for the models of \ttt{APPENDassoc@=}).
Let 
$M^t\,{=}\,\mttt{APPENDassoc@=}$ and
$p^t\,{=}\,\mttt{appendAssoc(l1@,l2@,l3@)}$
then 
$(M^t_{\mathtt{+e1}} \mdl p^t$ $\land$ 
$M^t_{\mathtt{+e2}} \mdl p^t)\:{\Ra}\:M^t \mdl p^t$
holds by {\bf (PR2)}, and the following CafeOBJ code \ttt{01:-09:}
\!can serve to check whether
$M^t_{\ttt{+e1}} \tsc p^t \land M^t_{\mathtt{+e2}} \tsc p^t$
holds, \ttt{01:-04:} for $\ttt{+e1}$ and
\ttt{05:-09:} for $\ttt{+e2}$.

\vspace*{0.5em}
\noindent
@\footnote{%
\ttt{01:-09:} is in the file
\ttt{list-append.cafe}
on the Web.}
\vspace*{-0.9em}
\begin{cosmall}
\begin{verbatim}
01:   open APPENDassoc@= .
02:   eq l1@ = nil .
03:   red appendAssoc(l1@,l2@,l3@) .
04:   close

05:   open APPENDassoc@= .
06:   op e$ : -> Elt . op l1$ : -> List .
07:   eq l1@ = e$ | l1$ .
08:   red appendAssoc(l1@,l2@,l3@) .
09:   close
\end{verbatim}
\end{cosmall}

The ``\ttt{open $M$ .}'' command 
creates a new tentative module where
all the $M$'s contents are imported,
new sorts/operators/equations can be declared,
and \ttt{reduce} commands can be executed.
\ttt{red} is a shorthand for \ttt{reduce}, and
if $M$ is opened, 
``\ttt{red $p$ .}'' stands for ``\ttt{red in $M+$ : $p$ .}''
where the module $M+$ includes
$M$'s contents plus the contents
declared in the opened tentative module
before the \ttt{red} command.
The command \ttt{close} deletes the created module.

If the \ttt{red} commands at \ttt{03:}
and \ttt{08:} would return \ttt{true},
$M^t_{\mathtt{+e1}} \tsc p^t \land M^t_{\mathtt{+e2}} \tsc p^t$
is proved,
and by {\bf Fact \ref{fact-crdmdl}}-{\bf (PR1)} 
$M^t \mdl p^t$ is proved.  
\ttt{03:} returns \ttt{true} but 
\ttt{08:} does not, and induction should be necessary.

\subsubsection{Structural Induction}
\label{sssec-stin}

The standard way of induction for algebraic specifications is
{\bf structural induction} \citep{burstall69} 
on initial algebras
(e.g., Theorem 6.5.9 and its Corollary of \citep{goguen06TPA}),
which can apply to any constructor algebra
(i.e., an $(S,\leq,F,F^c)$-algebra {\it \ref{sssec-cbosa}})
and formulated as follows.

Let $\vSg = (S,\le,F,F^c)$, and
$p(y_1,\cdots,y_n) \in T_\vSg(Y)\: (1 \le n)$ be a Boolean term
with finite variables $Y = \{y_1,\cdots,y_n\}$ that describes 
a property of interest for a module (specification) $M$.
The induction should be applied to a variable
$y_i \in Y$ of a constrained sort ({\it\ref{sssec-cboss}}),
and assume that the sort of $y_1$ is $\hat{s}$
and $\hat{s}$ is a constrained sort without loss of generality.
Let $G^{\hat{s}} \subseteq F^c$ be
the set of constructors ({\it\ref{sssec-cboss}}) for the sort $\hat{s}$,
i.e., $G^{\hat{s}}$ is the sorted sets
$G^{\hat{s}} = \{ G_{ws} \,|\, w {\in} S^*\!{,}\, s {\in} S{,}\, s {\le} \hat{s} \}$.
The principle of
structural induction is formulated as follows,
where 
$c_i$ is a fresh constant of the sort $s_i\:(1\,{\le}\,{i}\,{\le} m)$.

\vspace*{0.5em}
\noindent
\(
\hspace*{2.5em}[\,\forall G_{s_1{\cdots}s_m s}{\in}G^{\hat{s}}\,(0{\le}m)\\
\hspace*{2.9em}(\forall g{\in}G_{s_1{\cdots}s_m s}\\
\hspace*{3.3em}[\;(\land_{(1 \le i \le m)\,{\land}\,(s_i \le \hat{s})}
(M{\cup}\{c_1,{\cdots},c_m\}{\mdl}\,p(c_i,y_2,{\cdots},y_n)))\\
\hspace*{3.6em}\Ra 
M{\cup}\{c_1,{\cdots},c_m\}{\mdl}\,p(g(c_1,{\cdots},c_m),y_2,{\cdots},y_n)\;]\;)]\\
\hspace*{2.5em}\Ra M {\mdl}\,p(y_1,{\cdots},y_n)
\)
\vspace*{0.5em}

By 
$((p_1 \land \cdots \land p_l) \Ra q) \Lra (p_1 \Ra (\cdots (p_l \Ra q)\cdots))$
and {\bf Fact \ref{fact-msatp}}-{\it(iv)},
$[\;(\land_{(1 \le i \le m)\,{\land}\,(s_i \le \hat{s})}({\cdots})) \mdl
\cdots]$
in the premise can be rewritten as follows.
Note that 
$\{\mttt{eq}\cdots \,|\, (1{\le}i{\le}m)\,{\land}\cdots\}$ is empty
if $m = 0$.

\vspace*{0.5em}
\noindent
\(
\hspace*{3em}[M{\cup}\{c_1,{\cdots},c_m\}{\cup}
\{\mttt{eq}\;p(c_i,y_2,{\cdots},y_n)\;\mttt{=\,true\,{.}} 
\,|\, (1{\le}i{\le}m)\,{\land}\,(s_i{\le}\hat{s})\}\\
\hspace*{3.2em}\mdl{\;}p(g(c_1,{\cdots},c_m),y_2,{\cdots},y_n)]
\)
\vspace*{0.5em}

Let 
$y@_i$ be a fresh constant for the variable $y_i$ $(1 \le i \le n)$,
and $Y\!@ = \{y@_1,y@_2,\cdots,y@_n\}$.
By applying {\bf Fact \ref{fact-thcst}} 
to the above
$[M{\cup}\{c_1,{\cdots},c_m\}{\cup}{\cdots}]$,
with the correspondence $Y\,{\lrh}\,Y$,
the following is obtained.
Note that the variables $y_2,{\cdots},y_n$ in 
$\mttt{eq}\;p(c_i,y_2,{\cdots},y_n)\;\mttt{=\,true\,{.}}$
are bounded to the equation and not replaced with
the fresh constant $y@_2,\cdots,y@_n$.

\bfact \label{fact-stind}
({\it Structural Induction})\\
\(%
\hspace*{2em}[\,\forall G_{s_1{{\cdots}}s_m s}{\in}G^{\hat{s}}\,(0{\le}m)\\
\hspace*{2.2em}(\forall g{\in}G_{s_1{\cdots}s_m s}\\
\hspace*{2.8em}[M{\cup}Y\!@{\cup}\{c_1,{\cdots},c_m\}{\cup}
\{\mttt{eq}\;p(c_i,y_2,{\cdots},y_n)\;\mttt{= true\,{.}}%
 \,|\, (1{\le}i{\le}m)\,{\land}\,(s_i{\le}\hat{s})\}\\
\hspace*{3em}\mdl{\,}%
p(g(c_1,{\cdots},c_m),y@_2,{\cdots},y@_n)]\:\,)]\\
\hspace*{2em}\Ra M \mdl\,p(y_1,{\cdots},y_n)\)
\hfill\makebox[0pt][r]{\owari}
\efact

{\it\ref{sssec-cswee}}-\ttt{01:-04:} is a proof score
for the premise's conjunct $\forall g\,{\in}\,G_{s}(\cdots)$ 
(i.e., for $m = 0$), and 
the following \ttt{01:-07:} is a proof score for
the premise's conjunct $\forall g\,{\in}\,G_{s_1{\cdots}s_m s}(\cdots)$
(i.e., for $m > 0$)
of {\bf Fact \ref{fact-stind}}
with the following correspondences:\\
\hspace*{2em}${\hat{s}} \lrh \mttt{List}$, 
$G^{\hat{s}} \lrh \{\{\mttt{nil}\},\{\mttt{\us{|}\us}\}\}$, 
$G_s \lrh \{\mttt{nil}\}$,
$G_{s_1{\cdots}s_m s} \lrh \{\mttt{\us{|}\us}\}$,\\
\hspace*{2em}$p \lrh \mttt{appendAssoc}$,
$Y\!@ \lrh \{\mttt{l1@},\mttt{l2@},\mttt{l3@}\}$,
$M{\cup}Y\!@ \lrh \mttt{APPENDassoc@=}$,\\
\hspace*{2em}$\{c_1,\cdots,c_m\} \lrh \{\mttt{e\dl},\mttt{l1\dl}\}$,
$\{y_1,\cdots,y_n\} \lrh \{\mttt{L1},\mttt{L2},\mttt{L3}\}$.

\vspace*{0.2em}
\noindent
@\footnote{%
\ttt{01:-07:} is in the file
\ttt{list-append.cafe}
on the Web.}
\vspace*{-0.7em}
\begin{cosmall}
\begin{verbatim}
01:   open APPENDassoc@= .
02:   op e$ : -> Elt . op l1$ : -> List .
03:   eq l1@ = e$ | l1$ .
04:   eq ((l1$ # L2:List) # L3:List =
05:        l1$ # (L2:List # L3:List)) = true .
06:   red appendAssoc(l1@,l2@,l3@) .
07:   close
\end{verbatim}
\end{cosmall}

The left hand side of the equation \ttt{04:-05:} is
the reduced ({\it\ref{sssec-coex}}) form of\\
\hspace*{2em}\ttt{appendAssoc(l1\dl,L2:List,L3:List)},\\
because left hand sides should be reduced 
for being effective as reduction rules.
The \ttt{red} command \ttt{06:} returns \ttt{true} and
the proof score
({\it\ref{sssec-cswee}}-\ttt{01:-04:}\,+\,\ttt{01:- 07:}) is
effective for proving\\
\hspace*{2em}$\mttt{APPENDassoc@=} \mdl \mttt{appendAssoc(l1@,l2@,l3@)}$.\\
A proof score is called {\bf effective} 
if the score constructs an effective proof tree
(see {\it\ref{ssec-ptcalc}}).

\bex (\ttt{\us=\us} on \ttt{List})
\label{ex-listeq}

Suppose that the equality \ttt{\us=\us} on the sort \ttt{List}
is not enhanced like {\it\ref{ssec-spcsys}}-\ttt{08:-12:} and 
\ttt{APPENDassoc@} is used instead of 
\ttt{APPENDassoc@=} in \ttt{01:} above,
then 
``\verb!red appendAssoc(l1@,l2@,l3@) .!'' (\ttt{06:}) returns\\
\hspace*{1em}
\verb!((e$ | ((l1$ # l2@) # l3@)) = (e$ | (l1$ # (l2@ # l3@))))!\\
instead of \ttt{true}. 
This clearly shows that axioms 
{\it\ref{ssec-spcsys}}-\ttt{11:-12:} 
are necessary to get the effective proof score.
It is a nice example of a reduction result telling
necessary axioms.
This example also shows how module restructuring
is inspired by observing reduction results thanks to
powerful and flexible CafeOBJ's module system.
\hfill\makebox[0pt][r]{\owari}
\eex

\section{Advanced Proof Scores}
\label{sec-adprs}

In advanced proof scores, case-split is realized through
commands of the proof tree calculus (PTcalc), and
induction is realized through well-founded induction (WFI)
on argument tuples of a goal property predicate.
As a result, advanced proof scores are 
(i) succinct and transparent for case-split,
(ii) support a variety of induction schemes including
multi-arguments induction.

\subsection{Proof Tree Calculus (PTcalc) \citep{ptcalc-wp}}
\label{ssec-ptcalc}

PTcalc is a refined version of
a CafeOBJ version of CITP \citep{gainaZCA13},
and helps to prove $M{\mdl}p$ for a module $M = (\vSg,E)$.

We already have the following proof rule ({\bf Fact \ref{fact-crdmdl}}).\\
\hspace*{3em}{\bf (PR1)}\hspace*{1em}
{$M{\tsc}{p}\Rightarrow{M}{\mdl}{p}$}\\
\noindent
Usually {$M{\tsc}{p}$} is difficult to prove directly, and
we need to find exhaustive equations $e_1,\cdots,e_n$
and make use of the following
proof rule of case-split ({\bf Fact \ref{fact-csee}})
with the exhaustive equations.\\
\hspace*{3em}{\bf (PR2)}\hspace{1em}
$(M_{+e_1}{\!\mdl}{p}\land{M_{+e_2}}{\!\mdl}{p}\land
\cdots\land{M_{+e_n}}{\!\mdl}{p})\Rightarrow{M}{\!\mdl}{p}$\\
{$M_{+e_i}\!{\tsc}{p}$} would be still difficult to prove and
{\bf (PR2)} is applied repeatedly.  The repeated applications
of {\bf (PR2)} generate {\bf proof trees} successively.
Each of the generated proof trees has the {\bf root node}
$M\!{\mdl}{p}$ and each of other {\bf nodes} is of the form
$M_{+e_{i_1}{\cdots}+e_{i_m}}\!{\mdl}{p}$ ($1 \le m$)
that is generated as a {\bf child node} of 
$M_{+e_{i_1}{\cdots}+e_{i_{m-1}}}\!{\mdl}{p}$
by applying {\bf (PR2)}.
A {\bf leaf node}
(i.e.,{\;}a node without child nodes)
$M_{+e_{l_1}{\cdots}+e_{l_k}}\!{\mdl}{p}$ ($0 \le k$)
of a proof tree is
called {\bf effective} if $M_{+e_{l_1}{\cdots}+e_{l_k}}{\tsc}{p}$
holds.  A proof tree is called effective if all of whose
leaf nodes are effective.  PTcalc proves $M{\mdl}{p}$ by
constructing an effective proof tree whose root node is
$M{\mdl}{p}$.

\subsubsection{PTcalc Commands}
\label{sssec-ptcpsr}

PTcalc consists of the following commands with the keywords
with the head character {\tt :} that distinguishes them from
ordinary CafeOBJ commands.
{\tt :goal} declares goal propositions to be proved;
{\tt :def} defines a new command;
{\tt :apply} applies defined/pre-defined commands;
{\tt :csp} declares a case-split command;
{\tt :init} declares a initialize command;
{\tt :red}  does a reduction at the current goal;
{\tt :show} shows info on proof;
{\tt :describe} shows detailed info on proof;
{\tt :select} selects the goal (node) specified;
{\tt :set} sets parameters of PTcalc.

In PTcalc, a node in a proof tree is a special CafeOBJ module
and called a {\bf goal}.
Each goal {\it gl} has a {\bf name} like {\tt root} 
or {\tt 1-2-3} (i.e.,{\;}3rd child of 2nd child of 1st child of {\tt root}),
and consists of the following five items.
\vspace{-0.5em}
\begin{itemize}
\setlength{\itemsep}{-2pt}

\item[(1)] The {\bf next target (or default) goal} Boolean tag
{\tt NTG}({\it gl}\,) that indicates a goal where
a PTcalc command is executed.  
({\tt NTG}({\it gl}\,) = {\tt true}) holds for at most one
goal in a proof tree.

\item[(2)] The {\bf context module {\tt CTM}({\it gl}\,)} that 
is a CafeOBJ module and corresponds to $M$ of $M{\mdl}p$.
The goal {\it gl}\: inherits (imports)
all the contents of {\tt CTM}({\it gl}\,).  

\item[(3)] The set of {\bf introduced axioms (assumptions)
{\tt INA}({\it gl}\,)} that corresponds to 
${+e_{i_1}{\cdots}+e_{i_m}}$
of $M_{+e_{i_1}{\cdots}+e_{i_m}}{\mdl}{p}$,
i.e., {\tt INA}({\it gl}\,) = $\{e_{i_1},{\cdots},e_{i_m}\}$.
  
\item[(4)] The set of {\bf sentences (or equations) to be proved
{\tt STP}({\it gl}\,)} that corresponds to $p$ of 
$M_{+e_{i_1}{\cdots}+e_{i_m}}{\mdl}{p}$.

\item[(5)] The {\bf discharged} Boolean tag {\tt DCD}({\it gl}\,)
that indicates whether {\it gl} is already discharged
(i.e.,{\:}proved).
\end{itemize}
\vspace{-0.5em}
({\tt CTM}({\it gl}\,){$\cup$}{\tt INA}({\it gl}\,)){$\mdl$}%
{\tt STP}({\it gl}\,) corresponds to 
$M_{+e_{i_1}{\cdots}+e_{i_m}}$ ${\mdl}{p}$, where
({\tt CTM}({\it gl}\,){$\cup$} {\tt INA}({\it gl}\,))
is understood as 
the module obtained by adding all the equations in
{\tt INA}({\it gl}\,) to {\tt CTM}({\it gl}\,),
and {\tt STP}({\it gl}\,) 
is understood as the conjunction of its elements. 
{\it gl} sometimes means
({\tt CTM}({\it gl}\,){$\cup$}{\tt INA}({\it gl}\,)).

\citep{ptcalc-wp} defines the PTcalc commands
by specifying in which way the above five items
are changed by each command.
For example, the case-split command {\tt :csp}
is defined as follows.

Let (a) {\it tg} be a goal such that
({\tt NTG}({\it tg}\,) = {\tt true}) and
(b) {\it csid} be the name of a {\tt :csp} command
defined by 
``{\tt :def {\it csid} = :csp\{$eq_1$ $eq_2$ $\cdots$ $eq_n$\}}''
with
$n\,{\in}\,\{1,2,\cdots\}$ equations $eq_i$ 
($i\,{\in}\,\{1,2,\cdots,$ $n\}$).
Then executing the command
``{\tt :apply(\!{\it csid})}''
generates $n$ sub-goals (child goals)
{\it tg}{\tt-1},\;{\it tg}{\tt-2},\;$\cdots$,\;%
{\it tg}{\tt-n}
of {\it tg} as follows.

(1:1) Change {\tt NTG}({\it tg}\,) from {\tt true} to {\tt false}. 

(1:2) {\tt NTG}({\it tg}{\tt-1}) = {\tt true}.

(1:3) {\tt NTG}({\it tg}{\tt-$i$}) = {\tt false}
    ($i\,{\in}\,\{2,\cdots,n\}$).

(2) {\tt CTM}({\it tg}{\tt-$i$}) = {\tt CTM}({\it tg})\;\;
    ($i\,{\in}\,\{1,2,\cdots,n\}$).

(3) {\tt INA}({\it tg}{\tt-$i$}) = {\tt INA}({\it tg})$\cup$\{$eq_i$\}\;\;
    ($i\,{\in}\,\{1,2,\cdots,n\}$).

(4) {\tt STP}({\it tg}{\tt-$i$}) = {\tt STP}({\it tg})\;\;
    ($i\,{\in}\,\{1,2,\cdots,n\}$).

(5) {\tt DCD}({\it tg}{\tt-$i$}) = {\tt false}\;\;
    ($i\,{\in}\,\{1,2,\cdots,n\}$).

\subsubsection{Proof Scores with PTcalc}
\label{sssec-pscptc}

The following proof score \ttt{01:-12:} with
PTcalc commands \ttt{05:-12:} corresponds to
the \ttt{open...close} style proof score
({\it\ref{sssec-cswee}}-\ttt{01:-04:})+({\it\ref{sssec-stin}}-\ttt{01:-07:})
and proves 
$\mttt{APPENDassoc@=} \mdl \mttt{appendAssoc(l1@,l2@,l3@)}$.

\vspace*{0.2em}
\noindent
@\footnote{%
\ttt{01:-12:} is in the file
\ttt{list-append.cafe}
on the Web.}
\vspace*{-1em}
\begin{cosmall}
\begin{verbatim}
01:   mod APPENDassocPtc(X :: TRIV) {
02:   pr(APPENDassoc@=(X))
03:   op e$ : -> Elt . op l1$ : -> List . }

04:   select APPENDassocPtc .
05:   :goal{eq appendAssoc(l1@,l2@,l3@)= true .}
06:   :def l1@ = :csp{eq[e1]: l1@ = nil .
07:                   eq[e2]: l1@ = e$ | l1$ .}
08:   :apply(l1@ rd-)
09:   :def iHyp = 
10:     :init (eq appendAssoc(L1:List,L2:List,L3:List) = true .)
11:           by {L1:List <- l1$;}
12:   :apply(iHyp rd-)
\end{verbatim}
\end{cosmall}

\ttt{03:} corresponds to 
{\it\ref{sssec-stin}}-\ttt{02:} and 
the fresh constants
\ttt{e\dl} and \ttt{l1\dl} 
are added to the module \ttt{APPENDassoc@=} (\ttt{02:})
for defining the module \ttt{APPENDassocPtc},
which prepares for the execution of the PTcalc commands 
\ttt{05:-12:}.
The CafeOBJ's \ttt{select} command at \ttt{04:} selects
the module \ttt{APPENDassocPtc} as the current module.
The PTcalc's \ttt{:goal} command at \ttt{05:}
initiates a proof tree that consists only of the \ttt{root} node,
sets {\tt CTM}(\ttt{root}) =  \ttt{APPENDassocPtc} and
{\tt STP}(\ttt{root}) =
\{\ttt{eq appendAssoc(l1@,l2@,l3@)= true .}\}.
The \ttt{:def} command at \ttt{06:-07:} gives
the name \ttt{l1@} to the case-split (\ttt{:csp}) command 
with the exhaustive two equations \ttt{e1} and \ttt{e2}.
Note that the name \ttt{l1@} is overloaded to denote
the constant \ttt{l1@} of sort \ttt{LIST} and
the command name, defined with \ttt{:def}, for
refining the constant \ttt{l1@}.

The \ttt{:apply} command at \ttt{08:} applies the defined
case-split command \ttt{l1@} at the \ttt{root} node;
creates \ttt{root}'s two child nodes \ttt{1} and \ttt{2}
with {\tt STP}(\ttt{1}) = {\tt STP}(\ttt{2}) = {\tt STP}(\ttt{root}),
{\tt INA}(\ttt{1}) = {\tt INA}(\ttt{root}){$\cup$}\{\ttt{e1}\},
{\tt INA}(\ttt{2}) = {\tt INA}(\ttt{root}){$\cup$}\{\ttt{e2}\};
check $i$\,$\tsc$\,{\tt STP}($i$) for $i$ = \ttt{1,2}
by applying \ttt{rd-} command to the nodes
(i.e., proof modules) \ttt{1,2}.
The command \ttt{rd-} is builtin, and computes the reduced
form of each element of {\tt STP}(\textit{gl}) for checking
whether {\it gl}\,$\tsc$\,{\tt STP}({\it gl}) holds.
$i$\,$\tsc$\,{\tt STP}($i$) holds for $i$ = \ttt{1}
and the node (goal) \ttt{1} is discharged,
but not for $i$ = \ttt{2}.

\ttt{09:-11:} gives the name \ttt{iHyp} to 
the initialize (\ttt{:init}) command that
creates the induction hypothesis equation 
\ref{sssec-stin}-\ttt{04:-05:}.
An initialization command ``\ttt{:init {\it eq} by \{{\it subst}\}}''
declares initialization of an equation {\it eq} with a substitution
{\it subst}. {\it eq} is declared on-line as
\ttt{(eq\,{...}\,.)} (\ttt{10:})
or off-line as \ttt{[\,{\it lbl}\,]} 
({\it \ref{sssec-prswfisch}}-\ttt{21:})
where {\it lbl} is a label of
an already declared equation.

\ttt{12:} applies the command \ttt{iHyp} and
\ttt{rd-} to the node \ttt{2};
the application \ttt{iHyp} adds the created equation 
to {\tt INA}(\ttt{2});
the application \ttt{rd-} checks that
\ttt{2}\,$\tsc$\,{\tt STP}(\ttt{2}) holds,
and the proof is over.

\subsection{Well-Founded Induction (WFI)}
\label{ssec-wfi}

WFI is well recognized as the generic induction scheme that subsumes
a variety of induction schemes.
In CafeOBJ,
WFI nicely coordinates to PTcalc
and provides a universal and transparent
induction scheme \citep{futatsugi20}.
The WFI in CafeOBJ supports naturally
(i) induction with respect to multiple parameters,
(ii) simultaneous induction,
(iii) induction with associative/commutative constructors.

\subsubsection{Principle of WFI}
\label{sssec-prinwfi}

The principle of WFI is well established
(e.g., 14.1.5 of \citep{HrbaeekJech99}, A.1.6 \& A.1.7 of \citep{terese03}),
and formulated as follows.
\bfact {\it (Principle of WFI)}
\label{fact-powfi}
Let $T$ be a set and $\wfg$ be
a well-founded binary relation on $T$,
i.e., $\wfg\,\subseteq{T}\times{T}$
and there is no infinite sequence of
$t_i\in{T}$ ($i = 1, 2, \cdots$)
with $(t_{i},t_{i+1})\,{\in}\,{\wfg}$.
Let $p$ be a predicate on $T$
(a function from $T$ to truth values
$\{\ttt{true},\ttt{false}\}$),
then the following holds,
where $t\,{\wfg}\,t'$ means $(t,t'){\in}\,{\wfg}$.

\vspace*{0.5em}
\hspace*{3em}\(
\forall{t}{\in}{T}%
([\forall{t'}{\in}{T}%
(({t}\,{\wfg}\,t')\,{\Ra}\,{p(t')})]%
\,{\Ra}\,{p(t)})\,{\Ra}\,\forall{t}{\in}{T}(p(t))
\)
\vspace*{0.5em}

\noindent
That is,
if $p(t)$ whenever 
$p(t')$ for all ${t'} {\in} {T}$
such that ${t}\,{\wfg}\,t'$,
then $p(t)$ for all ${t}{\in}{T}$.
\hfill\makebox[0pt][r]{\owari}
\efact

It is easy to see {\bf Fact \ref{fact-powfi}}
holds. 
Assume there exists $t\in{T}$ such that $\lnot(p(t))$,
then there should be $t'\in{T}$ such that
$t\,{\wfg}\,t'$
and $\lnot(p(t'))$.
Repeating this produces an infinite
${\wfg}$-descending sequence.
It conflicts with well-foundedness of 
${\wfg}$. 

\subsubsection{WFI Scheme in CafeOBJ}
\label{sssec-wfisch}

Let 
$p(y_1,\cdots,y_n) \in T_\vSg(Y)_{\mathtt{Bool}}\: (1 \le n)$ be a Boolean term
with finite variables 
$Y = \{y_1,\cdots,y_n\}$ that describes 
a property of interest with $n$ parameters
for a module (specification)
$M = (\vSg,E)$.
The WFI can be applied to 
the tuple $(y_1,\cdots,y_n)$ of parameters
of $p(y_1,\cdots,y_n)$. 
Let the sort of $y_i$ be $s_i$ for $1 \le i \le n$,
and the set of the tuples 
$T_{\vSg}^{p} \df (T_\vSg)_{s_1}{\times}\cdots{\times}(T_\vSg)_{s_n}$
be the set $T$ of {\bf Fact \ref{fact-powfi}}.
$T_{\vSg}^{p}$ is called the set of $p$'s {\bf argument tuples}.
Let $\wfg$ be a well-founded binary relation on $T_{\vSg}^{p}$,
$\bar{y}$ be the tuple of variables $y_1,\cdots,y_n$, 
and $\bar{y'}$ be the tuple of variables $y'_1,\cdots,y'_n$.
The {\bf Fact \ref{fact-powfi}} is rewritten
as follows.
Note that $\bar{y}$ and $\bar{y'}$ in this context are tuples
of mathematical variables and not of CafeOBJ variables.
They are used as variables of both kinds interchangeably
in this subsubsection.

\vspace*{0.5em}
\(
\hspace*{1em}\forall{\bar{y}}{\in}{T_{\vSg}^{p}}%
([\forall{\bar{y'}}{\in}{T_{\vSg}^{p}}%
(({\bar{y}}\,{\wfg}\,\bar{y'})\,{\Ra}\,{p(\bar{y'})})]%
\,{\Ra}\,{p(\bar{y})})
\,{\Ra}\,\forall{\bar{y}}{\in}{T_{\vSg}^{p}}(p(\bar{y}))
\)
\vspace*{0.5em}

Let some appropriate binary relation
$\wfg \subseteq T_{\vSg}^{p}{\times}T_{\vSg}^{p}$
be defined on $M$, and $M_{\scriptsize \wfg}$ be the module with the defined
relation.  Then the following is obtained.

\vspace*{0.5em}
\(
\hspace*{1em}%
(M_{\scriptsize \wfg} \mdl
([\forall\bar{y'}(({\bar{y}}\,{\wfg}\,\bar{y'})\,\mttt{implies}\,{p(\bar{y'})})]
\mttt{implies}\,p(\bar{y})))
\;{\Ra}\;M{\mdl}p(\bar{y})
\)
\vspace*{0.5em}

By applying {\bf Fact \ref{fact-thcst}} 
to the premise $(M_{\scriptsize \wfg} \mdl \cdots)$
with the correspondence $\bar{y} \lrh Y$,
the following is obtained, where
$y@_i$ is a fresh constant for the variable $y_i$ $(1 \le i \le n)$,
$Y\!@ = \{y@_1,\cdots,y@_n\}$, $\overline{y@} = y@_1,\cdots,y@_n$.

\vspace*{0.5em}
\(
\hspace*{0.2em}%
(M_{\scriptsize \wfg}{\cup}Y\!@ \mdl
([\forall\bar{y'}(({\overline{y@}}\,{\wfg}\,\bar{y'})\,\mttt{implies}\,{p(\bar{y'})})]
\mttt{implies}\,p({\overline{y@}})))
\;{\Ra}\;M{\mdl}p(\bar{y})
\)
\vspace*{0.5em}

By applying {\bf Fact \ref{fact-msatp}} {\it (iii)}
with the correspondences:\\
\hspace*{3em}
$p$ $\lrh$
$[\forall\bar{y'}(({\overline{y@}}\,{\wfg}\,\bar{y'})\,
 \mttt{implies}\,{p(\bar{y'})})]$,
$q$ $\lrh$
$p({\overline{y@}})$\\
and seeing, via {\bf Fact \ref{fact-algsatis}} {\it (ii)},\\
\hspace*{3em}
\ttt{eq $[\forall\bar{y'}(({\overline{y@}}\,{\wfg}\,\bar{y'}) 
        \mttt{implies}\,{p(\bar{y'})})]$ = true .}\\
is equal to\\
\hspace*{3em}
\ttt{cq\;{$p(\bar{y'}$)}\,{=}\;{true}\;{if}\;%
  $\overline{y@}\;{\wfg}\;\bar{y'}$\,{.}\!}\\
the following is obtained.

\bfact {\it (WFI Scheme in CafeOBJ)}
\label{fact-wfishe}\\
\hspace*{2em}
$(M_{\scriptsize \wfg}{\cup}Y\!@{\cup}\{$%
\ttt{cq\;{$p(\bar{y'}$)}\,{=}\;{true}\;{if}\;%
  $\overline{y@}\;{\wfg}\;\bar{y'}$\,{.}\!}%
$\} {\mdl} p(\overline{y@}))
\;{\Ra}\; M{\mdl}p(\bar{y})$
\hfill\makebox[0pt][r]{\owari}
\efact

\subsubsection{Proof Scores with WFI Scheme}
\label{sssec-prswfisch}

The following CafeOBJ code \ttt{01:-23:} is
an effective proof score for the premise 
of {\bf Fact \ref{fact-wfishe}} with
the correspondences:\\
\hspace*{1em}$\mttt{APPENDassoc@(X,LIST=(X))} \lrh
M{\cup}Y\!@$,\\
\hspace*{1em}$\mttt{APPENDassocWfiHyp(X,LIST=(X))} \lrh\\
\hspace*{12em}M_{\scriptsize \wfg}{\cup}Y\!@{\cup}\{$%
\ttt{cq\;{$p(\bar{y'}$)}\,{=}\;{true}\;{if}\;%
  $\overline{y@}\;{\wfg}\;\bar{y'}$\,{.}\!\}}.

\vspace*{0.2em}
\noindent
@\footnote{%
\ttt{01:-23:} is in the file
\ttt{list-append.cafe}
on the Web.}
\vspace*{-0.7em}
\begin{cosmall}
\begin{verbatim}
01:   mod LLLwfRl(X :: TRIV,Y :: LIST(X)) {
02:   [Lll] 
03:   op t : List List List -> Lll {constr} .
04:   pred _wf>_ : Lll Lll .
05:   vars L1 L2 L3 : List . var E : Elt .
06:   eq t(E | L1,L2,L3) wf> t(L1,L2,L3) = true .}

07:   mod APPENDassocWfiHyp 
08:         (X :: TRIV,Y :: LIST(X)) {
09:   ex(APPENDassoc@(X,Y) + LLLwfRl(X,Y))
10:   vars L1 L2 L3 : List .
11:   cq[aaWfiHyp :nonexec]:
12:      appendAssoc(L1,L2,L3) = true
13:      if t(l1@,l2@,l3@) wf> t(L1,L2,L3) . }

14:   mod APPENDassocWfiPtc(X :: TRIV) {
15:   pr(APPENDassocWfiHyp(X,LIST=(X)))
16:   op e$ : -> Elt . op l1$ : -> List . }

17:   select APPENDassocWfiPtc .
18:   :goal{eq appendAssoc(l1@,l2@,l3@)= true .}
19:   :def l1@ = :csp{eq l1@ = nil .
20:                   eq l1@ = e$ | l1$ .}
21:   :def aaWfiHyp = :init [aaWfiHyp]
22:                   by {L1:List <- L1:List;}
23:   :apply(l1@ rd- aaWfiHyp rd-)
\end{verbatim}
\end{cosmall}

Module \ttt{LLLwfRl} (\ttt{01:-06:}) defined
the well-founded relation \ttt{\us{wf>}\us} (\ttt{04:})
on \ttt{Lll} (3-tuples of \ttt{List})
with the equation \ttt{06:}.
Like the equation \ttt{06:},
the majority of well-founded relations \ttt{\us{wf>}\us}
are defined based on the strict subterm relations
on constructor terms.
The equation \ttt{11:-13:} corresponds to
``\ttt{cq\;{$p(\bar{y'}$)}\,{=}\;{true}\;{if}\;%
  $\overline{y@}\;{\wfg}\;\bar{y'}$\,{.}}'' 
of {\bf Fact \ref{fact-wfishe}}.
\ttt{16:} declares two constants \ttt{e\dl}, \ttt{l1\dl}
for {\bf refining} the constant \ttt{l1@}
at \ttt{20:}.
\ttt{18:-23:} proves the premise of
{\bf Fact \ref{fact-wfishe}} with PTcalc.

\section{Proof Scores for Transition Systems}
\label{sec-prctrsy}

It is widely recognized that the majority of services and systems
in many fields can be modeled as transition systems.
A {\bf transition system }is defined as a 3-tuple
$(\mathit{St},\mathit{Tr},In)$.  $\mathit{St}$ is a set of states,
$\mathit{Tr}\,{\subseteq}\,\mathit{St}{\times}\mathit{St}$ is a set of
transitions on the states (i.e., a state transition relation),
and $In\,{\subseteq}\,\mathit{St}$ is a set of
initial states.
$(s,s')\,{\in}\,\mathit{Tr}$ denotes the transition from
a state $s$ to a state $s'$\footnote{%
Symbols like $s$, $s'$, $s_i$ were used to denote sorts
({\it\ref{sssec-sorts}}).
The same symbols denote states also, but
can be distinguished by the context.
}.
A finite or infinite sequence of states
$s_1 s_2 \cdots s_n$ ($1 \le n$) or
$s_1 s_2 \cdots$ with 
$(s_i,s_{i+1})\,{\in}\,\mathit{Tr}$ for each
$i\,{\in}\,\{1,\cdots\!,n{-}1\}$ or
$i\,{\in}\,\{1,2,\cdots\}$ 
is defined to be a {\bf transition sequence}.  
Note that any $s\,{\in}\,\mathit{St}$ is defined to be
a transition sequence of length 1.

\subsection{Invariant Properties}
\label{ssec-invprp}

Given a transition system
$\mathit{TS}\,{=}\,(\mathit{St},\mathit{Tr},\mathit{In})$,
a state $s^{r}{\in}\,\mathit{St}$ is defined to be
{\bf reachable} iff there exists a transition sequence 
$s_1 s_2 \cdots s_n$ ($1 \le n$) with $s_n{=}s^r$ 
such that $s_1\,{\in}\,\mathit{In}$.
A {\bf state predicate} (a function from $\mathit{St}$ to
\ttt{Bool}) $p$  is defined to be an {\bf invariant} (or an invariant
property) iff $p(s^{r})\,{=}\,\mttt{true}$ for any reachable state
$s^{r}$.
A state predicate $p$ may have an extra data argument with a declaration
like ``\ttt{pred\:$p$\:{:}\:{\it St}\:{\it Data}\:{.}}''.

\bfact{\it (Invariants)} \citep{futatsugi15}
\label{fact-inv}
Let {\it init} be a state predicate that specifies the
initial states, 
that is, $\forall{s}\,{\in}\,\mathit{St} (init(s)\,{\Lra}\,s{\in}\mathit{In})$.
Let 
$p_1,p_2,$ $\cdots$, $p_n (1 \le n )$ 
be  state predicates, and
$iinv(s) = (p_1(s)$ $\land$ $p_2(s)$ 
$\land$ $\cdots$ $\land$ $p_n(s))$ for $s$ $\in$ $\mathit{St}$.
The following two conditions are sufficient for $p_i (1 \le i \le n)$
to be an invariant.\\
\hspace*{2em}{\bf (INV1)}$\;\;\forall{s}\in\mathit{St}\,(init(s)\;{\Ra}\;iinv(s))$ \\
\hspace*{2em}{\bf (INV2)}$\;\;\forall{(s,s')}\in\mathit{Tr}\,(iinv(s)\;{\Ra}\;iinv(s'))$
\hfill\makebox[0pt][r]{\owari}
\efact

A predicate that satisfies the conditions {\bf (INV1)}
and {\bf (INV2)} like $iinv$ is called 
an {\bf inductive invariant}.
If a state predicate $p$ itself is an inductive invariant then 
taking $p_1 = p$ and $n {=} 1$ is enough.
However, $p_1, p_2, \cdots, p_n$ $(n > 1)$ are almost always needed
to be found for getting an inductive invariant,
and to find them is a most difficult part of 
the invariant verification.  

\subsection{Observational Transition Systems (OTS)}
\label{subsec-ots}

An equational specification $\mathit{SP} = (\vSg,E)$,
$\vSg = (S,\le,$ $F,F^c)$ 
is called an OTS specification if {\it SP} is described
as follows and defines a transition system
$\mathit{TS}\, =\,(\mathit{St},\mathit{Tr},\mathit{In})$.
An OTS is defined as an $(S,\le,F,F^c)$-algebra here
and differs from the original treatment in \citep{OgataF03}.

Assume 
(i) a unique state sort $\mathtt{St} \in S$ and
$S^d \df S {-} \{\mttt{St}\}$,
(ii) a unique initial state constant 
$\mttt{init} \in (T_\vSg)_{\mathtt{St}}$,
(iii) a set of operators called {\bf actions} $Ac \subseteq F^c$
of which each element $a_i \in Ac\,(1 \le i \le m_a)$ has a rank
$\mttt{St} s^{a_i}_1 {\cdots} s^{a_i}_{n_{a_i}} {\mttt{St}} 
\,(0 \le n_{a_i},\,s^{a_i}_{j} \in S^d,\,0 \le j \le n_{a_i})$,
(iv) a set of operators called {\bf observers} $Ob \subseteq F$
of which each element $o_i \in Ob\,(1 \le i \le m_o)$ has a rank
$\mttt{St} s^{o_i}_1 {\cdots} s^{o_i}_{n_{o_i}} s^{o_i}\, 
(0 \le n_{o_i},\,s^{o_i}_{j} \in S^d,\,0 \le j \le n_{o_i},\,s^{o_i} \in S^d)$.

The observers give observable values of a system,
and the behavior of the system is defined
by describing the changes of the observable values
before and after the actions with the 
conditional equations of the following form for
$1 \le i \le m_o$, $1 \le j \le m_a$, 
$1 \le k \le l_{{o_i}{a_j}}$:\\
\hspace*{0.3em}
{\tt
cq {$o_i(a_j(X^{\mathtt{St}},\overline{X^{a_j}}),\overline{X^{o_i}})$}
{=}
{$v^{o_{i}a_{j}}_k(X^{\mathtt{St}},\overline{X^{a_j}},\overline{X^{o_i}})$}
{if}
{$c^{{o_i}{a_j}}_k(X^{\mathtt{St}},\overline{X^{a_j}},\overline{X^{o_i}})$}
{.}}\\
where \\
\hspace*{1em}
$\overline{X^{o_i}}\,=\,X^{o_i}_1,{\cdots},X^{o_i}_{n_{o_i}}$,
$\overline{X^{a_j}}\,=\,X^{a_j}_1,{\cdots},X^{a_j}_{n_{a_j}}$,\\
each $X$ is a variable of the following sort\\
\hspace*{1em}
$X^{\mathtt{St}}${:}\ttt{St},
$X^{o_i}_h${:}$s^{o_i}_h$\,($1 \le h \le n_{o_i}$),
$X^{a_j}_h${:}$s^{a_j}_h$\,($1 \le h \le n_{a_j}$),\\
and\\
\hspace*{1em}$v^{o_{i}a_{j}}_k$($X^{\mathtt{St}},\overline{X^{a_j}},\overline{X^{o_i}}$) 
$\in
T_{\vSg}(X^{\mathtt{St}},\overline{X^{a_j}},\overline{X^{o_i}})_{s^{o_i}}$,\\ 
\hspace*{1em}$c^{o_ia_j}_k$($X^{\mathtt{St}},\overline{X^{a_j}},\overline{X^{o_i}}$)
$\in
T_{\vSg}(X^{\mathtt{St}},\overline{X^{a_j}},\overline{X^{o_i}})_{\mathtt{Bool}}$.\\
Then the transition system 
$\mathit{TS}\, =\,(\mathit{St},\mathit{Tr},\mathit{In})$
is defined as follows.%
\begin{enumerate}[(i)]
\setlength{\itemsep}{-1pt}

\item $\mathit{St} \df (T_{\vSg})_{\mathtt{St}}$,

\item $\mathit{Tr} \df \{\,(t_0,a_i(t_0,t_1,{\cdots},t_{n_{a_i}})) \mid
a_i(t_0,t_1,{\cdots},t_{n_{a_i}})\,{\in}\,(T_\vSg)_{\mathtt{St}}\}$,

\item $\mathit{In} \df \{\mathtt{init}\}$.

\end{enumerate}

The proof scores for 
the verification conditions of invariant
properties ({\bf Fact \ref{fact-inv}} {\bf (INV1) (INV2)})
on OTS specifications are well studied 
from the early age of CafeOBJ
\citep{OgataF03}.
Quite a few significant cases are developed as basic proof scores
\citep{futatsugi06,futatsugiGO08,futatsugi10},
and fairly non-trivial automatic proofs with
CITP (\citep{gainaZCA13,RiescoO18}) are achieved
\citep{gainaLOF14,RiescoO22}.

\subsection{({\bf\it p} leads-to {\bf\it q}) Properties}
\label{subsec-pltqpr}

Invariants are fundamentally important properties of
transition systems.  They are asserting that something bad
will not happen (safety).  However, it is
sometimes also important to assert that something good will
surely happen (liveness).  
\citep{OgataF08,PreiningOF14} studied
proof scores for liveness properties.  
This subsection focuses on ($p$ leads-to $q$) properties
and provides a concrete and unified approach 
for constructing proof scores for liveness properties.

For two state predicates $p$ and $q$\footnote{%
$p$ and $q$ may be with {\it Data} argument,
i.e., ``\ttt{preds} $p$ $q$ \ttt{:} {\it State} {\it Data} \ttt{.}''.
}, %
the {\bf ($p$ leads-to $q$) property} is a liveness property
asserting that a transition system will get into
a state $\hat{s}$ with $q(\hat{s})\,{=}\,\mttt{true}$ 
whenever the system gets into a state $s$ with
$p(s)\,{=}\,\mttt{true}$ no matter what transition sequence
is taken.
More formally, a transition system is defined to have
the ($p$ leads-to $q$) property as follows.
A finite transition sequence $s_1 \cdots s_n$ ($1 \le n$) is
called {\bf terminating} if there is no $s'$ such that
$(s_n,s'){\in}Tr$.
A transition system is defined to have
the ($p$ leads-to $q$) property 
iff 
for any terminating or infinite transition sequence
starting from the state $s$ such that
$s$ is reachable and
$p(s)\,{=}\,\mttt{true}$,
there exists a state $\hat{s}$ 
such that $q(\hat{s})\,{=}\,\mttt{true}$
somewhere in the sequence.


The ($p$ leads-to $q$) property is adopted from the UNITY logic
\citep{ChandyM89ppd}, the above definition is, however, not
the same as the original one.  In the UNITY logic, the basic
model is the parallel program with parallel assignments, and
($p$ leads-to $q$) is defined through applications of inference
rules.

\bfact {\it ($p$ leads-to $q$)} \citep{futatsugi15,futatsugi17}
\label{fact-pltq}
Let
$\mathit{TS}\,{=}\,(\mathit{St},\mathit{Tr},\mathit{In})$
be a transition system,
$p$, $q$, $\mathit{inv}$ be state predicates
with
$\mathit{inv}$ being a $\mathit{TS}$'s invariant property,
and $m$ be a function from $\mathit{St}$ to 
the set of natural numbers.
If the following {\bf (LT1)} and {\bf (LT2)}
are proved by choosing
$\mathit{inv}$ and $m$ appropriately,
then $\mathit{TS}$ has the ($p$ leads-to $q$) property.

{\bf (LT1)}\hspace*{0.5em}%
$\forall(s,s'){\in}\mathit{Tr}\\
\hspace*{5.5em}((\mathit{inv}(s)\,{\land}\,p(s)\,{\land}\,{\lnot}q(s))%
\,{\Ra}\,
((p(s')\,{\lor}\,q(s'))\,{\land}\,(m(s)\,{>}\,m(s'))))$

{\bf (LT2)}\hspace*{0.5em}%
$\forall{s}{\in}\mathit{St}\,%
((\mathit{inv}(s)\,{\land}\,p(s)\,{\land}\,{\lnot}q(s))\,{\Ra}\,%
(\exists{s'}{\in}\mathit{St}((s,s'){\in}\mathit{Tr})))$%
\hfill\makebox[0pt][r]{\owari}
\efact

A ($p$ leads-to $q$) property sometimes needs to be proved in
multiple steps by making use of the following facts which are
easily obtained from the definition of ($p$ leads-to $q$).

\bfact {\it (Multiple leads-to)}
\label{fact-mlpltq}
Let ($p$ lt$\rhd$ $q$) stand for ($p$ leads-to $q$)
and $p$,{\,}$q$,{\,}$r$ be state predicates,
then the followings hold.

{\bf (MLT1)}\hspace*{0.5em}%
(($p$ lt$\rhd$ $q$)\,{$\land$}\,($q$ lt$\rhd$ $r$))%
\,{$\Ra$}\,($p$ lt$\rhd$ $r$)

{\bf (MLT2)}\hspace*{0.5em}%
(($p$ lt$\rhd$ $q$)\,{$\lor$}\,($p$ lt$\rhd$ $r$))%
\,{$\Ra$}\,($p$ lt$\rhd$ $q{\lor}r$)

{\bf (MLT3)}\hspace*{0.5em}%
(($p$ lt$\rhd$ $r$)\,{$\land$}\,($q$ lt$\rhd$ $r$))%
\,{$\Ra$}\,($p{\lor}q$ lt$\rhd$ $r$)%
\hfill\makebox[0pt][r]{\owari}
\efact

\subsection{Transition Specifications and Builtin Search
  Predicates \citep{futatsugi15}}
\label{subsec-trrbtnsp}

A natural way to define a transition system 
$\mathit{TS}\,{=}$ $(\mathit{St},\mathit{Tr},\mathit{In})$
is to define transitions $\mathit{Tr}$
with rewrite rules ({\it\ref{sssec-rwrd}}) on
$\mathit{St}$.
The CafeOBJ's builtin search predicates facilitate
verifications of the transition systems 
with the rewrite rules\footnote{%
The ideas underlying this subsection were first presented in
\citep{futatsugi15}.  The content of this subsection is much
more elaborated based on the unified theories described in
Section {\bf\ref{sec-fth}}.
}.

\subsubsection{Transition Pre-Spec}
\label{sssec-trnpspc}

Let \ttt{State} be a special sort that satisfies (i) there is
no supersort ({\it\ref{sssec-sorts}}) of \ttt{State} (ii) no
term of the sort \ttt{State} has a strict subterm of the sort
\ttt{State}.
A sort like \ttt{State} is called {\bf topmost sort}%
\footnote{%
In \citep{Meseguer12rwl20y} a topmost sort $\mathit{ts}$ is defined to be
(i) a topmost sort of a connected component
of order-sorted sorts, such that
(ii) no operator has {\it ts} as the sort of any of its arguments.
}%
, and the quotient set $(T_{\vSg,E{\cup}AC})_{\mathtt{State}}$
({\it\ref{subsec-initalg}})
is supposed to model the state space of 
the system specified, where {\it AC} is the AC-equations
for the AC operators in $\vSg$ ({\it\ref{ssec-ddrdcex}}).

Let 
``\ttt{cq{\;}$l(X)${\;}={\;}$r(X)${\;}if{\;}$c(X)${\;}.}'' be
a rewrite rule on the sort \ttt{State},
i.e., $l(X), r(X)\in (T_{\vSg}(X))_{\mathtt{State}}$ ({\it\ref{sssec-rwrd}}),
then 
``\ttt{ctr{\;}$l(X)${\;}=> $r(X)${\;}if $c(X)${\;}.}''
is called a {\bf transition rule}.
\ttt{ctr} stands for conditional transition
and \ttt{ctrans} or \ttt{cr} can be used instead.
If $c(X)$ = \ttt{true}, a transition rule is written as
``\ttt{tr{\;}$l(X)${\;}=> $r(X)${\;}.}'';
\ttt{trans} can be used instead of \ttt{tr}.

Let $\mathit{Tl}$ be a set of transition rules
then a 3-tuple $(\vSg,E,\mathit{Tl})$ is called
a {\bf transition pre-spec} and defines
{\bf one-step transition relation}
${\tr{Tl/E/AC}}\,{\subseteq}$
$(T_\vSg)_{\mathtt{State}}{\times}$ $(T_\vSg)_{\mathtt{State}}$
and {\bf transition relation} 
${\trstr{Tl/E/AC}}\,{\subseteq}$
$(T_\vSg)_{\mathtt{State}}{\times}$ $(T_\vSg)_{\mathtt{State}}$
by the following inference rules,
where $t_l, t_r, t_m,\,{\in}\,(T_\vSg)_{\mathtt{State}}$ and
${\dedstr{E/AC}}$ is defined by $(\vSg,E)$ ({\it\ref{sssec-eqded}}).
\begin{enumerate}

\item[(a)] $(t_l\,{\dedstr{E/AC}}\,t_r) \Ra t_l\,\trstr{Tl/E/AC}\,t_r$.

\item[(b)] $t_l\,{\tr{Tl/E/AC}}\,t_m \land t_m\,{\trstr{Tl/E/AC}}\,t_r \Ra 
t_l\,{\trstr{Tl/E/AC}}\,t_r$.

\item[(c1)] For a transition rule%
``\ttt{ctr{\,}$l(X)${\,}=>{\,}$r(X)${\,}if{\,}$c(X)${\,}.}''%
${\in}\,\mathit{Tl}$,
a valuation $\theta:X \ra T_\vSg$,
if $t_l\,{\dedstr{E/AC}}\,\theta(l(X))$ $\land$
   $t_r\,{\dedstr{E/AC}}\,\theta(r(X))$,\\
\hspace*{4em}$\theta(c(X))\,{\dedstr{E/AC}}\,\ttt{true}
\:\Ra\: t_l\;{\tr{Tl/E/AC}}\;t_r$.

\end{enumerate}  

Another one-step transition relation
${\tr{Tl/E{$\cup$}AC}}\,{\subseteq}$
$(T_\vSg)_{\mathtt{State}}{\times}$ $(T_\vSg)_{\mathtt{State}}$
and transition relation
${\trstru{Tl/E{$\cup$}AC}}\,{\subseteq}$
$(T_\vSg)_{\mathtt{State}}{\times}$ $(T_\vSg)_{\mathtt{State}}$
are defined via the same inference rules (a), (b), (c1)
by using ${\dedus{E}{AC}}$ ({\it\ref{sssec-eqded}})
instead of ${\dedstr{E/AC}}$.
Note that ${\tr{Tl/E{$\cup$}AC}}$ and 
${\trstru{Tl/E{$\cup$}AC}}$
are transition relations on the ideal state space composed of
the quotient set $(T_{\vSg,E{\cup}AC})_{\mathtt{State}}$
({\it\ref{subsec-initalg}}), and,
since {\it\ref{sssec-eqded}}-{\bf (E/AC)} asserts
(${\dedus{E}{AC}}$ = ${\dedstr{E/AC}}$),
we have 
(${\tr{Tl/E{$\cup$}AC}}$ = ${\tr{Tl/E/AC}}$) and
(${\trstru{Tl/E{$\cup$}AC}}$ = ${\trstr{Tl/E/AC}}$).

CafeOBJ implements
one-step transition relation
${\tr{Tl/E/AC}}$ 
with {\bf one-step weak transition relation}
${\tr{Tl/E,AC}}$,
and transition relation
${\trstr{Tl/E/AC}}$
with {\bf weak transition relation}
${\trstr{Tl/E,AC}}$,
via the following inference rules
using \ttt{\us}$^{\mathtt{red}}$ ({\it\ref{sssec-coex}})
and \ttt{\us}$=_{\!\mtnit{AC}}$\ttt{\us} ({\it\ref{ssec-ddrdcex}})
defined by $(\vSg,E)$.
Note that these inference rules can be execution/computation rules.
That is, you can compute transition sequence
$st_0\,{\tr{Tl/E,AC}}\,st_1\,{\tr{Tl/E,AC}}\,\cdots\,st_n\,{\tr{Tl/E,AC}}\,\cdots$
for a given state $st_0$.
\begin{enumerate}

\item[(a)]
  $(t_l^{\mathtt{red}}\,{=_{\!\mtnit{AC}}}\,t_r^{\mathtt{red}}) \Ra
  t_l\,\trstr{Tl/E,AC}\,t_r$.

\item[(b)] $t_l\,{\tr{Tl/E,AC}}\,t_m \land t_m\,{\trstr{Tl/E,AC}}\,t_r \Ra 
t_l\,{\trstr{Tl/E,AC}}\,t_r$.

\item[(c1)] For a transition rule%
``\ttt{ctr{\,}$l(X)${\,}=>{\,}$r(X)${\,}if{\,}$c(X)${\,}.}''%
${\in}\,\mathit{Tl}$,
a valuation $\theta:X \ra T_\vSg$,
if $(t_l)^{\mathtt{red}}\,{=_{\!\mtnit{AC}}}\,\theta(l(X))$,\\
\hspace*{4em}$(\theta(c(X)))^{\mathtt{red}}\,{=_{\!\mtnit{AC}}}\,\ttt{true}
\:\Ra\: t_l\;{\tr{Tl/E,AC}}\;\theta(r(X))$.

\end{enumerate}  

A transition rule in $\mathit{Tl}$ of a transition pre-spec
$(\vSg,E,\mathit{Tl})$ is called {\bf left reduced} if its
left hand side $l(X)$ is ${\rdc{E,AC}}$-reduced modulo AC in $(\vSg,E)$
({\it\ref{sssec-coex}}) by
considering $X$ as a set of fresh constants.
That is, $(l(X))^{\mathtt{red}} = l(X)$.
A set of transition rules {\it Tl} is called
{\bf left reduced} if each element is so.
Each effective transition rule is assumed to be left reduced in CafeOBJ.
This assumption is natural, for 
``\ttt{st(5) => final}'' is better and safer than
``\ttt{st(2 + 3) => final}''.

\subsubsection{Transition Specifications}
\label{sssec-trnspc}

Let $(\varSigma,E,\mathit{Tl})$ be a transition pre-spec and
$\mathit{In}\,{\subseteq}\,(T_\vSg)_{\mathtt{State}}$
then a 4-tuple
$(\varSigma,E,\mathit{Tl},\mathit{In})$ is called
a {\bf transition specification}.
A transition specification defines the following two transition systems
({\it\ref{sec-prctrsy}})
based on the one-step weak transition relation $\tr{Tl/E,AC}$.\\
\hspace*{3em}
(TS1)
($(T_{\vSg,E{\cup}AC})_{\mathtt{State}},{\tr{[Tl/E,AC]}},
\{ [st]_{\equiv^{E{\cup}AC}} | st \in In \}$),\\
\hspace*{3em}
(TS2) ($(T_\vSg)_{\mathtt{State}},\tr{Tl/E,AC},\mathit{In}$).\\
The transition relation:\\
\hspace*{4em}
${\tr{[Tl/E,AC]}}\,{\subseteq}$
$(T_{\vSg,E{\cup}AC})_{\mathtt{State}}{\times}(T_{\vSg,E{\cup}AC})_{\mathtt{State}}$\\
on $(T_{\vSg,E{\cup}AC})_{\mathtt{State}}$ is defined as follows.\\
\hspace*{4em}
$([st_1]_{\equiv^{E{\cup}AC}}\,{\tr{[Tl/E,AC]}}\,[st_2]_{\equiv^{E{\cup}AC}})$ iff\\
\hspace*{4.2em}
$\exists st'_1\,st'_2
(st_1\,{\dedus{E}{AC}}\, st'_1 \land
 st_2\,{\dedus{E}{AC}}\,st'_2 \land
 st'_1\,{\tr{Tl/E,AC}}\,st'_2)$

(TS1) is an ideal system and can not be implemented directly
and CafeOBJ implements (TS2).
$[st]_{\equiv^{E{\cup}AC}}$ consists of all the terms $st'$
such that $st'\,{\dedus{E}{AC}}\,st$, and
if $(\vSg,E)$ is red-complete ({\bf Fact \ref{fact-cansys}}),
each $[st]_{\equiv^{E{\cup}AC}}$ can be represented by 
the unique term $st'^{\mathtt{red}}$ for any $st' \in [st]_{\equiv^{E{\cup}AC}}$
modulo AC. 
In this case, it is easy to see from the rule (c1)
for $\tr{Tl/E,AC}$ ({\it\ref{sssec-trnpspc}}) and 
{\it\ref{sssec-compspec}}-{\bf (E,AC)}
that ${\tr{Tl/E,AC}}$ is complete enough such that\\
\hspace*{0.2em}{\bf (Tl/E,AC)}
\hspace*{0.1em}
$\forall t, u\,{\in}\,T_{\vSg} (t\,{\tr{Tl/E/AC}}\,u \Ra
\exists u'\,{\in}\,T_{\vSg} (t\,{\tr{Tl/E,AC}}\,u' \land
u^{\mathtt{red}}\,{=_{\!\mtnit{AC}}}\,u'^{\mathtt{red}}))$ .\\
Hence, the following bi-simulation relation is obtained.\\
\hspace*{0.2em}{\bf (BS)}
$([st_1]_{\equiv^{\!E{\cup}AC}}\,{\tr{[Tl/E,AC]}}\,[st_2]_{\equiv^{\!E{\cup}AC}})$%
\,{$\Lra$}\,$\exists st'_2%
(st_1\,{\tr{Tl/E,AC}}\,st'_2\,{\land}\,%
st^{\mathtt{red}}_{2}\,{=_{\! \mathit{AC}}}\,st'^{\mathtt{red}}_{2})$\\
Invariant/leads-to properties of
a transition system (TS1) are verified
by constructing proof scores for the corresponding (TS2).

A verification condition
{\bf Fact \ref{fact-inv}} {\bf (INV1)}
can be proved without using the builtin search predicates.
Verification conditions
{\bf Fact \ref{fact-inv}} {\bf (INV2)},
{\bf Fact \ref{fact-pltq}} {\bf (LT1) (LT2)}, 
however, need builtin search predicates
that search all possible transitions in (TS2), 
and that implies to search all possible transitions
in (TS1) thanks to {\bf (BS)}.

\subsubsection{Builtin Search Predicates}
\label{sssec-btnsp}

The primary builtin search predicate is declared as follows.\\
{\verb|   pred _=(*,1)=>+_if_suchThat_{_} : State State Bool Bool Info .|}\\
\noindent 
\ttt{Info} is a sort for outputting information;
the 1st argument is the current state
$s \in (T_\vSg)_{\mathtt{State}}$;
the 2nd and 3rd arguments are variables 
\ttt{SS:State} and \ttt{CC:Bool} for binding 
the next state and the condition to be found, respectively;
the 4th argument is a predicate \ttt{p($s$,SS,CC)}
whose validity is to be checked;
the 5th argument is a term \ttt{i($s$,SS,CC)} of sort \ttt{Info}
for outputting the information.

For a state term $s \in (T_\vSg)_{\mathtt{State}}$
the CafeOBJ's reduction of a Boolean term:\\
\hspace*{3em}\ttt{$s$ =(*,1)=>+ SS:State if CC:Bool\\
\hspace*{8.8em}suchThat p($s$,SS,CC) \ob{i($s$,SS,CC)}\cb }\\
in a transition pre-spec $M$ = ($\vSg,E,\mathit{Tl}$) behaves as follows.

\begin{itemize}
  \setlength{\itemsep}{-2pt}
  
\item[(i)] Search for every pair $(\mathit{tl},\theta)$ of 
a transition rule 
$\mathit{tl} =$ ``\ttt{ctr{\;}$l(X)${\;}=>{\;}$r(X)${\;}if $c(X)${\;}.}''
in $\mathit{Tl}$ and 
a valuation $\theta : X \ra T_{\varSigma}$
such that $s^{\mathtt{red}} =_{\!\mtnit{AC}} \theta(l(X))$. 

\item[(ii)] For each found pair $(tl,\theta)$, 
let ($\mttt{SS}\,{=}\,\theta(r(X))$) and ($\mttt{CC}\,{=}\,\theta(c(X))$)
and print out 
\ttt{i($s$,SS,CC)} 
and $tl$
if
(\ttt{p($s$,SS,CC)})$^{\mathtt{red}}$
= \ttt{true}\,.

\item[(iii)] Returns \ttt{true} if some print out exists,
and returns \ttt{false} otherwise.

\end{itemize}

\ttt{p($s$,SS,CC)} can be defined to
check whether a relation between current state $s$ and next state
\ttt{SS} holds.
Let ``\ttt{pred cnr : State State Data .}''
(Current Next Relation)
be a predicate stating some relation like one in 
{\bf Fact \ref{fact-inv}} {\bf (INV2)} or
{\bf Fact \ref{fact-pltq}} {\bf (LT1)}
of the current and next states.
\ttt{Data} may be necessary if it appears
in some state predicates involved 
in the current/next state relation.

Let \ttt{check-cnr} be defined as follows.
Note that an argument \ttt{D:Data} is added to 
\ttt{p(S,SS,CC)} and \ttt{i(S,SS,CC)}, and
\ttt{p(S,SS,CC,D)} corresponds to
``\ttt{not((CC implies cnr(S,SS,D)) == true)}''.

\begin{verbatim}
  pred check-cnr : State Data .
  eq check-cnr(S:State,D:Data) =
      not(S =(*,1)=>+ SS:State if CC:Bool
          suchThat not((CC implies cnr(S,SS,D)) == true)
          {i(S,SS,CC,D)}) .
\end{verbatim}

\noindent
Let $s {\in} (T_{\Sigma})_{\mathtt{State}}$,
$d {\in} (T_{\Sigma})_{\mathtt{Data}}$, and assume
``(\ttt{check-cnr($s$,$d$)})$^{\mathtt{red}}$ = \ttt{true}''
in the transition pre-spec $M$,
i.e., $M{\tsc}$\ttt{check-cnr($s$,$d$)}.
The reduction of ``\ttt{check-cnr ($s$,$d$)}'' involves
the reduction of\\
\hspace*{3em}\ttt{$s$ =(*,1)=>+ SS:State if CC:Bool\\
\hspace*{3em}suchThat not((CC implies cnr($s$,SS,$d$)) == true)\\
\hspace*{3em}\{i($s$,SS,CC,$d$)\}}\\
and searches all the 3-tuples ($s$, \ttt{SS}, \ttt{CC})
for the pairs $(tl,\theta)$ 
such that ($s^{\mathtt{red}} =_{\!\mtnit{AC}} \theta(l(X))$),
($\mttt{SS}\,{=}\,\theta(r(X))$), and ($\mttt{CC}\,{=}\,\theta(c(X))$)
(see (i), (ii) above).
Therefore,\\
\hspace*{2em}(\ttt{check-cnr($s$,$d$)})$^{\mathtt{red}}$ = \ttt{true}\\
implies\\
\hspace*{2em}for any searched 3-tuples ($s$, \ttt{SS}, \ttt{CC})\\
\hspace*{3em}(\ttt{((CC implies cnr($s$,SS,$d$)) == true)} = \ttt{true}),\\
and it implies\\
\hspace*{2em}for any one-step transition
\ttt{($s$,SS)}${\in}\,{\tr{Tl/E,AC}}$
(i.e., for the case \ttt{CC} = \ttt{true})\\
\hspace*{3em}(\ttt{cnr($s$,SS,$d$)} = \ttt{true}),\\
and the following Fact is obtained.
Note that the definition of the returned values
given in (iii) and the two \ttt{not}\hspace{0.1em}s in the definition of
\ttt{check-cnr($s$,$d$)} delicately support the above implications.

\bfact \label{fact-validq}
For $s {\in} (T_{\Sigma})_{\mathtt{State}}$,
$d {\in} (T_{\Sigma})_{\mathtt{Data}}$
of a transition pre-spec $M\,{=}\,(\vSg,E,\mathit{Tl})$,\\
\hspace*{2em}$M{\tsc}$\ttt{check-cnr($s$,$d$)}
$\Ra$ $\forall (s,s'){\in}\,{\tr{Tl/E,AC}}$%
$($\ttt{cnr($s,s',d\,$)} $=$ \ttt{true}$)$ .
\hfill\makebox[0pt][r]{\owari}
\efact

There is another builtin search predicate:\\
\hspace*{3em}\ttt{pred {\us=(1,1)=>+\us}\;{:}\;State\;State\;.}\\
Reducing 
(\ttt{$s$ =(1,1)=>+ SS:State})
searches all the possible one-step transitions from a state $s$
and returns \ttt{true} if found.  Hence the following fact.

\bfact \label{fact-cont}
For $s \in (T_{\Sigma})_{\mathtt{State}}$ of 
a transition pre-spec $M = (\vSg,E,\mathit{Tl})$,\\
\hspace*{2em}$M{\tsc}\!($\ttt{$s$ =(1,1)=>+ SS:State}$)$
$\Ra$ $\exists (s,s'){\in}\,{\tr{Tl/E,AC}}$ .%
\hfill\makebox[0pt][r]{\owari}
\efact

Proof scores for verification conditions
{\bf (INV2)}, {\bf (LT1)} and {\bf (LT2)}
can be constructed by making use of 
{\bf Fact \ref{fact-validq}} and
{\bf Fact \ref{fact-cont}} respectively.

PTcalc+WFI proof scores for leads-to properties
of a practical size cloud protocol in
transition rules
have been developed by making use of
{\bf Fact \ref{fact-inv}},
{\bf Fact \ref{fact-pltq}}, 
{\bf Fact \ref{fact-validq}},
{\bf Fact \ref{fact-cont}} 
in a follow up research of \citep{YoshidaOF15}.

\section{QLOCK Example}
\label{sec-qlkex}

This section is devoted to show
how to construct proof scores for
transition systems based on theories and methods
presented in Sections {\bf\ref{sec-fth}}, {\bf\ref{sec-bscprs}}, 
{\bf\ref{sec-adprs}}, {\bf\ref{sec-prctrsy}}
by making use of 
a simple but non-trivial example QLOCK (Queue LOCKing) 
of mutual exclusion protocols.
Mutual exclusion protocol is described as follows:

\vspace{0.5em}
\textrm{
  \parbox{110mm} {Assume that many agents (or processes) are competing
    for a common equipment (e.g., a printer or a file system), but at
    any moment of time only one agent can use the equipment.  That is,
    the agents are mutually excluded in using the equipment.  A
    protocol (distributed mechanism or algorithm) which can achieve
    the mutual exclusion is called a {\bf mutual exclusion protocol}.  }
}\vspace{0.5em}

QLOCK is realized by using a
global queue (first in first out storage) of agent names 
(or identifiers) as follows:

\begin{itemize}

\item Each of unbounded number of agents who participates in the protocol
  behaves as follows:

\begin{itemize}

\item[{\bf[want]}] If the agent wants to use the common equipment and its name is
  not in the queue yet, put its name at the bottom of the queue.

\item[{\bf [try]}] If the agent wants to use the common equipment and its name is
  already in the queue, check if its name is on the top of the queue.
  If its name is on the top of the queue, start to use the common equipment.  
  If its name is not on the top of the queue, wait until its name is
  on the top of the queue.

\item[{\bf [exit]}] If the agent finishes to use the common equipment, remove its
  name from the top of the queue.

\end{itemize}

\item The protocol starts from the state with the empty queue.

\end{itemize}

\subsection{QLOCK as an Observational Transition System (OTS)}
\label{ssec-qlkots}

QLOCK can be described as an OTS specification ({\it\ref{subsec-ots}})
and the specification is called \ttt{QLOCK/OTS}.  
For an OTS specification,
(i) no need to design structure for state space,
(ii) no need to make some part red-complete for proof scores to be valid,
but to prove leads-to properties is difficult.

\subsubsection{\ttt{QLOCK/OTS}: System Specification of QLOCK as an OTS}
\label{sssec-sspqlkots}

For \ttt{QLOCK/OTS} and proof scores for that,
the specification \ttt{QUEhdTlPt} (\ttt{24:- 29:}) of  
generic (or parameterized) queues
with hd (head), tl (tail), pt (put) operators
is necessary.
\ttt{QUEhdTlPt} adds the definitions of operators
\ttt{tl\us} and \ttt{pt\us} to \ttt{QUEhd} (\ttt{18:-22:})
that adds the definition of operator \ttt{hd\us} to
\ttt{QUE} (\ttt{14:-16:}).
\ttt{QUE} is defined by renaming a sort and operators of
\ttt{SEQ=} (\ttt{07:-12:}) with \verb%*{...}% (\ttt{15:-16:}).
\ttt{SEQ=} enhances the builtin operator \ttt{\us{=}\us}
of \ttt{SEQ} that defines generic sequences.
Note that \ttt{hd\us} takes the left-most element of a queue,
and \ttt{pt\us} puts a element to a queue from the right.

\vspace*{0.2em}
\noindent
@\footnote{
\ttt{01:-76:} is in the files
\ttt{seq-spc.cafe},
\ttt{que-spc.cafe},
\ttt{ots-sys-spc.cafe}
on the Web:\\
\hspace*{2em}
\url{https://cafeobj.org/~futatsugi/misc/apsco-220907/}\,.
}
\vspace*{-1em}
\begin{cosmall}
\begin{verbatim}
01:   --> generic SEQuences
02:   mod! SEQ (X :: TRIV) {
03:   [Elt < Seq] -- an element is a sequence
04:   op nil : -> Seq {constr} .
05:   op __ : Seq Seq -> Seq {constr assoc id: nil} . }

06:   --> enhancing the builtin operator _=_ of SEQ
07:   mod! SEQ= (X :: TRIV) {
08:   pr(SEQ(X))
09:   eq (nil = (E:Elt S2:Seq)) = false .
10:   cq ((E1:Elt S1:Seq) = (E2:Elt S2:Seq)) =
11:      ((E1 = E2) and (S1 = S2)) 
12:      if not((S1 == nil) and (S2 == nil)) . }

13:   --> generic queues
14:   mod! QUE (X :: TRIV) {
15:   pr(SEQ=(X)*{sort Seq -> Que,
16:               op nil -> nilQ, op __ -> _|_}) }

17:   --> generic queues with hd (head) operator
18:   mod! QUEhd (X :: TRIV) {
19:   pr(QUE(X)) [Elt < EltEr]
20:   op hd_ : Que -> EltEr .
21:   eq hd(X:Elt | Q:Que) = X .
22:   eq (hd(nilQ) = X:Elt) = false . }

23:   --> generic queues with hd (head), tl (tail), pt (put) operators
24:   mod! QUEhdTlPt (X :: TRIV) {
25:   pr(QUEhd(X))
26:   op tl_ : Que -> Que .
27:   eq tl(X:Elt | Q:Que) = Q .
28:   op pt : Elt Que -> Que .
29:   eq pt(X:Elt,Q:Que) = Q | X . }
\end{verbatim}
\end{cosmall}

The module \ttt{LABEL} (\ttt{31:-37:}) specifies the set 
\{\ttt{rm wt cs}\}\footnote{%
These strange abbreviations are historical;
\ttt{rm} stands for remainder section,
\ttt{wt} stands for waiting section,
\ttt{cs} stands for critical section.
}%
of three objects of the sort 
\ttt{LabelLtl}.  
\ttt{\us{==}\us} is another builtin equality predicate of
CafeOBJ defined for $t_1, t_2\in\,T_{\vSg}$ as follows.
(\ttt{$t_1$\,{==}\,$t_2$})$^{\mathtt{red}}$ is \ttt{true}
if $t_1^{\mathtt{red}}\,{=_{\!\mtnit{AC}}}\,t_1^{\mathtt{red}}$,
and is \ttt{false} otherwise
(see {\it\ref{ssec-ddrdcex}} for
\ttt{\us{=}}$_{\!\mtnit{AC}}$\ttt{\us},
and {\it\ref{sssec-coex}} for \ttt{\us}$^{\mathtt{red}}$).
In contrast, for \verb|_=_| ({\it\ref{sssec-bibpred}}),
(\ttt{$t_1$\,{=}\,$t_2$})$^{\mathtt{red}}$ is \ttt{true}
if $t_1^{\mathtt{red}}\,{=_{\!\mtnit{AC}}}\,t_2^{\mathtt{red}}$,
and is ``\ttt{$t_1^{\mathtt{red}}$\,{=}\,$t_2^{\mathtt{red}}$}''
otherwise.
Hence, (\ttt{rm\,{=}\,wt})$^{\mathtt{red}}$ = \ttt{false},
but for a fresh constant ``\ttt{op\,l1\,{:}\,{->}\,Label\,{.}}'',
(\ttt{rm\,{=}\,l1})$^{\mathtt{red}}$ = ``\ttt{rm\,{=}\,l1}''.

\begin{cosmall}
\begin{verbatim}
30:   --> Label Literal (LabelLtl) and Label
31:   mod! LABEL {
32:   [LabelLtl < Label]
33:   ops rm wt cs : -> LabelLtl {constr} .
34:   -- var L : Label .
35:   -- eq (L = L) = true . -- instance of builtin equation
36:   vars Ll1 Ll2 : LabelLtl .
37:   eq (Ll1 = Ll2) = (Ll1 == Ll2) . }
\end{verbatim}
\end{cosmall}

The module \ttt{QLOCK/OTS} (\ttt{40:-76:})
has a parameter module \ttt{X\;{::}\;AID} (\ttt{41}),
and \ttt{AID} specifies any collection of objects 
without any constraints (\ttt{39:}).
It declares that \ttt{QLOCK/OTS} can apply
to any collection of agent identifiers
(or objects of sort \ttt{Aid}).
\ttt{QLOCK/OTS} imports the module \ttt{LABEL} (\ttt{42:})
and the module \ttt{QUEhdTlPt} with the
parameter \ttt{X} that is the parameter of 
\ttt{QLOCK/OTS} itself (\ttt{41:,43:}).

The state space of \ttt{QLOCK/OTS} is specified as
the set of constructor terms of the sort \ttt{St}
(\ttt{45:-50:}).
Each term of the sort \ttt{St},
i.e., a point in the state space,
is the term constructed from 
the actions \ttt{init, want, try, exit}
(see {\it\ref{subsec-ots}}).

\begin{cosmall}
\begin{verbatim}
38:   --> Agent Identifiers
39:   mod* AID {[Aid]}

40:   --> a system specification of QLOCK in OTS
41:   mod! QLOCK/OTS (X :: AID) {
42:   pr(LABEL)
43:   pr(QUEhdTlPt(X)*{sort EltEr -> AidEr})

44:   -- state space of system
45:   [St]
46:   -- actions of system
47:   op init : -> St {constr} .
48:   op want : St Aid -> St {constr} .
49:   op try  : St Aid -> St {constr} .
50:   op exit : St Aid -> St {constr} .
\end{verbatim}
\end{cosmall}

QLOCK is specified by
defining the values of the observers
\ttt{pc} and \ttt{que} (\ttt{52:-53:})
for each term of the sort \ttt{St}.
\ttt{55:-56:} defines the values of
\ttt{pc} and \ttt{que} for 
the term \ttt{init} representing initial states.
\ttt{60:-64:} defines the values of
\ttt{pc} and \ttt{que} for each
term with the top operator \ttt{want}.
\ttt{66:-70:} is for \ttt{try} and
\ttt{72:-76:} is for \ttt{exit}.

\begin{cosmall} 
\begin{verbatim}
51:   -- observers of system
52:   op pc    : St Aid -> Label .
53:   op que : St -> Que .

54:   -- for each initial state
55:   eq pc(init,I:Aid)  = rm .
56:   eq que(init) = nilQ .

57:   -- variables
58:   var S : St . vars I J : Aid .

59:   -- for want
60:   ceq pc(want(S,I),J) = (if I = J then wt else pc(S,J) fi)
61:       if (pc(S,I) = rm) .
62:   ceq pc(want(S,I),J) = pc(S,J) if not(pc(S,I) = rm) .
63:   ceq que(want(S,I)) = pt(I,que(S)) if (pc(S,I) = rm) .
64:   ceq que(want(S,I)) = que(S) if not(pc(S,I) = rm) .

65:   -- for try
66:   ceq pc(try(S,I),J) = (if I = J then cs else pc(S,J) fi)
67:       if (pc(S,I) = wt and hd(que(S)) = I) .
68:   ceq pc(try(S,I),J) = pc(S,J)
69:       if not(pc(S,I) = wt and hd(que(S)) = I) .
70:   eq  que(try(S,I)) = que(S) .

71:   -- for exit
72:   ceq pc(exit(S,I),J) = (if I = J then rm else pc(S,J) fi)
73:       if (pc(S,I) = cs) .
74:   ceq pc(exit(S,I),J) = pc(S,J) if not(pc(S,I) = cs) .
75:   ceq que(exit(S,I)) = tl(que(S)) if (pc(S,I) = cs) .
76:   ceq que(exit(S,I)) = que(S) if not(pc(S,I) = cs) . }
\end{verbatim}
\end{cosmall}
 
\subsubsection{Properties of Interest on \ttt{QLOCK/OTS}
 and WFI Hypotheses for Them}
\label{sssec-piwfihypoqlkots}

The main property of interest on \ttt{QLOCK/OTS} is
the mutual exclusion property \ttt{mtx(S:St,I:Aid,J:Aid)}
defined in \ttt{07:-09:}.
However, \\
\hspace*{3em}\ttt{OTSmxInvPrp} $\mdl$ \ttt{mtx(S:St,I:Aid,J:Aid)}\\
is difficult to prove by induction, and 
\ttt{qtp(S,K:Aid)} (\ttt{11:-12:}) is needed to be found to make\\
\hspace*{3em}\ttt{OTSmxInvPrp} $\mdl$
(\ttt{mtx(S:St,I:Aid,J:Aid) and qtp(S,K:Aid)})\\
provable by induction.
That is, (\ttt{mtx(S:St,I:Aid,J:Aid) and qtp(S,K:Aid)}) is
an inductive invariant (see {\it\ref{ssec-invprp}}),
but \ttt{mtx(S:St,I:Aid,J:Aid)} is not.

\vspace*{0.2em}
\noindent
@\footnote{
\ttt{01:-32:} is in the files
\ttt{ots-mx-inv-prp.cafe},
\ttt{ots-mx-wfi-hypo.cafe}
on the Web.}
\vspace*{-1em}
\begin{cosmall} 
\begin{verbatim}
01:   --> module defining predicates mtx and qtp
02:   mod OTSmxInvPrp {
03:   pr(QLOCK/OTS)
04:   -- variables
05:   vars S :  St . vars I J K : Aid . 

06:   --  predicate defining mutual exclusion (mtx) property
07:   pred mtx : St Aid Aid .
08:   eq mtx(S,I,J) =
09:      ((pc(S,I) = cs) and (pc(S,J) = cs)) implies (I = J) .

10:   -- predicate defining queue top (qtp) property
11:   pred qtp : St Aid .
12:   eq qtp(S,K) = (pc(S,K) = cs) implies (hd(que(S)) = K) . }
\end{verbatim}
\end{cosmall}

The module \ttt{OTSmxWfiHypo} (\ttt{13:-32:}) corresponds to
the module in the premise of {\bf Fact \ref{fact-wfishe}}
with the following correspondences:\\
\hspace*{2em}(1) \ttt{OTSmxInvPrp} $\lrh$ $M$,\\
\hspace*{2em}(2) \ttt{S:St,I:Aid,J:Aid,K:Aid} $\lrh$ $\bar{y}$, $\bar{y'}$,\\
\hspace*{2em}(3) \ttt{s@,i@,j@,k@} $\lrh$ $\overline{y@}$,\\
\hspace*{2em}(4) \ttt{mtx(S:St,I:Aid,J:Aid) and qtp(S,K:Aid)} $\lrh$
$p(\bar{y})$,\\
\hspace*{2em}(5) \ttt{16:-26:} $\lrh$ $M_{\scriptsize \wfg}{\cup}Y\!@$,\\
\hspace*{2em}(6) \ttt{28:\,{+}\,30:} $\lrh$
\ttt{cq\;{$p(\bar{y'})$}\,{=}\;{true}\;{if}\;%
$\overline{y@}\;{\wfg}\;\bar{y'}$\,{.}} ,\\
\hspace*{2em}(7) \ttt{OTSmxWfiHypo} $\lrh$
$M_{\scriptsize \wfg}{\cup}Y\!@{\cup}\{$%
\ttt{cq\;{$p(\bar{y'})$}\,{=}\;{true}\;{if}\;%
$\overline{y@}\;{\wfg}\;\bar{y'}$\,{.}\!\}}.\\
(4) shows that
$p(\bar{y})$ is a conjunction of
two conjuncts
\ttt{mtx(S:St,I:Aid,J:Aid)} and \ttt{qtp(S,K:Aid)},
and (6) shows that the WFI hypothesis for $p(\bar{y})$
is divided into two WFI hypotheses 
\ttt{28:} for \ttt{mtx} and \ttt{30:} for \ttt{qtp}.
Note that \ttt{wf>} on the argument tuples of $p(\bar{y})$\\
\hspace*{2em}\ttt{(s@,i@,j@,k@) wf> (S:St,I:Aid,J:Aid,K:Aid)}\\
is defined by ``\ttt{s@ wf> S:St}''
that is defined in \ttt{24:-26:}.

\ttt{OTSmxWfiHypo} is a typical example of 
a module defining WFI hypotheses for {\bf simultaneous induction}.

{\bf Fact \ref{fact-wfishe}} shows that\\
\hspace*{2em}\ttt{OTSmxWfiHypo} $\mdl$
(\ttt{mtx(s@,i@,j@) and qtp(s@,k@)})\\
is a sufficient condition for\\
\hspace*{2em}\ttt{OTSmxInvPrp} $\mdl$
(\ttt{mtx(S:St,I:Aid,J:Aid) and qtp(S,K:Aid)}).

\begin{cosmall} 
\begin{verbatim}
13:   --> proof module defining WFI hypotheses 
14:   mod OTSmxWfiHypo {
15:   -- the module on which this WFI applies
16:   pr(OTSmxInvPrp)
17:   -- fresh constants for declaring goal and WFI hypotheses
18:   -- goal proposition: "mtx(s@,i@,j@) and qtp(s@,k@)"
19:   op s@ : -> St . ops i@ j@ k@ : -> Aid .
20:   -- variables
21:   var S :  St . vars H I J K : Aid . 
22:   -- well-founded binary relation on St
23:   pred _wf>_ : St St .
24:   eq want(S,H) wf> S = true .
25:   eq try(S,H) wf> S = true .
26:   eq exit(S,H) wf> S = true .
27:   -- WFI hypotheses for mtx
28:   ceq[mtx-hypo :nonexec]: mtx(S,I,J) = true if s@ wf> S .
29:   -- WFI hypotheses for qtp
30:   ceq[qtp-hypo :nonexec]: qtp(S,K) = true if s@ wf> S .
31:   -- fresh constants for refinements
32:   op s$  : -> St . ops h$ g$ : -> Aid . op q$ : -> Que . }
\end{verbatim}
\end{cosmall}

\subsubsection{PTcalc Proof Scores on \ttt{OTSmxWfiHypo}}
\label{sssec-pscqlkots}

\ttt{01:-29:} is a proof score with PTcalc for\\
\hspace*{2em}\ttt{OTSmxWfiHypo} $\mdl$ \ttt{mtx(s@,i@,j@)}\\
and a similar proof score\footnote{%
In the file
\ttt{ots-mx-psc.cafe}
on the Web.}%
is made for\\
\hspace*{2em}\ttt{OTSmxWfiHypo} $\mdl$ \ttt{qtp(s@,k@)}\\
and together prove\\
\hspace*{2em}\ttt{OTSmxWfiHypo} $\mdl$ (\ttt{mtx(s@,i@,j@) and qtp(s@,k@)}).

\ttt{04:-18:} define user defined commands for PTcalc,
and \ttt{20:-29:} apply them for constructing
an effective proof tree (see {\it\ref{ssec-ptcalc}}) for
proving ``\ttt{mtx(s@,i@,j@) = true}'' (\ttt{03:}).
Instead of \ttt{03:}, the declaration of the goal\\
\hspace*{2em}
\ttt{:goal\{eq mtx(s@,i@,j@) = true . eq qtp(s@,k@) = true .\}}\\
is possible for proving\\
\hspace*{2em}\ttt{OTSmxWfiHypo} $\mdl$ (\ttt{mtx(s@,i@,j@) and qtp(s@,k@)})\\
all at once, but the proof score\footnote{%
In the file \ttt{ots-mx-2cj-psc.cafe} on the Web.
}%
tends to be complex.

\ttt{04:-13:} define case-split commands.
\ttt{c-s@iwte} (\ttt{04:-07:}) 
classifies the terms of sort \ttt{St} via 
the four action operators (constructors),
and is exhaustive (see {\it\ref{ssec-csi}}).
\ttt{:ctf\{eq $l$ = $r$ .\}} is an abbreviation
of \ttt{:csp\{eq $l$ = $r$ . eq ($l$ = $r$) = false .\}}
and is exhaustive for any $l$, $r$.
The equations defining the case-split in \ttt{08:-13:}
appear in the conditions of the conditional equations 
or \ttt{if{\us}then{\us}else{\us}fi} operators
in \ttt{QLOCK/OTS} 
({\it \ref{sssec-sspqlkots}}-\ttt{59:-76:}).


\ttt{14:-18:} define initialization commands
each of them initializing a WFI hypothesis
\ttt{[mtx-hypo]} ({\it\ref{sssec-piwfihypoqlkots}}-\ttt{28:})
or \ttt{[qtp-hypo]} ({\it\ref{sssec-piwfihypoqlkots}}-\ttt{30:}).


\vspace*{0.2em}
\noindent
@\footnote{
\ttt{01:-29:} is in the file
\ttt{ots-mx-psc.cafe}
on the Web.}
\vspace*{-1em}
\begin{cosmall} 
\begin{verbatim}
01:   --> executing proof of "mtx(s@,i@,j@) = true" with PTcalc
02:   select OTSmxWfiHypo .
03:   :goal{eq mtx(s@,i@,j@) = true .}
04:   :def c-s@iwte = :csp{eq s@ = init .
05:                        eq s@ = want(s$,h$) .
06:                        eq s@ = try(s$,h$) .
07:                        eq s@ = exit(s$,h$) .}
08:   :def c-s$h$rm = :ctf{eq pc(s$,h$) = rm .}
09:   :def c-s$h$wt = :ctf{eq pc(s$,h$) = wt .}
10:   :def c-s$h$cs = :ctf{eq pc(s$,h$) = cs .}
11:   :def c-i@=h$- = :ctf{eq i@ = h$ .}
12:   :def c-j@=h$- = :ctf{eq j@ = h$ .}
13:   :def c-tqs$h$ = :ctf{eq hd(que(s$)) = h$ .}
14:   :def i-s$I-J- = :init as mxs$I-J- [mtx-hypo] by {S:St <- s$;}
15:   :def i-qts$i@ = :init as qts$i@ [qtp-hypo]
16:                   by {S:St <- s$; K:Aid <- i@;}
17:   :def i-qts$j@ = :init as qts$j@ [qtp-hypo]
18:                   by {S:St <- s$; K:Aid <- j@;}
19:   -- root
20:   :apply(c-s@iwte rd- i-s$I-J-)
21:   -- 1 (s@ = init) 
22:   -- 2-1 (s@ = want(s$,h$))
23:   :apply(c-s$h$rm rd- c-i@=h$- rd- c-j@=h$- rd-)
24:   -- 3-1 (s@ = try(s$,h$))
25:   :apply(c-s$h$wt rd- c-tqs$h$ rd- c-i@=h$- c-j@=h$- rd-)
26:   :apply(i-qts$j@ rd-)
27:   :apply(i-qts$i@ rd-)
28:   -- 4-1 (s@ = exit(s$,h$))
29:   :apply(c-s$h$cs rd- c-i@=h$- c-j@=h$- rd-)
\end{verbatim}
\end{cosmall}

Note that the proof has been done just by using WFI
({\bf Fact \ref{fact-wfishe}}) without using
{\bf Fact \ref{fact-inv}}. 

\bex (Guessing Next Command)
\label{ex-gussncom}

Construction of {\em effective} 
(i.e., able to discharge all goals)
proof scores is basically by trial and error.
However, the next command to be applied has a good chance to be
indicated by the information gotten from the next target goal
({\it\ref{sssec-ptcpsr}}).
After executing \ttt{:apply} command at
\ttt{25:}, for example, 
the reduced form of the goal proposition 
\ttt{mtx(s@,i@,j@)} (\ttt{03:}) is detected to be
\verb!(true xor (pc(s$,j@) = cs))!
by using PTcalc's
``\ttt{:show goal}'', \ttt{:red} commands,
that is exactly the proposition gotten by
applying the \ttt{:init} command 
as defined by \verb!i-qts$j@! (\ttt{17:-18:,26:}).
Similar guessing is possible at each ``next target goal''.
\hfill\makebox[0pt][r]{\owari}
\eex

After executing \ttt{01:-29:}, inputting the PTcalc command
``\ttt{:show proof}'' gives the following output \ttt{c01:-c34:}
showing the effective proof tree constructed.
\ttt{root} (\ttt{c01:}) is the root node of the proof tree.
\ttt{1} (\ttt{c01:}), \ttt{2} (\ttt{c03:}),
\ttt{3} (\ttt{c11:}), \ttt{4} (\ttt{c25:}) are four
child nodes (goals) (see {\it\ref{sssec-ptcpsr}}) 
constructed with the case-split command
\ttt{c-s@iwte}.
Each node of the tree except \ttt{root}
is represented by a line containing
a command name that creates the node and the
node name like ``\verb|[i-s$I-J-] 2-1*|'' (\ttt{c04:}).
The symbol \ttt{*} after a node name indicates that
the node is discharged.

For example, it can be seen that
the node \ttt{2-1-2} (\ttt{c10:}) is created as
2nd child of the case-split \verb|c-s$h$rm| of
the child of the initialization \verb|i-s$I-J-| (\ttt{2-1}, c04:) of
the 2nd child of the case-split \ttt{c-s@iwte} (\ttt{2}, c03:),
and discharged by \ttt{rd-} after \verb|c-s$h$rm| in the arguments of 
the \ttt{:apply} command in \ttt{23:}.
It can also be seen that
\ttt{c18:} is discharged with \ttt{26:}
and 
\ttt{c21:} is discharged with \ttt{27:}.

The output \ttt{c13:-c18:} shows a major utility of PTcalc. \\
{\small \verb!c18:  [i-qts$j@] 3-1-1-1-1-2-1*!} \\
tells that 
from the goal \ttt{3-1} after 4 case-splits 
\verb![c-s$h$wt]-1!,
\verb![c-tqs$h$]-1!,
\verb![c-i@=h$-]-1!,
\verb![c-j@=h$-]-2!,
and 1 instantiation
\verb![i-qts$j@]-1!,
the goal \ttt{3-1-1-1-1-2-1} is discharged.  
The checks with the 4 case-splits are achieved with one \ttt{:apply}
command:\\
{\small \verb!25:   :apply(c-s$h$wt rd- c-tqs$h$ rd- c-i@=h$- c-j@=h$- rd-)!.}\\
The argument of this \ttt{:apply} command is the sequence of 4
\ttt{:csp} commands and 3 \ttt{rd-} commands.  
This \ttt{:apply} command has the power of ``generating and checking'' all
of the $16 = 2^4$ cases determined by the 4 \ttt{:csp} commands. 
The full generation (i.e., generation of $2^4$ goals)
is exponentially costly and, of course, should be avoided.
That is, try to make each \ttt{rd-} command discharge at least
one goal.
\ttt{25:-27:} constructs the effective proof tree 
\ttt{c13:-c24:}.
PTcalc commands are fairly simple and succinct comparing with
\ttt{open...close} style,
and are great help for constructing effective proof trees
by trial and error.

\noindent
\begin{minipage}[t]{3.7cm}
\begin{coscript}\begin{verbatim}
c01:  root*
c02:  [c-s@iwte] 1*
c03:  [c-s@iwte] 2*
c04:  [i-s$I-J-] 2-1*
c05:  [c-s$h$rm] 2-1-1*
c06:  [c-i@=h$-] 2-1-1-1*
c07:  [c-i@=h$-] 2-1-1-2*
c08:  [c-j@=h$-] 2-1-1-2-1*
c09:  [c-j@=h$-] 2-1-1-2-2*
c10:  [c-s$h$rm] 2-1-2*
\end{verbatim}\end{coscript}
\end{minipage}
\begin{minipage}[t]{4.3cm}
\begin{coscript}\begin{verbatim}
c11:  [c-s@iwte] 3*
c12:  [i-s$I-J-] 3-1*
c13:  [c-s$h$wt] 3-1-1*
c14:  [c-tqs$h$] 3-1-1-1*
c15:  [c-i@=h$-] 3-1-1-1-1*
c16:  [c-j@=h$-] 3-1-1-1-1-1*
c17:  [c-j@=h$-] 3-1-1-1-1-2*
c18:  [i-qts$j@] 3-1-1-1-1-2-1*
c19:  [c-i@=h$-] 3-1-1-1-2*
c20:  [c-j@=h$-] 3-1-1-1-2-1*
c21:  [i-qts$i@] 3-1-1-1-2-1-1*
c22:  [c-j@=h$-] 3-1-1-1-2-2*
c23:  [c-tqs$h$] 3-1-1-2*
c24:  [c-s$h$wt] 3-1-2*
\end{verbatim}\end{coscript}
\end{minipage}  
\begin{minipage}[t]{3.8cm}
\begin{coscript}\begin{verbatim}
c25:  [c-s@iwte] 4*
c26:  [i-s$I-J-] 4-1*
c27:  [c-s$h$cs] 4-1-1*
c28:  [c-i@=h$-] 4-1-1-1*
c29:  [c-j@=h$-] 4-1-1-1-1*
c30:  [c-j@=h$-] 4-1-1-1-2*
c31:  [c-i@=h$-] 4-1-1-2*
c32:  [c-j@=h$-] 4-1-1-2-1*
c33:  [c-j@=h$-] 4-1-1-2-2*
c34:  [c-s$h$cs] 4-1-2*
\end{verbatim}\end{coscript}
\end{minipage}  
\vspace*{0.5em}

(1) The reduced Boolean value of the goal proposition,
e.g., \ttt{mtx(s@,i@,j@)} in \ttt{03:},
at each node of a proof tree gives many good hints
on which command should be applied next
(see {\bf Example \ref{ex-gussncom}}).
(2) Besides, the possible commands,
e.g., commands defined in \ttt{04:-18:},
can be constructed systematically once the set of
fresh constants,
e.g., {\it\ref{sssec-piwfihypoqlkots}}-\ttt{19:,32:},
is fixed.
\citep{RiescoO22} reported a system that 
fully automated proofs of invariant properties
on OTS based on (1) and (2).
The full automation techniques 
could be applied for automatically generating
effective advanced proof scores like \ttt{01:-29:}.

\subsection{QLOCK in Transition Specification (TSP)}
\label{ssec-qlktsp}

QLOCK is described as a transition specification ({\it\ref{sssec-trnspc}})
and the specification is called \ttt{QLOCK/TSP}.
As a matter of fact, 
\ttt{QLOCK/TSP} does not include 
the specification of initial states {\it In} of
a transition specification
($\vSg,E,\mathit{Tl},\mathit{In}$) and is just
a transition pre-spec ($\vSg,E,\mathit{Tl}$) ({\it\ref{sssec-trnspc}}). 
The initial states are specified later.
For a transition pre-spec ($\vSg,E,\mathit{Tl}$),
(i) data structure for the sort \ttt{State} needs to be designed,
(ii) ($\vSg,E$) needs to be red-complete
({\bf Fact \ref{fact-cansys}}) for 
proof scores with builtin search predicates to be valid,
but proof scores for complex properties like
leads-to properties can be constructed
thanks to transparent specifications of transitions
in transition rules.

\subsubsection{\ttt{QLOCK/TSP}: System Specification of QLOCK in TSP}
\label{sssec-sspqlktsp}

For describing \ttt{QLOCK/TSP}, the sort \ttt{State} should be defined
so that state transitions can be defined with transition rules
on \ttt{State}.
For defining \ttt{State}, generic sets with the operators
\ttt{\us{in}\us}, \ttt{\us{=}\us}, \verb|_^_| are specified
as the module \verb|SET=^| (\ttt{24:-31:})
after the modules \ttt{SET} (\ttt{01:-06:}),
\ttt{SETin} (\ttt{07:-14:}),
\ttt{SET=} (\ttt{15:-23:}).
The \ttt{assoc} and \ttt{comm} attributes of
the operator \verb|_^_| can be proved (e.g., \citep{futatsugi17})
and declaration of \ttt{\{assoc comm\}} (\ttt{27:}) is justified.
\ttt{SETlem} (\ttt{32:-40:}) declares necessary lemmas\footnote{%
Proof scores for the lemmas are in the file
\ttt{set-lem.cafe} 
on the Web:\\
\hspace*{2em}
\url{https://cafeobj.org/~futatsugi/misc/apsco-220907/}%
\,.}%
on \verb|SET=^|.

\vspace*{0.2em}
\noindent
@\footnote{
\ttt{01:-80:} is in the files
\ttt{set-psc.cafe},
\ttt{tsp-state-psc.cafe},
\ttt{tsp-sys-psc.cafe}
on the Web.}
\vspace*{-1em}
\begin{cosmall}\begin{verbatim}
01:   --> generic sets
02:   mod! SET(X :: TRIV) {
03:   [Elt < Set]
04:   op empty : -> Set {constr} .
05:   op __ : Set Set -> Set {constr assoc comm id: empty} .
06:   ceq (S:Set S) = S if not(S == empty) . }

07:   --> generic sets with _in_ predicate
08:   mod! SETin (X :: TRIV) {
09:   pr(SET(X))
10:   pred _in_ : Elt Set .
11:   eq (E:Elt in empty) = false .
12:   eq E1:Elt in (E2:Elt S:Set) = (E1 = E2) or (E1 in S) .
13:   eq[in-fls :nonexec]:
14:     ((E:Elt in S:Set) and (ES:Set = E S)) = false .

15:   --> generic sets with _=_ predicate
16:   mod! SET= (X :: TRIV) {
17:   pr(SETin(X))
18:   pred _=<_ : Set Set . -- equal of less
19:   eq S:Set =< S = true .
20:   eq empty =< S:Set = true .
21:   eq (E:Elt S1:Set) =< S2:Set = (E in S2) and (S1 =< S2) .
22:   cq (S1:Set = S2:Set) = (S1 =< S2) and (S2 =< S1)
23:      if not((S1 :is Elt) and (S2 :is Elt)) . }

24:   --> generic sets with intersection operator
25:   mod! SET=^ (X :: TRIV) {
26:   pr(SET=(X))
27:   op _^_ : Set Set -> Set {assoc comm} .
28:   eq empty ^ S2:Set = empty .
29:   eq (E:Elt S1:Set) ^ S2:Set =
30:      if E in S2 then E (S1 ^ S2) else (S1 ^ S2) fi .
31:   eq (S:Set ^ S) = S . }

32:   --> module declaring proved lemmas on SET=^
33:   mod! SETlem {
34:   pr(SET=^)
35:   cq (A:Elt in (S1:Set S2:Set)) = (A in S1) or (A in S2)
36:      if not(S1 == empty or S2 == empty) .
37:   eq ((S1:Set =< S2:Set) and (S1 =< (A:Elt S2))) = (S1 =< S2) .
38:   cq (S1:Set =< (A:Elt S2:Set)) = S1 =< S2 if (not(A in S1)) .
39:   cq[es1s2]: (E:Elt in S1:Set) and (S1 =< S2:Set) = false
40:              if not(E in S2) . }
\end{verbatim}\end{cosmall}

\ttt{AID-QU} (\ttt{43:-45:}) specifies queues of agent identifiers
by renaming sorts of \ttt{QUEhd}
({\it\ref{sssec-sspqlkots}}-\ttt{17:-22:}).
\ttt{AID-SET} (\ttt{46:-51:}) specifies sets of agent identifiers
by renaming sorts and operator of \ttt{SETlem}
(\ttt{32:-40:}), and adds the operator
\ttt{\us{-as}\us}.
\ttt{STATE} (\ttt{52:-59:}) specifies the state space
of QLOCK as the terms of the sort \ttt{State} that
are constructed with a constructor \verb|[_r_w_c_]|,
and adds the operator \ttt{q->s\us}.

\begin{cosmall}\begin{verbatim}
41:   --> agent identifiers
42:   mod* AID {[Aid]}

43:   --> Queues of Aid (agent identifiers)
44:   mod! AID-QU (X :: AID) {
45:   pr(QUEhd(X{sort Elt -> Aid})*{sort Que -> Aq,sort EltEr -> AidEr})

46:   --> Sets of Aid
47:   mod! AID-SET (X :: AID) {
48:   pr(SETlem(X{sort Elt -> Aid}) 
49:      *{sort Set -> As, op empty -> empS})
50:   op _-as_ : As Aid -> As {prec: 38} .
51:   eq (A:Aid S:As) -as A:Aid = S . eq empS -as A:Aid = empS . }

52:   --> States of QLOCK/TSP
53:   mod! STATE (X :: AID) {
54:   pr(AID-QU(X) + AID-SET(X))
55:   [State]
56:   op [_r_w_c_] : Aq As As As -> State {constr} .
57:   op q->s_ : Aq -> As .
58:   eq q->s nilQ = empS .
59:   eq q->s (Q1:Aq | A:Aid | Q2:Aq) = A (q->s (Q1 | Q2)) . }
\end{verbatim}\end{cosmall}

The modules
\ttt{WTtr} (\ttt{60:-65:}),
\ttt{TYtr} (\ttt{66:-71:}),  
\ttt{EXtr} (\ttt{72:-78:}) specify
the {\bf want, try, exit} behaviors of QLOCK (Section {\bf\ref{sec-qlkex}})
as rewritings among terms of the sort \ttt{State}
with transition rules 
\ttt{[wt]} (\ttt{64:-65:}),
\ttt{[ty]} (\ttt{70:-71:}),  
\ttt{[ex]} (\ttt{76:-78:}) respectively.
The transition rule \ttt{[ex]} can be without condition,
but just for an example of conditional transition rules.
The module \ttt{QLOCK/TSP} (\ttt{80:})
is just the module sum (i.e., conjunction)
of the three modules \ttt{WTtr}, \ttt{TYtr}, \ttt{EXtr},
where ``\ttt{make $name$ ($modexp$)}'' is equal to
``\ttt{mod $name$ \{pr($modexp$)\}}''.

\begin{cosmall}\begin{verbatim}
60:   --> WT: want transition; get into waiting queue
61:   mod! WTtr {
62:   pr(STATE)
63:   var Q : Aq . var Ar : Aid . vars Sr Sw Sc : As .
64:   tr[wt]: [ Q       r (Ar Sr) w     Sw  c Sc] =>
65:           [(Q | Ar) r     Sr  w (Ar Sw) c Sc] .  }

66:   --> TY: try transition; try to use the common resource
67:   mod! TYtr {
68:   pr(STATE)
69:   var Q : Aq . vars A Ar : Aid . vars Sr Sw Sc : As .
70:   tr[ty]: [(A | Q) r Sr w (A Sw) c    Sc]  =>
71:           [(A | Q) r Sr w    Sw  c (A Sc)] . }

72:   --> EX: exit transition; end using the common resource
73:   mod! EXtr {
74:   pr(STATE)
75:   var Q : Aq . vars A Ar : Aid . vars Sr Sw Sc : As .
76:   ctr[ex]: [(A | Q) r    Sr  w Sw c  Sc] =>
77:            [     Q  r (A Sr) w Sw c (Sc -as A)]
78:            if (A in Sc) . }

79:   --> QLOCK in transition rules
80:   make QLOCK/TSP (WTtr + TYtr + EXtr)
\end{verbatim}\end{cosmall}
  
The majority of operators and equations in the module
\ttt{STATE} is for constructing proof scores.
The sort \ttt{State}'s reduction specification for \ttt{QLOCK/TSP},
which correspond to ($\vSg,E$) of ($\vSg,E,Tl$)
({\it\ref{sssec-trnspc}}),
is the following \ttt{STATE-b} that is easily shown to be
red-complete.

\vspace*{0.2em}
\noindent
@\footnote{
\ttt{81:-93:} is in the file
\ttt{tsp-bare-state-spc.cafe}
on the Web.}
\vspace*{-1em}
\begin{cosmall}\begin{verbatim}
81:   --> Bare Queues of Aid (agent identifiers)
82:   mod! AID-QU-b (X :: AID) {
83:   pr(QUE(X{sort Elt -> Aid})*{sort Que -> Aq}) }

84:   --> Bare Sets of Aid
85:   mod! AID-SET-b (X :: AID) {
86:   pr(SETin(X{sort Elt -> Aid})*{sort Set -> As, op empty -> empS})
87:   op _-as_ : As Aid -> As {prec: 38} .
89:   eq (A:Aid S:As) -as A:Aid = S . eq empS -as A:Aid = empS . }

90:   --> Bare States of QLOCK/TSP
91:   mod! STATE-b (X :: AID) {
92:   pr(AID-QU-b(X) + AID-SET-b(X))
93:   [State] op [_r_w_c_] : Aq As As As -> State {constr} . }
\end{verbatim}\end{cosmall}

\subsubsection%
{Property Specifications for Mutual Exclusion Property of \ttt{QLOCK/TSP}}
\label{sssec-pspmxqlktsp}

For proving the mutual exclusion property
(MX property for short) of \ttt{QLOCK/ TSP}},
three modules
\ttt{INITprp} (\ttt{01:-10:}),
\ttt{MXprp} (\ttt{11:-18:}),  
\ttt{HQ=Cprp} (\ttt{19:-24:})
are prepared to define the three predicates
\ttt{init\us} (\ttt{07:-08:}),
\ttt{mx\us} (\ttt{17:-18:}),  
\ttt{hq=c\us} (\ttt{22:-24:}) respectively.

\vspace*{0.2em}
\noindent
@\footnote{
\ttt{01:-24:} is in the files
\ttt{tsp-init-prp.cafe},
\ttt{tsp-mx-inv-prp.cafe}
on the Web.}
\vspace*{-1em}
\begin{cosmall}\begin{verbatim}
01:   --> initial state property of QLOCK/TSP
02:   mod! INITprp {
03:   pr(STATE)
04:   pred init_ : As .
05:   eq (init empS) = false .
06:   eq (init (A:Aid AS:As)) = true .
07:   pred init_ : State .
08:   eq (init [Q:Aq r ASr:As w ASw:As c ASc:As]) 
09:       = ((Q = nilQ) and (init ASr) and
10:          (ASw = empS) and (ASc = empS)) . }

11:   --> mutual exclusion property of QLOCK/TSP
12:   mod! MXprp {
13:   pr(STATE)
14:   pred mx_ : As .
15:   eq (mx empS) = true .
16:   eq (mx (A:Aid AS:As)) = (AS = empS) .
17:   pred mx_ : State .
18:   eq (mx [Q:Aq r ASr:As w ASw:As c ASc:As]) = (mx ASc) . }

19:   --> queue top property of QLOCK/TSP
20:   mod! HQ=Cprp {
21:   pr(STATE)
22:   pred hq=c_ : State  .
23:   eq (hq=c [Q:Aq r ASr:As w ASw:As c ASc:As]) = 
24:      (ASc = empS) or (not(Q = nilQ) and ((hd Q) = ASc)) . }
\end{verbatim}\end{cosmall}

\bex (Simulation with Builtin Search Predicates)
\label{ex-simbybsp}

CafeOBJ provides another builtin search predicate declared as follows.\\
{
\verb|    pred _=(*,*)=>*_suchThat_ : State State Bool .|}\\
Let $s{\in}(T_{\vSg})_{\mathtt{State}}$ and
$p$ be a predicate on \ttt{State},
when a term\\
\hspace*{2em}\ttt{$s$ =(*,*)=>* SS:State suchThat $p$(SS)}\\
is reduced in a transition pre-spec ($\vSg,E,\mathit{Tl}$),
it behaves as follows.
Note that the second argument is always a variable for 
binding found results. 
(i) Search for all the reachable states $\mathit{ss}$ from $s$
(i.e., $s\,\trstr{Tl,AC}\,\mathit{ss}$, {\it\ref{sssec-trnspc}})
such that \ttt{$p$($\mathit{ss}$)},
(ii) if such states exist then print out all the found states
and returns \ttt{true}, and 
(iii) if such states do not exist then returns \ttt{false}.

This builtin search predicate can be used to prove an invariant
property with respect to {\em finite} transition systems.
The following CafeOBJ code \ttt{25:-33}, for example,
proves that any \ttt{QLOCK/TSP} with up to 4 agents 
satisfies \ttt{mx} (mutual exclusion) property
by checking the reductions \ttt{28:-29:}, \ttt{30:-31:},
\ttt{32:-33:} return \ttt{false}.
The parameter module ``\ttt{X.STATE :: AID}''
of the module \ttt{(QLOCK/TSP + MXprp)}
is instantiated by the builtin module \ttt{NAT} 
with the view
\ttt{\{sort Aid -> Nat, op A1:Aid = A2:Aid -> A1:Nat == A2:Nat\}}
in \ttt{26:-27:}.
The initial state is specified with the term like 
\verb![nilQ r 1 2 3 w empS c empS]! (\ttt{30:}).

\vspace*{0.2em}
\noindent
@\footnote{
\ttt{25:-37:} is in the files
\ttt{tsp-mx-simulation-bsp.cafe}
on the Web.}
\vspace*{-1em}
\begin{cosmall}\begin{verbatim}
25:   open (QLOCK/TSP + MXprp)
26:        (NAT{sort Aid -> Nat,
27:             op A1:Aid = A2:Aid -> A1:Nat == A2:Nat}) .

28:   red [nilQ r 1 2 w empS c empS] =(*,*)=>* S:State 
29:       suchThat (not (mx S)) .
30:   red [nilQ r 1 2 3 w empS c empS] =(*,*)=>* S:State 
31:       suchThat (not (mx S)) .
32:   red [nilQ r 1 2 3 4 w empS c empS] =(*,*)=>* S:State 
33:       suchThat (not (mx S)) .

34:   red [nilQ r 1 2 w empS c empS] =(*,*)=>* 
35:       [Q:Aq r ASr:As w ASw:As c ASc:As] 
36:       suchThat ((ASr = empS) and (ASc = empS)) .

37:   close
\end{verbatim}\end{cosmall}

This builtin search predicate can also be used to find a counter example.
Note that \ttt{[Q:Aq r ASr:As w ASw:As c ASc:As]} denotes any term in
$(T_{\vSg})_{\mathtt{State}}$.
Assume we want to prove that
\ttt{(not\,((ASr\,{=}\,empS)\,and\,(ASc\,{=}\,empS)))}
is an invariant.
\ttt{34:-36:} searches all reachable states from the initial state
with 2 agents, and prints out two states
\ttt{[ (2 | 1) r empS w (2 1) c empS ]} and 
\ttt{[ (1 | 2) r empS w (1 2) c empS ]}
that satisfy \ttt{((ASr = empS) and (ASc = empS))}
(i.e., two counter examples).  

Simulation of finite transition systems with builtin search predicates,
together with verifications of infinite systems with proof scores,
is quite effective for specification analyses, verifications, and improvements
(see {\bf Example \ref{ex-newextr}}).%
\hfill\makebox[0pt][r]{\owari}
\eex

\subsubsection{Proof Scores for MX property of \ttt{QLOCK/TSP}}
\label{sssec-pscmxqlktsp}

If \ttt{(mx S:State)} is an invariant 
then the mutual exclusion property of \ttt{QLOCK/ TSP} holds.
\ttt{(mx S:State)} is not an inductive invariant,
and for using {\bf Fact \ref{fact-inv}} to
prove \ttt{(mx S:State)} is an invariant,
\ttt{(hq=c S:State)} is necessary to make
\ttt{((mx S:State) and (hq=c S))} an inductive invariant.

With the correspondences:\\
\hspace*{2em}{\it init(S:State)} $\lrh$ \ttt{(init S:State)} and\\
\hspace*{2em}{\it iinv(S:State)} $\lrh$ \ttt{((mx S:State) and (hq=c S))},\\
\ttt{01:-20:} is a proof score for {\bf INV1} and
\ttt{21:-79:} is a proof score for {\bf INV2} 
of {\bf Fact \ref{fact-inv}}.
Hence, \ttt{01:-79:} is a proof score for proving that
\ttt{(mx S:State)} is an invariant.

``\verb|st@ = [q$ r sr$ w sw$ c sc$]|'' is the most general
way to represent the fresh constant \ttt{st@} of the sort \ttt{State},
and the case-split \ttt{st} (\ttt{17:}) is justified.

\vspace*{0.2em}
\noindent
@\footnote{
\ttt{01:-79:} is in the files
\ttt{tsp-mx-init-psc.cafe},
\ttt{check-cnr.cafe},
\ttt{tsp-mx-iinv-psc.cafe}
on the Web.}
\vspace*{-1em}
\begin{cosmall}\begin{verbatim}
01:   --> defining mx initial condition 
02:   mod CHECK-mx-init {
03:   pr(INITprp + MXprp + HQ=Cprp)
04:   pred check-mx-init_ : State .
05:   eq check-mx-init S:State =
06:      (init S) implies ((mx S) and (hq=c S)) .
07:   -- fresh constant
08:   op st@ : -> State .
09:   -- goal proposition is "check-mx-init st@ = true"
10:   -- fresh constants for refinements
11:   ops a$ ar$ a$1 ac$1 ac$2 : -> Aid .
12:   ops q$ q$1 : -> Aq .
13:   ops sr$ sw$ sc$ sc$1 sc$2 : -> As . }

14:   --> executing proof of "check-mx-init st@ = true" with PTcalc
15:   select CHECK-mx-init .
16:   :goal{eq check-mx-init st@ = true .}
17:   :def st = :csp{eq st@ = [q$ r sr$ w sw$ c sc$] .}
18:   :def q=nil# = :ctf{eq q$ = nilQ .}
19:   :def sc=em# = :ctf{eq sc$ = empS .}
20:   :apply(st q=nil# rd- sc=em# rd-) 
\end{verbatim}\end{cosmall}

\ttt{21:-79:} implements {\bf Fact \ref{fact-inv} (INV2)}
by making use of {\bf Fact \ref{fact-validq}} 
with the following correspondences:\\
\hspace*{2em}(1) \ttt{mx-iinv} $\lrh$ {\it iinv} (\ttt{35:}),\\
\hspace*{2em}(2) \ttt{Aid} $\lrh$ \ttt{Data} (\ttt{41:}),\\
\hspace*{2em}(3) \ttt{cnr-mx-iinv} $\lrh$ \ttt{cnr} (\ttt{42:}),\\
\hspace*{2em}(4) \ttt{check-cnr-mx-iinv} $\lrh$ \ttt{check-cnr} (\ttt{45:}).\\
The definition of \ttt{check-cnr},
i.e., the equation just before {\bf Fact \ref{fact-validq}},
is in the module \ttt{CHECKcnr} 
\footnote{%
\ttt{...} (\ttt{24:,30:}) shows an abbreviation and the full description is in the file 
\ttt{check-cnr.cafe} on the Web.%
}%
(\ttt{23:-30:}).
\ttt{Aid} (\ttt{Data}) is not necessary and dropped from the arguments
of \ttt{check-cnr-mx-iinv} (\ttt{45:}) by using dummy element
\ttt{dummyAid} (\ttt{44:}).

\begin{cosmall}\begin{verbatim}
21:   --> Current and Next state Relation with a data argument
22:   mod* CNR { [State Data] pred cnr : State State Data . }

23:   --> module for defining check-cnr
24:   mod! CHECKcnr (X :: CNR) { ...
25:   -- predicate to check cnr against all 1 step transitions 
26:   pred check-cnr : State Data .
27:   eq check-cnr(S:State,D:Data) =
28:       not(S =(*,1)=>+ SS:State if CC:Bool
29:           suchThat not((CC implies cnr(S,SS,D)) == true)
30:           {i(S,SS,D,CC) ... }) . }

31:   --> module defining cnr-mx-iinv 
32:   mod CNR-mx-iinv {
33:   pr(MXprp + HQ=Cprp)
34:   pred mx-iinv : State .
35:   eq mx-iinv(S:State) = ((mx S) and (hq=c S)) .
36:   pred cnr-mx-iinv : State State Aid .
37:   eq cnr-mx-iinv(S:State,SS:State,A:Aid) =
38:      (mx-iinv(S) implies mx-iinv(SS)) . }

39:   --> proposition to check whether cnr-mx-iinv holds
40:   mod CHECK-cnr-mx-iinv {
41:   inc(CHECKcnr(CNR-mx-iinv{sort Data -> Aid,
42:                            op cnr -> cnr-mx-iinv}))
43:   pred check-cnr-mx-iinv : State .
44:   op dummyAid : -> Aid .
45:   eq check-cnr-mx-iinv(S:State) = check-cnr(S,dummyAid) .
46:   -- fresh constant
47:   op st@ : -> State .
48:   -- goal proposition is: "check-cnr-mx-iinv(st@) = true" 
49:   -- fresh constants for refinements
50:   ops a$ ar$ a$1 ac$1 ac$2 : -> Aid .
51:   ops q$ q$1 : -> Aq .
52:   ops sr$ sw$ sc$ sc$1 sc$2 : -> As . }
\end{verbatim}\end{cosmall}

{\bf Fact \ref{fact-validq}} says that to show\\
\hspace*{2em}\ttt{(CHECK-cnr-mx-iinv + QLOCK/TSP)} $\tsc$
\ttt{check-cnr-mx-iinv(st@)}\\
is sufficient for proving {\bf Fact \ref{fact-inv} (INV2)}.
\ttt{QLOCK/TSP} is \ttt{(WTtr + TYtr + EXtr)}
({\it \ref{sssec-sspqlktsp}}-\ttt{80:}) and\\
\hspace*{2em}(\ttt{(CHECK-cnr-mx-iinv + WTtr)} $\tsc$
\ttt{check-cnr-mx-iinv(st@)} $\land$\\
\hspace*{2.2em}\ttt{(CHECK-cnr-mx-iinv + TYtr)} $\tsc$
\ttt{check-cnr-mx-iinv(st@)} $\land$\\
\hspace*{2.2em}\ttt{(CHECK-cnr-mx-iinv + EXtr)} $\tsc$
\ttt{check-cnr-mx-iinv(st@)})\\
\hspace*{0.5em}$\Ra$ \ttt{(CHECK-cnr-mx-iinv + QLOCK/TSP)} $\tsc$
  \ttt{check-cnr-mx-iinv(st@)}.\\
\ttt{53:-60:}, \ttt{61:-69:}, \ttt{70:-79:} are three
PTcalc proof scores for the above three premises.
Hence, \ttt{21:-79:} is a proof score for
{\bf Fact \ref{fact-inv} (INV2)}.

The case-split \ttt{wt} (\ttt{56:}) refines
\ttt{st@} into the most general instance with fresh constants
of the left hand side of the transition rule
\ttt{wt} ({\it \ref{sssec-sspqlktsp}}-\ttt{64:})
in the module \ttt{WTtr}.
Any transition with the transition rule \ttt{wt}
should happen from a term of sort \ttt{State}
that is a refined instance of the left hand side,
and the case-split \ttt{wt} is justified.
The same argument is valid for the case-splits
\ttt{ty} (\ttt{64:}) and \ttt{ex} (\ttt{73:}).

\begin{cosmall}\begin{verbatim}
53:   --> executing proof mx-iinv WT 
54:   open (CHECK-cnr-mx-iinv + WTtr) .
55:   :goal{eq check-cnr-mx-iinv(st@) = true .}
56:   :def wt = :csp{eq st@ = [q$ r (ar$ sr$) w sw$ c sc$] .}
57:   :def q=nil = :csp{eq q$ = nilQ . eq q$ = (a$1 | q$1) .}
58:   :apply(wt q=nil rd-)
59:   :show proof
60:   close
61:   --> executing proof mx-iinv TY 
62:   open (CHECK-cnr-mx-iinv + TYtr) .
63:   :goal{eq check-cnr-mx-iinv(st@) = true .}
64:   :def ty = :csp{eq st@ = [(a$ | q$) r sr$ w (a$ sw$) c sc$] .}
65:   :def sc=em = :csp{eq sc$ = empS . eq sc$ = (ac$1 sc$1) .}
66:   :def a=ac1 = :ctf{eq a$ = ac$1 .}
67:   :apply(ty sc=em rd- a=ac1 rd-)
68:   :show proof
69:   close
70:   --> executing proof mx-iinv EX 
71:   open (CHECK-cnr-mx-iinv + EXtr) .
72:   :goal{eq check-cnr-mx-iinv(st@) = true .}
73:   :def ex = :csp{eq st@ = [(a$ | q$) r sr$ w sw$ c sc$] .}
74:   :def sc=em = :csp{eq sc$ = empS . eq sc$ = (ac$1 sc$1) .}
75:   :def sc1=em = :csp{eq sc$1 = empS . eq sc$1 = (ac$2 sc$2) .}
76:   :def a=ac1 = :ctf{eq a$ = ac$1 .}
77:   :apply(ex sc=em rd- sc1=em rd- a=ac1 rd-)
78:   :show proof
79:   close
\end{verbatim}\end{cosmall}

Four ``\ttt{:show proof}'' commands for the four pieces of code
\ttt{14:-20:}, \ttt{53:-60:}, \ttt{61:-69:}, \ttt{70:-79:} 
print out the following \ttt{c01:-c24:} that show
the effectiveness of the proof trees constructed.
(``\ttt{:show proof}'' for \ttt{14:-20:} is done
after \ttt{20:}.)
A proof score\footnote{%
In the file \ttt{tsp-mx-iinv-allrl-psc.cafe} on the Web.
}%
for the module \ttt{(CHECK-cnr-mx-iinv + QLOCK/TSP)}
is possible instead of \ttt{53:-79:} but tends to be complex.

\noindent
\begin{minipage}[t]{2.9cm}
\begin{coscript}\begin{verbatim}
c01: root*
c02: [st]  1*
c03: [q=nil#] 1-1*
c04: [sc=em#] 1-1-1*
c05: [sc=em#] 1-1-2*
c06: [q=nil#] 1-2*
\end{verbatim}\end{coscript}
\end{minipage}
\begin{minipage}[t]{2.5cm}
\begin{coscript}\begin{verbatim}
c07: root*
c08: [wt]  1*
c09: [q=nil] 1-1*
c10: [q=nil] 1-2*
\end{verbatim}\end{coscript}
\end{minipage}
\begin{minipage}[t]{2.8cm}
\begin{coscript}\begin{verbatim}
c11: root*
c12: [ty]  1*
c13: [sc=em] 1-1*
c14: [sc=em] 1-2*
c15: [a=ac1] 1-2-1*
c16: [a=ac1] 1-2-2*
\end{verbatim}\end{coscript}
\end{minipage}
\begin{minipage}[t]{3cm}
\begin{coscript}\begin{verbatim}
c17: root*
c18: [ex]  1*
c19: [sc=em] 1-1*
c20: [sc=em] 1-2*
c21: [sc1=em] 1-2-1*
c22: [a=ac1] 1-2-1-1*
c23: [a=ac1] 1-2-1-2*
c24: [sc1=em] 1-2-2*
\end{verbatim}\end{coscript}
\end{minipage}\\ 
\bex  (Improving \ttt{EXtr})
\label{ex-newextr}


Suppose we want to
change the \ttt{EXtr}'s transition rule
{\it\ref{sssec-sspqlktsp}}-\ttt{76:-78:} as follows.\\
{\small
\verb!   tr[ex]: [(A | Q) r Sr w Sw c (Ac Sc)] => [Q  r (A Sr) w Sw c Sc ] .!
}\\
This rule does not require to check (\ttt{A} = \ttt{Ac}) and
provides a more flexible protocol,
and it is worthwhile to check whether it also satisfies the \ttt{mx}
property.
The proof can be achieved by adding \verb!ac$! to \ttt{50:} and
replacing \ttt{73:} as follows.\\
{\small
\verb!50:   ops a$ ar$ a$1 ac$ ac$1 ac$2 : -> Aid .!\\
\verb!73:   :def ex = :csp{eq st@ = [(a$ | q$) r sr$ w sw$ c (ac$ sc$)] .}!
}\\
The new rule \ttt{tr[ex]} is not conditional and the proof
becomes simpler, and \ttt{c22:-c23:} are omitted.
\hfill\makebox[0pt][r]{\owari}
\eex

\subsubsection{Property Specifications for Leads-to Property of \ttt{QLOCK/TSP}}
\label{sssec-prpwcqlktsp}

\ttt{03:-05:} defines two state predicates
\verb|_inw_|,\verb|_inc_| for defining a leads-to property
(\verb|_inw_| leads-to \verb|_inc_|) (WC property for short)
which asserts that if an agent gets into \ttt{ASw:As}
(Waiting section)
then the agent will surely get into \ttt{ASc:As}
(Critical section).

\vspace*{0.2em}
\noindent
@\footnote{
\ttt{01:-23:} is in the files
\ttt{tsp-wc-prp.cafe},
\ttt{tsp-wc-inv-prp.cafe}
on the Web.}
\vspace*{-1em}
\begin{cofootnote}\begin{verbatim}
01:   --> defining (_inw_) (_inc_) for WC property
02:   mod! WCprp { pr(STATE)
03:   preds (_inw_) (_inc_) : Aid State .
04:   eq A:Aid inw [Q:Aq r ASr:As w ASw:As c ASc:As] = A in ASw .
05:   eq A:Aid inc [Q:Aq r ASr:As w ASw:As c ASc:As] = A in ASc . }
\end{verbatim}\end{cofootnote}

For constructing proof scores for the WC property
(\verb|_inw_| leads-to \verb|_inc_|),
the state predicates
\verb|r^w_, w^c_, r^c_, q=wc_, qvr_, qnd_|
(\ttt{08:-23:})
need to be used as invariant properties
{\it inv(s)} in {\bf Fact \ref{fact-pltq} (LT1)} and {\bf (LT2)}.

\begin{cofootnote}\begin{verbatim}
06:   --> defining invariant properties for WC property
07:   mod! WCinvs { pr(STATE)
08:   preds (r^w_) (w^c_) (r^c_) (q=wc_) (qvr_) (qnd_) : State .
09:   eq r^w [Q:Aq r ASr:As w ASw:As c ASc:As] =
10:      ((ASr ^ ASw) = empS) .
11:   eq w^c [Q:Aq r ASr:As w ASw:As c ASc:As] =
12:      ((ASw ^ ASc) = empS) .
13:   eq r^c [Q:Aq r ASr:As w ASw:As c ASc:As] =
14:      ((ASr ^ ASc) = empS) .
15:   eq q=wc [Q:Aq r ASr:As w ASw:As c ASc:As] =
16:      ((q->s Q) = (ASw ASc)) .
17:   eq qvr [Q:Aq r ASr:As w ASw:As c ASc:As] =
18:      not(((q->s Q) ASr) = empS) .
19:   pred qnd_ : Aq .
20:   eq qnd nilQ = true .
21:   eq qnd (Q1:Aq | A:Aid | Q2:Aq) =
22:      not(A in ((q->s Q1) (q->s Q2))) and (qnd Q1) and (qnd Q2) .
23:   eq qnd [Q:Aq r ASr:As w ASw:As c ASc:As] = qnd Q . }
\end{verbatim}\end{cofootnote}

\subsubsection{Proof Scores for WC property of \ttt{QLOCK/TSP}}
\label{sssec-pscwcqlktsp}

The proof scores\footnote{%
In the files
\ttt{tsp-wc-init-psc.cafe},
\ttt{tsp-wc-iinv-psc.cafe}
on the Web.
}%
for proving that the state predicates
\verb|r^w_|, \verb|w^c_|, \verb|r^c_|, \verb|q=wc_|, \verb|qvr_|, \verb|qnd_|
({\it\ref{sssec-prpwcqlktsp}}-\ttt{08:-23:})
are invariant properties of \ttt{QLOCK/TSP}
can be constructed in a similar way as the proof scores
({\it \ref{sssec-pscmxqlktsp}}-\ttt{01:-79:})
for the state predicates \verb|mx_|, \verb|hq=c_|.
Since each of the state predicates
\verb|mx_|, \verb|hq=c_|,
\verb|r^w_|, \verb|w^c_|, \verb|r^c_|, \verb|q=wc_|, \verb|qvr_|, \verb|qnd_|
is invariant property, 
{\it inv(s)} in {\bf Fact \ref{fact-pltq} (LT1)} and {\bf (LT2)}
can be specified with \ttt{03:-11:}.
As a matter of fact, \ttt{08:}, \ttt{11:} are not necessary
in the following proof scores.

\vspace*{0.2em}
\noindent
@\footnote{
\ttt{01:-87:} is in the files
\ttt{tsp-wc-inv-lem.cafe},
\ttt{tsp-wc-dms-prp.cafe},
\ttt{tsp-wc-daq-lem. cafe},
\ttt{tsp-wc-wc1-psc.cafe},
\ttt{tsp-wc-wc2-psc.cafe}
on the Web.}
\vspace*{-1em}
\begin{cofootnote}\begin{verbatim}
01:   --> module declaring inv_ for WC property of QLOCK/TSP
02:   mod! INVlem { pr(MXprp + HQ=Cprp + WCinvs)
03:   pred inv : State .
04:   cq inv(S:State) = false if not(mx S) .
05:   cq inv(S:State) = false if not(hq=c S) .
06:   cq inv(S:State) = false if not(r^w S) .
07:   cq inv(S:State) = false if not(w^c S) .
08:   cq inv(S:State) = false if not(r^c S) .
09:   cq inv(S:State) = false if not(q=wc S) .
10:   cq inv(S:State) = false if not(qvr S) .
11:   cq inv(S:State) = false if not(qnd S) . }
\end{verbatim}\end{cofootnote}

\ttt{35:-38:} indicates that
\ttt{12:-60:} implements {\bf Fact \ref{fact-pltq} (LT1)}
by making use of {\bf Fact \ref{fact-validq}} 
with the following correspondences:\\
\hspace*{2em}(1) \ttt{inv} (\ttt{03:}) $\lrh$ {\it inv},\\
\hspace*{2em}(2) \verb|_inw_| ({\it\ref{sssec-prpwcqlktsp}}-\ttt{03:})
   $\lrh$ {\it p},\\
\hspace*{2em}(3) \verb|_inc_| ({\it\ref{sssec-prpwcqlktsp}}-\ttt{03:})
   $\lrh$ {\it q},\\
\hspace*{2em}(4) \verb|#dms| (\ttt{24:}) $\lrh$ {\it m},\\
\hspace*{2em}(5) \ttt{Aid} $\lrh$ \ttt{Data} (\ttt{41:}),\\
\hspace*{2em}(6) \ttt{cnr-wc1} $\lrh$ \ttt{cnr} (\ttt{41:}),\\
\hspace*{2em}(7) \ttt{check-cnr-wc1} $\lrh$ \ttt{check-cnr} (\ttt{42:}).

\ttt{15:-26:} defines \verb|#dms| and its subsidiary operators
\verb|#_|, \verb|#daq|, \verb|#c_|,
and \ttt{29:-30:} declares a lemma\footnote{%
The proof score for this lemma is in the file
\ttt{tsp-wc-daq-lem.cafe}
on the Web.
}%
on \verb|#daq| that is used at \ttt{51:}.
\ttt{PNAT*ac} (\ttt{14:}) specifies Peano natural numbers
with \verb|_=_|, \verb|_+_|, \verb|_*_|, and 
\ttt{PNAT*ac>} (\ttt{33:}) specifies \ttt{PNAT*ac} plus
\verb|_>_|\footnote{%
The specifications are in the file 
\ttt{pnat-spc.cafe}
on the Web.
}.

The PTcalc proof scores\footnote{%
In the file
\ttt{tsp-wc-wc1-psc.cafe}
on the Web.%
}%
on the modules
\ttt{(CHECK-cnr-wc1 + TYtr)} and \ttt{(CHECK-cnr-wc1 + EXtr)}
are constructed as the one on \ttt{(CHECK-cnr-wc1 + WTtr)}
(\ttt{49:-60:}).

\begin{cofootnote}\begin{verbatim}
12:   -- defining decreasing measure function #dms
13:   mod* DMS { pr(STATE)
14:   pr(PNAT*ac)
15:   op #_ : As -> Nat .
16:   eq # empS = 0 .
17:   eq # (A:Aid AS:As) = s (# AS) .
18:   op #daq : Aq Aid -> Nat .
19:   cq #daq(A1:Aid | Q:Aq,A2:Aid) =
20:      (if (A1 = A2) then 0 else (s #daq(Q,A2)) fi)
21:      if (A2 in (q->s (A1 | Q))) .
22:   op #c_ : Nat -> Nat .
23:   eq (#c N:Nat) = if N = 0 then (s 0) else 0 fi .
24:   op #dms : State Aid -> Nat .
25:   eq #dms([Q:Aq r ASr:As w ASw:As c ASc:As],A:Aid) =
26:      ((s s s 0) * #daq(Q,A)) + (# ASr) + (#c (# ASc)) . }
\end{verbatim}\end{cofootnote}

\begin{cofootnote}\begin{verbatim}
27:   --> module declaring lemma [dag-lem]
28:   mod! DAQ-lem { pr(DMS)
29:   cq[dag-lem]: #daq((Q:Aq | A1:Aid),A2:Aid) = #daq(Q,A2)
30:                if (A2 in (q->s Q)) . }
\end{verbatim}\end{cofootnote}

\begin{cofootnote}\begin{verbatim}
31:   --> module defining cnr for wc1
32:   mod CNR-wc1 {
33:   pr(MXprp + HQ=Cprp + WCinvs + DMS + PNAT*ac> + WCprp)
34:   pr(INVlem)
35:   pred cnr-wc1 : State State Aid .
36:   eq cnr-wc1(S:State,SS:State,A:Aid) = 
37:      (inv(S) and (A inw S) and not(A inc S)) implies
38:      (((A inw SS) or (A inc SS)) and (#dms(S,A) > #dms(SS,A))) . }
39:   --> proof module checking condition wc1
40:   mod CHECK-cnr-wc1 {
41:   inc(CHECKcnr(CNR-wc1{sort Data -> Aid,op cnr -> cnr-wc1})
42:                *{op check-cnr -> check-cnr-wc1})
43:   -- fresh constant
44:   op st@ : -> State .  op aa@ : -> Aid .
45:   -- goal proposition is "check-cnr-wc1(st@,aa@) = true"
46:   ops a$ ar$ a$1 ac$1 ac$2 : -> Aid .
47:   ops q$ q$1 : -> Aq .
48:   ops sr$ sw$ sw$1 sc$ sc$1 sc$2 : -> As . }
49:   --> executing proof of check-cnr-wc1-wt with PTcalc
50:   open (CHECK-cnr-wc1 + WTtr) .
51:   pr(DAQ-lem)
52:   :goal{eq check-cnr-wc1(st@,aa@) = true .}
53:   :def wt = :csp{eq st@ = [q$ r (ar$ sr$) w sw$ c sc$] .}
54:   :def sc=em = :csp{eq sc$ = empS . eq sc$ = ac$1 sc$1 .}
55:   :def aa%sw = :csp{eq sw$ = aa@ sw$1 . eq (aa@ in sw$) = false .}
56:   :def aa=ar = :ctf{eq aa@ = ar$ .}
57:   :def aa!q = :ctf{eq (aa@ in (q->s q$)) = true .}
58:   :apply(wt sc=em rd- aa%sw rd- aa=ar rd- aa!q rd-)
59:   :show proof
60:   close
\end{verbatim}\end{cofootnote}

\ttt{65:-68:} indicates that
\ttt{61:-87:} implements {\bf Fact \ref{fact-pltq} (LT2)}
by making use of {\bf Fact \ref{fact-cont}}.

\begin{cofootnote}\begin{verbatim}
61:   --> module defining the 2nd verification condition wc2 of wc
62:   mod WC2 { inc(RWL)
63:   pr(WCprp)
64:   pr(INVlem)
65:   pred wc2 : State Aid .
66:   eq wc2(S:State,A:Aid) =
67:      (inv(S) and (A inw S) and not(A inc S)) implies
68:      (S =(1,1)=>+ SS:State) . }
69:   --> proof module for checking condition wc2 with PTcalc
70:   mod CHECK-wc2 { inc(WC2)
71:   pr(WTtr + TYtr + EXtr)
72:   -- fresh constants
73:   op st@ : -> State . op aa@ : -> Aid .
74:   -- goal proposition: wc2(st@,aa@) 
75:   ops a$1 ar$1 : -> Aid .
76:   ops q$ q$1 : -> Aq .
77:   ops sr$ sw$ sc$ sr$1 sw$1 sc$1 : -> As . }
78:   --> executing proof of wc2 with PTcalc
79:   select CHECK-wc2 .
80:   :goal{eq wc2(st@,aa@) = true .}
81:   :def st = :csp{eq st@ = [q$ r sr$ w sw$ c sc$] .}
82:   :def sr=em = :csp{eq sr$ = empS . eq sr$ = (ar$1 sr$1) .}
83:   :def q=nil = :csp{eq q$ = nilQ . eq q$ = (a$1 | q$1) .}
84:   :def a1%sw = :csp{eq sw$ = a$1 sw$1 . eq (a$1 in sw$) = false .}
85:   :def a1%sc = :csp{eq sc$ = a$1 sc$1 . eq (a$1 in sc$) = false .}
86:   :apply(st sr=em rd- q=nil rd- a1%sw rd- a1%sc rd-)
87:   :show proof
\end{verbatim}\end{cofootnote}

\section{Discussion}
\label{sec-dscs}

\subsection{Related Works}
\label{ssec-rwk}

\subsubsection{Specification languages}
\label{sssec-speclang}

Many algebraic specification languages/systems have been
developed, among them are 
OBJ \citep{futatsugiGJM85},
HISP \citep{futatsugio80}, ASL \citep{Wirsing86},
ASF+SDF \citep{BrandDHJJKKMOSVVV01}, 
Larch \citep{GuttagHGJMW93},
CafeOBJ \citep{futatsugiN97},
CASL \citep{AstesianoBKKMST02}, and
Maude \citep{maude-wp}.
Some of them, including Larch,
CASL, and Maude, have their own verification tools, and only OBJ
and CafeOBJ adopt the proof score approach just with
equational reduction engines.
Maude is a sibling language of CafeOBJ
and the two languages share many important features,
but Maude's Inductive Theorem Prover (MITP)
\citep{ClavelPR06-jucs}
does not take the proof score approach.

Set theory based formal specification languages,
like VDM \citep{vdm-wp}, Z \citep{z-wp}, B \citep{b-wp},
are popular and widely used.
The verification of specifications in these languages 
can be achieved with proof assistants 
like
PVS \citep{pvs-wp},
Isabelle/HOL \citep{IsabelleHOL-wp},
Rodin \citep{rodin-wp}
by making use of stepwise refinements
and seem to be practically effective.
Proof scores with module reuses and refinements in CafeOBJ
have a potential for realizing
the proofs via stepwise refinements.

\subsubsection{Theorem Provers}
\label{sssec-thopro}

Interactive theorem provers or proof assistants have similar
motivation as proof scores.  Among the most noted proof assistants are
ACL2 \citep{acl2-wp},
Coq \citep{coq-wp}, 
Isabelle/HOL \citep{IsabelleHOL-wp},
PVS \citep{pvs-wp}.
All of them are not algebraic and
not based on equational deduction/reduction.

SAT/SMT solvers, e.g., MiniSAT \citep{minisat-wp}, Z3 \citep{z3-wp}, and
Yices \citep{yices-wp},
have a nice power to fully automate proofs in some specific classes,
and are used in several interactive theorem provers.
This kind of fully automated theorem provers
can be used to discharge a sub-goal in a PTcalc proof tree.
\citep{RiescoO18,RiescoO22} reported CiMPA/CiMPG/CiMPG-F
for CafeInMaude (CiM) \citep{RiescoOF17}
that 
fully automated some proof scores for OTS's invariant properties.
CiMPA/CiMPG/CiMPG-F plus SAT/SMT have a potential to
automate the PTcalc+WFI proof scores in a significant way.

\subsubsection{CafeOBJ Related}
\label{sssec-corelated}

The logical semantics of CafeOBJ is structured by the concept of
institution \citep{cafesem02}.  Recent studies inheriting this
tradition include \citep{Gaina20,GainaNOF20}.

Recent focus on applications of specification verification with
CafeOBJ is multitask, hybrid, real-time system.  Recent publications
in this category include \citep{NakamuraSOO21,NakamuraHSO22}.  Proof
scores for cloud protocols and automotive software standards still continue to
be investigated.  \citep{YoshidaOF15} reported an interesting rare
achievement on proof scores for leads-to properties, 
and develops to domain specific reuse of proof scores.

CafeInMaude (CiM) \citep{RiescoOF17} is the recent second
implementation of CafeOBJ on the Maude system.  Promising proof
automation/assistant tools CiMPA/CiMPG /CiMPG-F have already been
developed \citep{RiescoO18,RiescoO22}
by making use of the Maude meta-programming facility.

\subsection{Distinctive Features of Proof Scores}

\subsubsection{ADT and Model Based Proof}
\label{sssec-adtmbp}

Comparing with the notable theorem provers like
ACL2, Coq, Isabelle/HOL, PVS,
the specification verification with proof scores
in CafeOBJ has the following characteristics.
\begin{itemize}
\setlength{\itemsep}{-2pt}

\item
Systems/services are modeled
via Algebraic Abstract Data Types (ADT) with equations
and an appropriate higher abstraction level
can be settled for each system/property specification.
Moreover, the equations are used directly
as reduction/rewriting
rules for proofs at the higher abstraction level.

\item
The major two proof rules 
{\bf Fact \ref{fact-crdmdl}} {\bf (PR1)}, 
{\bf Fact \ref{fact-csee}} {\bf (PR2)}
are simple and transparent,
formalized at the level of specification satisfaction
\textit{SP}$\mdl${\it p}
(i.e., any model of \textit{SP} satisfies $p$),
and supporting model based proofs.
Another important proof rule for inductions
formalized as {\bf Fact \ref{fact-wfishe}} is also succinct and 
nicely harmonizes with the two rules
{\bf (PR1)}, {\bf (PR2)}.

\end{itemize}

\subsubsection{Reduction Based Proof}
\label{ssec-rbp}

A proof score succeeds if each of the reductions involved returns the
expected result, usually \ttt{true}.  Each reduction is with the
equations in the current module.  Hence, by observing the current
equations and the failed reduction result, the equations to be added for
getting desirable reduction result could be guessed.
The following three items are all declared as equations
that are going to be used in reductions,
and have a good chance to be predicted through
the current and expected reduction results.
This is a nice and important feature based on
the transparent reduction (rewriting) based proof.

\begin{itemize}
\setlength{\itemsep}{-2pt}

\item Necessary missing axioms and lemmas
  (see {\bf Example \ref{ex-listeq}}).

\item Exhaustive equations for case-splits.

\item Necessary induction hypotheses.

\end{itemize}

\subsubsection{Initiality, Termination, Confluence,
  Sufficient Completeness}
\label{sssec-itcs}

The properties proof scores prove are
the semantic properties of interest on specifications.
Each of the properties is supposed to be expressed
in a CafeOBJ's predicate ({\it\ref{sssec-bibpred}}), 
an invariant property ({\bf Fact \ref{fact-inv}}),
or a leads-to property ({\bf Fact \ref{fact-pltq}}).
These properties are 
the primary indexes of the quality of specifications.
That is, if a proof score succeeds in proving a property
on a specification then
the equations (i.e., axioms) in the specification are shown to be
sufficient enough to imply the property.

Initiality ({\it\ref{subsec-initalg}}),
termination,
confluence,
and sufficient completeness  ({\it\ref{sssec-coex}})
(ITCS for short) are 
another properties of specifications.
They are also important indexes of the quality of
specifications.
Each specification is better to satisfy 
these properties as much as possible depending
on each context.  
There are many studies on these properties,
e.g.,
\citep{goguen06TPA, terese03, LucasMM05, NakamuraOF14, Meseguer17},
that help to make a specification satisfy them.

($\vSg,E$) part of a transition specification need to be
red-complete for proof scores (with builtin search
predicates) to be valid ({\it\ref{sssec-trnspc}}).
For other kind of specifications,
proof scores do not assume that each specification
satisfies any of ITCS properties, 
but the followings are observed.
\begin{itemize}
\setlength{\itemsep}{-1pt}

\item
A high quality specification with respect to   
ITCS properties has the high possibility of success 
in proofs with proof scores.
Actually, a property relating to initial models
could not be proved without some equations that
are implied by the initiality.
The last sentence of {\it\ref{ssec-spcsys}}
explains that the equations
{\it\ref{ssec-spcsys}}-{\ttt{11:-12:}}
are of this kind.

\item 
The lack of an ITCS property of a specification
is sometimes detected 
in a middle of constructing proof scores
on the specification.
For example, subtle non-terminating reduction rules
(i.e., equations) in a specification 
could be detected by encountering
a non-terminating proof score.

\end{itemize}

\subsubsection{PTcalc+WFI Proof Scores}
\label{sssec-ptcwfi}

The case-split with equations in
a PTcalc command \ttt{:csp\{...\}} has the following
merits comparing to
the one with the \ttt{open\,{...}\,close} constructs.
\begin{itemize}
\setlength{\itemsep}{-1pt}

\item The exhaustive equations are
declared intensively inside a \ttt{:csp} command and
the user's intention is clearly expressed
in an easy to check style.

\item Although 
CafeOBJ does not support yet,
the exhaustiveness can be checked
automatically in the majority of cases based on
constructor declarations.

\end{itemize}

PTcalc is more fundamental than WFI
and there are many PTcalc proof scores without WFI.
However, {\bf term refinements}
(instances of {\bf Fact \ref{fact-csee}}-{\bf (PR2)},
e.g., {\it\ref{sssec-prswfisch}}-\ttt{19:-20:})
are needed to do WFI, and 
WFI is always PTcalc+WFI.

Many kinds of induction schemes
including structural induction 
were coded into proof scores,
and many cases were developed
before PTcalc+WFI was prepared
\citep{futatsugi06,futatsugiGO08,futatsugi10}.
PTcalc+WFI subsumes all of these various induction schemes
in a uniform and transparent way.
This contributes to applicability
and flexibility of PTcalc+WFI.

The conditions for PTcalc+WFI proof scores
to be correct (sound) are intensively localized into
the following three points.

\begin{itemize}
\setlength{\itemsep}{-1pt}

\item
Exhaustiveness of the equations declared
in a \ttt{:csp\{...\}} command.

\item Validity of the equation declared 
in an \ttt{:init} command with \ttt{(...)},
e.g., {\it\ref{sssec-pscptc}}-\ttt{10:}.

\item Well-foundedness of the binary relation \ttt{\us{wf>}\us}
used to declare an induction hypothesis.

\end{itemize}

Although no automatic support from CafeOBJ yet,
the well-foundedness of a binary relation 
\ttt{\us{wf>}\us} on 
the argument tuples of a goal predicate can be
established, in the majority of cases,
via the strict subterm relations
on constructor terms.

Full automation of PTcalc+WFI would be difficult because of 
at least the followings. 

\begin{itemize}
\setlength{\itemsep}{-1pt}

\item
A \ttt{:csp} command could be used to declare the interested class of
models for which the equations in the command are exhaustive.  
The exhaustiveness can not be checked automatically in this case.

\item 
It would be difficult to automate the procedure for finding (i)
a predicate $p$ and (ii) a well-founded relation on the argument tuples of
$p$ that lead to successful induction.

\end{itemize}

The transparent and flexible structure of PTcalc+WFI, however, could
help to design effective interactive tools.

\subsection{Conclusions}
\label{ssec-cnclsn}


\begin{itemize}
  \setlength{\itemsep}{-2pt}

\item 
A minimum of theories for justifying the practice of
the proof score constructions in CafeOBJ is presented
in a unified style, and important results are formulated
in 19 Facts.

\item
Proof scores for case-split and/or induction are shown to be
represented as the highly stylized proof scores with the proof tree
calculus (PTcalc) and the well-founded induction (WFI).

\item
Concrete proof scores in the PTcalc+WFI style
for non-trivial specifications of transition systems are 
demonstrated to be justified by the theories and the Facts.

\end{itemize}

\noindent
{\it [Acknowledgments]}\,
The author appreciates comments given by
D. {G{\u a}in{\u a}},
N. Hirokawa,
M. Nakamura,
K. Ogata,
N. Preining,
A. Riesco,
T. Sawada,
H. Yatsu,
H. Yoshida.

Comments from anonymous referees were great help to improve the
quality of the paper.

This work was supported in part by Grant-in-Aid for
Scientific Research (S) 23220002 from Japan Society for the
Promotion of Science (JSPS).

\bibliography{apscor}

\newpage

\begin{appendices}

\section{Proof Score Construction}
\label{apps:gpcps}

This appendix describes 
{\it a generic procedure for constructing proof scores}
in {\it \ref{appss:constps}} with brief descriptions of necessary concepts 
in {\it \ref{appss:sysspec}-\ref{appss:ptcalc}}
by referring to related parts of the paper.
It intends to be self-contained and
gives another illuminating view of the paper.

\subsection{System Specifications} 
\label{appss:sysspec}

In CafeOBJ,
a static system is specified as an {\bf abstract data type}
(ADT for short),
and a dynamic system is specified as 
a state transition system
(or simply {\bf transition system}).  
ADT is defined algebraically with possibly conditional
{\bf equations} ({\it\ref{sssec-seneq}}).
A transition system is described as (i) a special ADT
called {\bf OTS (observational transition system)}
({\it\ref{subsec-ots}})
or (ii) a {\bf transition specification}
({\it\ref{sssec-trnspc}}).

A transition specification specifies the following three entities.

\begin{itemize}

\item ADT for defining {\bf state configurations} of the transition system.

\item {\bf Transitions} on the state configurations that define the behavior
  of the transition system. 
  The transitions on the state configurations are defined with
  possibly conditional {\bf transition rules};
  a transition rule is a special equation on state configurations
  that defines transitions from an 
  instance of the rule's left hand side to the corresponding instance
  of the rule's right hand side.

\item {\bf Initial states} as a subset of the state configurations.

\end{itemize}

\subsection{Models of a CafeOBJ Specification} 
\label{appss:modspec}

ADT and transition system ({\it\ref{appss:sysspec}}) are well established
concepts in software engineering and programming.
They are not formal enough, however, to do
formal proofs on them, and we need to define mathematical models of 
a CafeOBJ module (i.e., a CafeOBJ specification).

A mathematical system that consists of a collection of sets and functions
on the sets is called an {\bf algebra}.
Each CafeOBJ module denotes a class of algebras
each of which interprets the {\bf sorts} and {\bf operators} 
declared in the module as sets and functions on the sets
(i.e., the sorts denote the sets and the operators denote
the functions on the sets) ({\it\ref{sssec-sigalg}}).

If equations are declared in a module, each model (algebra) $A$ 
of the module satisfies the equations
(i.e., any assignment of $A$'s values to variables 
makes the left and right hand sides of each equation equal)
({\it\ref{subsubsec-algst}}).

If a special sort {\tt State} and
a set $\mathit{Tl}$ of transition rules on the sort {\tt State}
are declared in a module,
for each model (algebra) $A$ of the module,
$\mathit{Tl}$ defines the {\bf transition relation} 
$\mathit{TR}_{\mathit{Tl}} \subseteq A_{{\tt State}} \times A_{{\tt State}}$,
where $A_{{\tt State}}$ is the set the sort {\tt State} denotes
({\it\ref{sssec-trnspc}} with 
the correspondence $\mathit{TR}_{\mathit{Tl}}\,\lrh\,\tr{Tl/E,AC}$).

CafeOBJ has a builtin module {\tt BOOL}, with a sort {\tt Bool},
that implements
the {\bf propositional calculus} as an executable Boolean algebra
({\it\ref{sssec-bibpred}}).  
A standard CafeOBJ module includes {\tt BOOL} automatically and every model of
the module includes the Boolean algebra.

For a CafeOBJ module $M$ and a ground term (i.e., a term without variables)
$\mathit{bt}$ of the sort {\tt Bool},
let $M {\models} \mathit{bt}$ mean that every model of $M$ {\bf satisfies} $\mathit{bt}$
(i.e., $\mathit{bt}$ is {\tt true} for every model of $M$).
A ground term can contain {\bf fresh constants} each of which denotes,
like a variable,
any element of the corresponding sort
({\bf Fact \ref{fact-thcst}}).

\subsection{Property Specifications and Proof Modules}
\label{appss:pspm}

Let $M {\models_{\rm prp}} \pi$ mean that 
a property $\pi$ holds for a module $M$
(i.e., $\pi$ holds for every model of $M$).

If $M$ specifies an ADT, 
the property $\pi$ is supposed to be such that
a Boolean ground term $\mathit{bt}$ and a module $M^{\pi}$,
which satisfies the following implication, 
can be constructed.\\
\hspace*{2em}
{\bf (pm)}\;\;$M^{\pi} {\models} \mathit{bt} \Rightarrow M {\models_{\rm prp}} \pi$\\
Let $M_{bt}$ be the module defining the operators 
(functions including predicates)
needed to express the term $bt$, 
then ``$M^{\pi}$ = $M {+} M_{bt}$''
(i.e., the module $M^{\pi}$ is obtained by adding $M_{bt}$ to $M$).

Modules for proof like $M^{\pi}$ or $M_{bt}$ are called
{\bf proof modules}, and modules like $M_{bt}$ are called
{\bf property specifications} in particular.

If $M$ is a transition specification and specifies 
a transition system,
$\pi$ is supposed to be
an {\bf invariant property} ({\it\ref{ssec-invprp}})
or a {\bf leads-to property} ({\it\ref{subsec-pltqpr}}).   
An invariant property is a safety property and is specified with 
a state predicate $p$ that holds for all reachable states.
A leads-to property is a liveness property and is specified with
two state predicates $p$ and $q$.
($p$ leads-to $q$) holds for a transition system if the system's
entry to the state where $p$ holds implies that the system will surely
get into the state where $q$ holds no matter what transition sequence
is taken.  

For a transition specification (module) $M$ and its property $\pi$,
modules $M^{\pi}_i$ and Boolean ground terms $bt_i$
$(1 \le i \le m)$ that satisfy the following implication
can be constructed by making use of
{\bf builtin search predicates} 
({\it\ref{sssec-btnsp}}).\\
\hspace*{2em}
{\bf (PM)}\;\;$(M^{\pi}_1 {\models} \mathit{bt}_1 \land 
  M^{\pi}_2 {\models} \mathit{bt}_2 \land
\cdots \land 
  M^{\pi}_m {\models} \mathit{bt}_m)
\Rightarrow M {\models_{\rm prp}} \pi$\\
Let $M_{bt_i}$ be the property specification defining the operators 
needed to express the term $bt_i$ $(1 \le i \le m)$, 
then $M^{\pi}_i = M {+} M_{bt_i}$.
Predicates $p$, $q$, which are needed to describe the 
invariant or leads-to properties of a transition system
are typical examples
of predicates to be specified in the property specifications.

\subsection{Model Based Deduction Rules}
\label{appss:mbdr}

CafeOBJ system has a reduction command
``{\tt reduce in $M$ : $\mathit{bt}$ .}'' for getting a reduced form of $\mathit{bt}$
in the module $M$ by using $M$'s equations as reduction rules from
left to right.  
Let $M {{\vdash}\!{}_{\rm c}} \mathit{bt}$ means that ``{\tt reduce in $M$ : $\mathit{bt}$ .}''
returns {\tt true}, then 
the following primary proof rule says that 
$\mathit{bt}$ is {\tt true} for every model of $M$ if 
$\mathit{bt}$ reduces to {\tt true} in $M$.

{\bf (PR1)}\;\;({\bf Fact \ref{fact-crdmdl}})\hspace*{4em}%
$M {{\vdash}\!{}_{\rm c}} \mathit{bt} \Rightarrow M {\models} \mathit{bt}$ 

Let $M$ be a module with a signature $\varSigma$ and a set of equations $E$
($M = (\varSigma,E)$ in symbols).
Equations $e_1,{\cdots},e_n$ ($1 \le n$) 
are defined to be {\bf exhaustive} for $M$
iff for any model $A$ of $M$ there exists some $i\,(1 \le i \le n)$
such that $e_i$ holds in $A$.
Let $e_1,{\cdots},e_n$ ($1 \le n$) be exhaustive equations
and $M_{+e_i}\,{=}\,(\varSigma {\cup} Y_i, E {\cup} \{e_i\})$
($1 \le i \le n$, $Y_i$ is the set of fresh constants in $e_i$),
then the following proof rule holds on which 
every case-split in CafeOBJ can be based.

{\bf (PR2)}\;\;({\bf Fact \ref{fact-csee}})\hspace*{1em}%
$(M_{+e_1} {\models} \mathit{bt} \land M_{+e_2} {\models} \mathit{bt} \land
\cdots \land M_{+e_n} {\models} \mathit{bt}) \Rightarrow M {\models} \mathit{bt}$

A {\bf constractor} of a sort is an operator that constructs elements of the sort
({\it\ref{sssec-cboss}}, {\it\ref{sssec-cbosa}}).
%
A sort is called {\bf constrained} if
there exists a constructor of the sort.
Let $\mathit{bt}(y@_1,\cdots,y@_n)$ be a ground term of the sort {\tt Bool} 
containing fresh constants 
$y@_i$ of constrained sorts
$s_i$ $(1 \le i \le n)$.
Let $Y\!@$ be the set $\{y@_1,{\cdots},y@_n\}$, 
$\overline{y@}$ be the $n$-tuple $y@_1,{\cdots},y@_n$,
$y_i $ be variables of sorts $s_i (1 \le i \le n)$,
and $\overline{y}$ be the $n$-tuple $y_1,{\cdots},y_n$.
Let {\tt \us}${\mbox{\it wf}\!\cab}${\tt \us} be a well-founded binary
relation on the set of $n$-tuples $ct_1,{\cdots},ct_n$ of
ground constructor terms of sorts $s_1,{\cdots},s_n$,
and $M_{\scriptsize {\mbox{\it wf}\!\cab}}$ be the module 
adding the definition of
{\tt \us}${\mbox{\it wf}\!\cab}${\tt \us} to $M$.
The following {\bf well-founded induction (WFI)} rule can be used for
deducing $M {\models} \mathit{bt}(y@_1,{\cdots},y@_n)$.

{\bf (WFI)}\;\;({\bf Fact \ref{fact-powfi}})\\
\hspace*{3.5em}%
($M_{\scriptsize {\mbox{\it wf}\!\cab}}{\cup}Y\!@{\cup}$%
\{{\tt cq}\;{$\mathit{bt}(\overline{y})$}\,{\tt =}\;{\tt true}\;{\tt if}\;%
$\overline{y@}\,{{\mbox{\it wf}\!\cab}}\,\overline{y}$\,{\tt .}\!\})\!
${\models} \mathit{bt}(\overline{y@})
\Rightarrow M {\models} \mathit{bt}(\overline{y@})$

\subsection{PTcalc
  (Proof Tree Calculus, {\it\ref{ssec-ptcalc}})}
\label{appss:ptcalc}

PTcalc is a subsystem of CafeOBJ  system to support proving
$M {\models} \mathit{bt}$ with the proof rules {\bf (PR1)} and {\bf (PR2)}.

$M{{\vdash}\!{}_{\rm c}}\mathit{bt}$ does not always holds, and
$M{\models}\mathit{bt}$ is usually difficult to duduce directly
using the proof rule {\bf (PR1)}.
Exhaustive equations $e_1,{\cdots},e_n$ need to be found
and the proof rule {\bf (PR2)} should be used.
$M_{+e_i}{\models}\mathit{bt}$ would be still difficult to prove 
with the proof rule {\bf (PR1)}, and {\bf (PR2)} should be
applied repeatedly.  The repeated applications
of {\bf (PR2)} generate {\bf proof trees} successively.
Each of the generated proof trees has the {\bf root node}
$M{\models}\mathit{bt}$ and each of other {\bf nodes} is of the form
$M_{+e_{i_1}{\cdots}+e_{i_m}}{\models}\mathit{bt}$ ($1 \le m$)
that is generated as a {\bf child node} of 
$M_{+e_{i_1}{\cdots}+e_{i_{m-1}}}{\models}\mathit{bt}$ by applying {\bf (PR2)}.
A {\bf leaf node} (i.e.,{\;}a node without child nodes)
$M_{+e_{l_1}{\cdots}+e_{l_k}}{\models}\mathit{bt}$ ($0 \le k$)
of a proof tree is called {\bf effective} if 
$M_{+e_{l_1}{\cdots}+e_{l_k}}{{\vdash}\!{}_{\rm c}}\mathit{bt}$ holds.
A proof tree is called effective if all of whose
leaf nodes are effective.  PTcalc proves $M{\models}\mathit{bt}$ by
constructing an effective proof tree whose root node is
$M{\models}\mathit{bt}$.

\subsection{Generic Procedure for Constructing Proof Scores}
\label{appss:constps}

A collection of CafeOBJ codes (i.e., pieces of CafeOBJ code)
for deducing 
that a property $\pi$ holds for every model of a module $M$
($M {\models}_{\rm prp} \pi$ in symbols)
is called a proof score.
A generic procedure to construct a proof score 
for deducing $M {\models}_{\rm prp} \pi$ is as follows.

\begin{enumerate}


\item[(1)] Construct proof modules $M^{\pi}_i$ and
  Boolean ground terms $\mathit{bt}_i$
$(1 \le i \le m)$ satisfying the following implication
({\it\ref{appss:pspm}} {\bf (pm)}, {\bf (PM)}).

\hspace*{2em}
$(M^{\pi}_1 {\models} \mathit{bt}_1 \land 
  M^{\pi}_2 {\models} \mathit{bt}_2 \land
\cdots \land 
  M^{\pi}_m {\models} \mathit{bt}_m)
\Rightarrow M {\models_{\rm prp}} \pi$\;.

\item[(2)] For each
$M^{\pi}_i {\models} \mathit{bt}_i\,(1 \le i \le m)$
do as follows.

\begin{enumerate}

\item[(a)] Construct a root module $M^{{\rm rt}}_i$ as follows depending on
whether {\bf WFI} is used or not.

\begin{enumerate}

\item[(i)] If $\mathit{bt}_i$ is expressed as
  $\mathit{bt}_i(\overline{y@})$ with fresh constants $\overline{y@}$
  of constrained sorts and the $\mathit{bt}_i(\overline{y@})$'s proof
  can be achieved with WFI,  then define a well-founded relation
  {\tt \us}${\mbox{\it wf}\!\cab}${\tt \us} appropriately and
  construct the module $M^{{\rm rt}}_i$ as follows
  with variables $\overline{y}$ of the corresponding constrained sorts
 ({\it\ref{appss:mbdr}} {\bf (WFI)}). 

\hspace*{3em}%
$M^{{\rm rt}}_i$ =
$({M^{\pi}_i})_{\scriptsize {\mbox{\it wf}\!\cab}}{\cup}Y\!@{\cup}$%
\{{\tt cq}\;{$\mathit{bt}_i(\overline{y})$}\,{\tt =}\;{\tt true}\;{\tt if}\;%
$\overline{y@}\,{{\mbox{\it wf}\!\cab}}\,\overline{y}$\,{\tt .}\!\}

\item[(ii)] If the condition of (i) is not satisfied, 
then construct the module $M^{{\rm rt}}_i$ as follows.

\hspace*{3em}$M^{{\rm rt}}_i$ = ${M^{\pi}_i}$

\end{enumerate}

\item[(b)] Construct a CafeOBJ code with PTcalc commands
that constructs an effective proof tree with the root node 
$M^{{\rm rt}}_i {\models} \mathit{bt}_i$.
If a lemma (a property) $\lambda$ is found necessary to be proved
with respect to a module $M^{\lambda}$
(i.e., $M^{\lambda} {\models_{\rm prp}} \lambda$ needs to be proved),
then set $\pi$ = $\lambda$, $M$ = $M^{\lambda}$ 
and construct a proof score for $M {\models_{\rm prp}} \pi$
using the procedure (1)-(2).
Note that constructions of proof scores for lemmas may be
recursively invoked.

\end{enumerate}

\end{enumerate}

The procedure (1)-(2) does not always terminate.
Interactive trial and error are needed to reach to successful termination.
If terminates, the CafeOBJ codes constructed, including the codes
for proving necessary lemmas, constitute a proof score 
for deducing $M {{\models}_{\rm prp}} \pi$.\\

\end{appendices}

\end{document}

%% file: kf-pre.tex
\usepackage{amsmath,amssymb,amsfonts}

\usepackage{xcolor}

\usepackage{hyperref}

\usepackage{appendix}

\newcommand{\sm}[1]{{\small{#1}}}

\newcommand{\ttt}[1]{{\tt {#1}}}

\newcommand{\mttt}[1]{\mbox{\tt {#1}}}

\newcommand{\mtnit}[1]{\mbox{\tiny{\it {#1}}}}

\newcommand{\df}{\stackrel{\rm def}{=}}

\newcommand{\vSg}{\varSigma}
\newcommand{\Sg}{\mbox{$\varSigma$}}

\newcommand{\La}{\Leftarrow}
\newcommand{\ra}{\rightarrow}
\newcommand{\lra}{\leftrightarrow}
\newcommand{\Lra}{\Leftrightarrow}
\newcommand{\Ra}{\Rightarrow}
\newcommand{\lrh}{\leftrightharpoons}
\newtheorem{theorem}{Theorem} 
\newtheorem{definition}[theorem]{Definition}
\newtheorem{proposition}[theorem]{Proposition}
\newtheorem{corollary}[theorem]{Corollary}
\newtheorem{lemma}[theorem]{Lemma}
\newtheorem{fact}[theorem]{Fact}
\newtheorem{example}{Example}

\newtheorem{notation}{Notation}
\newtheorem{remark}{Remark}
\newcommand{\bdfn}{\begin{definition} \begin{rm}}
\newcommand{\edfn}{$\Box$ \end{rm} \end{definition}}
\newcommand{\bthm}{\begin{theorem} \begin{rm}}
\newcommand{\ethm}{$\Box$ \end{rm} \end{theorem}}
\newcommand{\bprop}{\begin{proposition} \begin{rm}}
\newcommand{\eprop}{\end{rm} \end{proposition}}
\newcommand{\bcor}{\begin{corollary}\begin{rm}}
\newcommand{\ecor}{$\Box$ \end{rm} \end{corollary}}
\newcommand{\blem}{\begin{lemma} \begin{rm}}
\newcommand{\elem}{$\Box$ \end{rm} \end{lemma}}

\newcommand{\bfact}{\begin{fact}}
\newcommand{\efact}{\end{fact}}
\newcommand{\bex}{\begin{example} \begin{rm}}
\newcommand{\eex}{\end{rm} \end{example}}
\newcommand{\bntn}{\begin{notation}\begin{rm}}
\newcommand{\entn}{$\Box$ \end{rm} \end{notation}}
\newcommand{\brem}{\begin{remark}\begin{rm}}
\newcommand{\erem}{$\Box$ \end{rm} \end{remark}}

\newcommand{\dedu}[2]{\lra\!\!_{\mtnit{#1}\,{\cup}\mtnit{#2}}}
\newcommand{\dedus}[2]{\lra\!\!_{\mtnit{#1}\,{\cup}\mtnit{#2}}\hspace{-19pt}^*\hspace{14pt}}
\newcommand{\ded}[1]{\lra\!\!_{\mtnit{#1}}}

\newcommand{\dedstr}[1]{{\lra\!\!_{\mtnit{#1}}\hspace{-15pt}^{\star}}\hspace{11pt}}
\newcommand{\rdc}[1]{\ra\!\!_{\mtnit{#1}}}

\newcommand{\rdcstr}[1]{{\ra\!\!_{\mtnit{#1}}\hspace{-15pt}^{\star}}\hspace{11pt}}

\newcommand{\tr}[1]{\ra\!\!_{\mtnit{#1}}}
\newcommand{\trstr}[1]{{\ra\!\!_{\mtnit{#1}}\hspace{-24pt}^{\star}}\hspace{20pt}}
\newcommand{\trstru}[1]{{\ra\!\!_{\mtnit{#1}}\hspace{-26pt}^{\star}}\hspace{22pt}}

\newcommand{\Sign}{\mathbb{S}ign}
\newcommand{\Mod}{\mathbb{M}od}
\newcommand{\Sen}{\mathbb{S}en}

\newenvironment{cosmall}{\small\baselineskip = 9pt}

\newenvironment{cofootnote}{\footnotesize\baselineskip = 8pt}

\newenvironment{coscript}{\scriptsize\baselineskip = 7pt}

\newcommand{\owari}{$\square$}

\newcommand{\spd}{\hspace{0.2em}}   
   
\newcommand{\tsr}{\spd{\vdash}\!\!{}_{\rm r}\spd}
\newcommand{\tse}{\spd{\vdash}\!\!{}_{\rm e}\spd}
\newcommand{\mdl}{\spd{\vDash}\spd}

\newcommand{\tsc}{\spd{\vdash}\!\!{}_{\rm c}\spd}
\newcommand{\wfg}{\mbox{\it wf}\!\!\cab}

\newcommand{\cab}{\symbol{"3E}} 
\newcommand{\us}{\symbol{"5F}}  
\newcommand{\ob}{\symbol{"7B}}  
\newcommand{\cb}{\symbol{"7D}}  
\newcommand{\dl}{\symbol{"24}}  
\newcommand{\shp}{\symbol{"23}}  
\newcommand{\tl}{\symbol{"7E}}  


%% file: apscor.bbl
\begin{thebibliography}{60}
\expandafter\ifx\csname natexlab\endcsname\relax\def\natexlab#1{#1}\fi
\providecommand{\url}[1]{\texttt{#1}}
\providecommand{\href}[2]{#2}
\providecommand{\path}[1]{#1}
\providecommand{\DOIprefix}{doi:}
\providecommand{\ArXivprefix}{arXiv:}
\providecommand{\URLprefix}{URL: }
\providecommand{\Pubmedprefix}{pmid:}
\providecommand{\doi}[1]{\href{http://dx.doi.org/#1}{\path{#1}}}
\providecommand{\Pubmed}[1]{\href{pmid:#1}{\path{#1}}}
\providecommand{\bibinfo}[2]{#2}
\ifx\xfnm\relax \def\xfnm[#1]{\unskip,\space#1}\fi
\bibitem[{ACL2(2022.09 accessed)}]{acl2-wp}
\bibinfo{author}{ACL2}, \bibinfo{year}{2022.09 accessed}.
\newblock \bibinfo{title}{Web page}.
\newblock \URLprefix \url{https://www.cs.utexas.edu/users/moore/acl2/}.
\bibitem[{Astesiano and et~al.(2002)}]{AstesianoBKKMST02}
\bibinfo{author}{Astesiano, E.}, \bibinfo{author}{et~al.},
  \bibinfo{year}{2002}.
\newblock \bibinfo{title}{{CASL:} the common algebraic specification language}.
\newblock \bibinfo{journal}{Theor. Comput. Sci.} \bibinfo{volume}{286},
  \bibinfo{pages}{153--196}.
\newblock \DOIprefix\doi{10.1016/S0304-3975(01)00368-1}.
\bibitem[{B(2022.09 accessed)}]{b-wp}
\bibinfo{author}{B}, \bibinfo{year}{2022.09 accessed}.
\newblock \bibinfo{title}{Web page}.
\newblock \URLprefix \url{https://formalmethods.fandom.com/wiki/B-Method}.
\bibitem[{van~den Brand and et~al.(2001)}]{BrandDHJJKKMOSVVV01}
\bibinfo{author}{van~den Brand, M.}, \bibinfo{author}{et~al.},
  \bibinfo{year}{2001}.
\newblock \bibinfo{title}{The {Asf+Sdf} meta-environment: a component-based
  language development environment}.
\newblock \bibinfo{journal}{Electron. Notes Theor. Comput. Sci.}
  \bibinfo{volume}{44}, \bibinfo{pages}{3--8}.
\newblock \DOIprefix\doi{10.1016/S1571-0661(04)80917-4}.
\bibitem[{Burstall(1969)}]{burstall69}
\bibinfo{author}{Burstall, R.}, \bibinfo{year}{1969}.
\newblock \bibinfo{title}{Proving properties of programs by structural
  induction}.
\newblock \bibinfo{journal}{Computer Journal} \bibinfo{volume}{12(1)},
  \bibinfo{pages}{41--48}.
\newblock \DOIprefix\doi{10.1093/comjnl/12.1.41}.
\bibitem[{{CafeOBJ}(2022.09 accessed)}]{cafeobj-wp}
\bibinfo{author}{{CafeOBJ}}, \bibinfo{year}{2022.09 accessed}.
\newblock \bibinfo{title}{Web page}.
\newblock \URLprefix \url{{https://cafeobj.org/}}.
\bibitem[{Chandy and Misra(1989)}]{ChandyM89ppd}
\bibinfo{author}{Chandy, K.M.}, \bibinfo{author}{Misra, J.},
  \bibinfo{year}{1989}.
\newblock \bibinfo{title}{Parallel program design - a foundation}.
\newblock \bibinfo{publisher}{Addison-Wesley}.
\bibitem[{Clavel et~al.(2006)Clavel, Palomino and Riesco}]{ClavelPR06-jucs}
\bibinfo{author}{Clavel, M.}, \bibinfo{author}{Palomino, M.},
  \bibinfo{author}{Riesco, A.}, \bibinfo{year}{2006}.
\newblock \bibinfo{title}{Introducing the {ITP} tool: a tutorial}.
\newblock \bibinfo{journal}{J. UCS} \bibinfo{volume}{12},
  \bibinfo{pages}{1618--1650}.
\newblock \DOIprefix\doi{10.3217/jucs-012-11-1618}.
\bibitem[{{Coq}(2022.09 accessed)}]{coq-wp}
\bibinfo{author}{{Coq}}, \bibinfo{year}{2022.09 accessed}.
\newblock \bibinfo{title}{Web page}.
\newblock \URLprefix \url{http://coq.inria.fr/}.
\bibitem[{Diaconescu and Futatsugi(1998)}]{rd-kf1998}
\bibinfo{author}{Diaconescu, R.}, \bibinfo{author}{Futatsugi, K.},
  \bibinfo{year}{1998}.
\newblock \bibinfo{title}{{CafeOBJ Report}}. volume~\bibinfo{volume}{6} of
  \textit{\bibinfo{series}{{AMAST} Series in Computing}}.
\newblock \bibinfo{publisher}{World Scientific}.
\newblock \DOIprefix\doi{10.1142/3831}.
\bibitem[{Diaconescu and Futatsugi(2002)}]{cafesem02}
\bibinfo{author}{Diaconescu, R.}, \bibinfo{author}{Futatsugi, K.},
  \bibinfo{year}{2002}.
\newblock \bibinfo{title}{Logical foundations of {CafeOBJ}}.
\newblock \bibinfo{journal}{Theor. Comput. Sci.} \bibinfo{volume}{285},
  \bibinfo{pages}{289--318}.
\newblock \DOIprefix\doi{10.1016/S0304-3975(01)00361-9}.
\bibitem[{Futatsugi(2006)}]{futatsugi06}
\bibinfo{author}{Futatsugi, K.}, \bibinfo{year}{2006}.
\newblock \bibinfo{title}{Verifying specifications with proof scores in
  {CafeOBJ}}, in: \bibinfo{booktitle}{Proc. 21st {IEEE/ACM ASE}},
  \bibinfo{publisher}{IEEE}. pp. \bibinfo{pages}{3--10}.
\newblock \DOIprefix\doi{10.1109/ASE.2006.73}.
\bibitem[{Futatsugi(2010)}]{futatsugi10}
\bibinfo{author}{Futatsugi, K.}, \bibinfo{year}{2010}.
\newblock \bibinfo{title}{Fostering proof scores in {CafeOBJ}}, in:
  \bibinfo{booktitle}{Proc. 12th {ICFEM} (LNCS-6447)},
  \bibinfo{publisher}{Springer}. pp. \bibinfo{pages}{1--20}.
\newblock \DOIprefix\doi{10.1007/978-3-642-16901-4\_1}.
\bibitem[{Futatsugi(2015)}]{futatsugi15}
\bibinfo{author}{Futatsugi, K.}, \bibinfo{year}{2015}.
\newblock \bibinfo{title}{Generate {\&} check method for verifying transition
  systems in {CafeOBJ}}, in: \bibinfo{booktitle}{Software, Services, and
  Systems (LNCS-8950)}, \bibinfo{publisher}{Springer}. pp.
  \bibinfo{pages}{171--192}.
\newblock \DOIprefix\doi{10.1007/978-3-319-15545-6\_13}.
\bibitem[{Futatsugi(2017)}]{futatsugi17}
\bibinfo{author}{Futatsugi, K.}, \bibinfo{year}{2017}.
\newblock \bibinfo{title}{Introduction to Specification Verification in
  {CafeOBJ} (in Japanese)}.
\newblock \bibinfo{publisher}{Saiensu-Sha, Tokyo}.
\newblock \URLprefix \url{https://cafeobj.org/iprog/}.
\bibitem[{Futatsugi(2020)}]{futatsugi20}
\bibinfo{author}{Futatsugi, K.}, \bibinfo{year}{2020}.
\newblock \bibinfo{title}{Well-founded induction via term refinement in
  {CafeOBJ}}, in: \bibinfo{booktitle}{PreProc. WRLA 2020}, pp.
  \bibinfo{pages}{64--78}.
\newblock \URLprefix \url{http://wrla2020.webs.upv.es/pre-proceedings.pdf}.
\bibitem[{Futatsugi(2021)}]{futatsugi21}
\bibinfo{author}{Futatsugi, K.}, \bibinfo{year}{2021}.
\newblock \bibinfo{title}{Advances of proof scores in {CafeOBJ} : Invited
  paper}, in: \bibinfo{booktitle}{Intl. Symp. on Theoretical Aspects of
  Software Engineering {TASE} 2021}, \bibinfo{publisher}{{IEEE}}. pp.
  \bibinfo{pages}{3--12}.
\newblock \DOIprefix\doi{10.1109/TASE52547.2021.00012}.
\bibitem[{Futatsugi(2022)}]{Futatsugi22}
\bibinfo{author}{Futatsugi, K.}, \bibinfo{year}{2022}.
\newblock \bibinfo{title}{Advances of proof scores in {CafeOBJ}}.
\newblock \bibinfo{journal}{Sci. Comput. Program.} \bibinfo{volume}{224},
  \bibinfo{pages}{102893}.
\newblock \DOIprefix\doi{10.1016/J.SCICO.2022.102893}.
\bibitem[{Futatsugi et~al.(2012)Futatsugi, G{\u a}in{\u a} and
  Ogata}]{futatsugiGo12}
\bibinfo{author}{Futatsugi, K.}, \bibinfo{author}{G{\u a}in{\u a}, D.},
  \bibinfo{author}{Ogata, K.}, \bibinfo{year}{2012}.
\newblock \bibinfo{title}{Principles of proof scores in {CafeOBJ}}.
\newblock \bibinfo{journal}{Theor. Comput. Sci.} \bibinfo{volume}{464},
  \bibinfo{pages}{90--112}.
\newblock \DOIprefix\doi{10.1016/j.tcs.2012.07.041}.
\bibitem[{Futatsugi et~al.(1985)Futatsugi, Goguen, Jouannaud and
  Meseguer}]{futatsugiGJM85}
\bibinfo{author}{Futatsugi, K.}, \bibinfo{author}{Goguen, J.A.},
  \bibinfo{author}{Jouannaud, J.P.}, \bibinfo{author}{Meseguer, J.},
  \bibinfo{year}{1985}.
\newblock \bibinfo{title}{Principles of {OBJ2}}, in: \bibinfo{booktitle}{Proc.
  12th ACM POPL (POPL85)}, \bibinfo{publisher}{ACM}. pp.
  \bibinfo{pages}{52--66}.
\newblock \DOIprefix\doi{10.1145/318593.318610}.
\bibitem[{Futatsugi et~al.(2008)Futatsugi, Goguen and Ogata}]{futatsugiGO08}
\bibinfo{author}{Futatsugi, K.}, \bibinfo{author}{Goguen, J.A.},
  \bibinfo{author}{Ogata, K.}, \bibinfo{year}{2008}.
\newblock \bibinfo{title}{Verifying design with proof scores}, in:
  \bibinfo{booktitle}{Proc. VSTTE (LNCS-4171)}, \bibinfo{publisher}{Springer}.
  pp. \bibinfo{pages}{277--290}.
\newblock \DOIprefix\doi{10.1007/978-3-540-69149-5}.
\bibitem[{Futatsugi and Nakagawa(1997)}]{futatsugiN97}
\bibinfo{author}{Futatsugi, K.}, \bibinfo{author}{Nakagawa, A.T.},
  \bibinfo{year}{1997}.
\newblock \bibinfo{title}{An overview of {CAFE} specification environment}, in:
  \bibinfo{booktitle}{Proc. First {IEEE ICFEM}}, \bibinfo{publisher}{IEEE}. pp.
  \bibinfo{pages}{170--182}.
\newblock \DOIprefix\doi{10.1109/ICFEM.1997.630424}.
\bibitem[{Futatsugi and Okada(1980)}]{futatsugio80}
\bibinfo{author}{Futatsugi, K.}, \bibinfo{author}{Okada, K.},
  \bibinfo{year}{1980}.
\newblock \bibinfo{title}{Specification writing as construction of
  hierarchically structured clusters of operators}, in:
  \bibinfo{editor}{Lavington, S.H.} (Ed.), \bibinfo{booktitle}{The 8th {IFIP}
  Congress 1980}, \bibinfo{publisher}{North-Holland/IFIP}. pp.
  \bibinfo{pages}{287--292}.
\bibitem[{G{\u a}in{\u a}(2020)}]{Gaina20}
\bibinfo{author}{G{\u a}in{\u a}, D.}, \bibinfo{year}{2020}.
\newblock \bibinfo{title}{Forcing and calculi for hybrid logics}.
\newblock \bibinfo{journal}{J. {ACM}} \bibinfo{volume}{67},
  \bibinfo{pages}{25:1--25:55}.
\newblock \DOIprefix\doi{10.1145/3400294}.
\bibitem[{G{\u a}in{\u a} and Futatsugi(2015)}]{gainaF15}
\bibinfo{author}{G{\u a}in{\u a}, D.}, \bibinfo{author}{Futatsugi, K.},
  \bibinfo{year}{2015}.
\newblock \bibinfo{title}{Initial semantics in logics with constructors}.
\newblock \bibinfo{journal}{J. Log. Comput.} \bibinfo{volume}{25},
  \bibinfo{pages}{95--116}.
\newblock \DOIprefix\doi{10.1093/logcom/exs044}.
\bibitem[{G{\u a}in{\u a} et~al.(2014)G{\u a}in{\u a}, Lucanu, Ogata and
  Futatsugi}]{gainaLOF14}
\bibinfo{author}{G{\u a}in{\u a}, D.}, \bibinfo{author}{Lucanu, D.},
  \bibinfo{author}{Ogata, K.}, \bibinfo{author}{Futatsugi, K.},
  \bibinfo{year}{2014}.
\newblock \bibinfo{title}{On automation of {OTS/CafeOBJ} method}, in:
  \bibinfo{booktitle}{Specification, Algebra, and Software (LNCS-8373)},
  \bibinfo{publisher}{Springer}. pp. \bibinfo{pages}{578--602}.
\newblock \DOIprefix\doi{10.1007/978-3-642-54624-2\_29}.
\bibitem[{G{\u a}in{\u a} et~al.(2020)G{\u a}in{\u a}, Nakamura, Ogata and
  Futatsugi}]{GainaNOF20}
\bibinfo{author}{G{\u a}in{\u a}, D.}, \bibinfo{author}{Nakamura, M.},
  \bibinfo{author}{Ogata, K.}, \bibinfo{author}{Futatsugi, K.},
  \bibinfo{year}{2020}.
\newblock \bibinfo{title}{Stability of termination and sufficient-completeness
  under pushouts via amalgamation}.
\newblock \bibinfo{journal}{Theor. Comput. Sci.} \bibinfo{volume}{848},
  \bibinfo{pages}{82--105}.
\newblock \DOIprefix\doi{10.1016/j.tcs.2020.09.024}.
\bibitem[{G{\u a}in{\u a} et~al.(2013)G{\u a}in{\u a}, Zhang, Chiba and
  Arimoto}]{gainaZCA13}
\bibinfo{author}{G{\u a}in{\u a}, D.}, \bibinfo{author}{Zhang, M.},
  \bibinfo{author}{Chiba, Y.}, \bibinfo{author}{Arimoto, Y.},
  \bibinfo{year}{2013}.
\newblock \bibinfo{title}{Constructor-based inductive theorem prover}, in:
  \bibinfo{booktitle}{Proc. 5th {CALCO} (LNCS-8089)},
  \bibinfo{publisher}{Springer}. pp. \bibinfo{pages}{328--333}.
\newblock \DOIprefix\doi{10.1007/978-3-642-40206-7\_26}.
\bibitem[{Goguen(2006)}]{goguen06TPA}
\bibinfo{author}{Goguen, J.}, \bibinfo{year}{2006}.
\newblock \bibinfo{title}{Theorem Proving and Algebra}.
\newblock \URLprefix \url{https://arxiv.org/abs/2101.02690}.
\bibitem[{Goguen and Meseguer(1992)}]{goguenM92OSA}
\bibinfo{author}{Goguen, J.A.}, \bibinfo{author}{Meseguer, J.},
  \bibinfo{year}{1992}.
\newblock \bibinfo{title}{Order-sorted algebra {I}: Equational deduction for
  multiple inheritance, overloading, exceptions and partial operations.}
\newblock \bibinfo{journal}{Theor. Comput. Sci.} \bibinfo{volume}{105},
  \bibinfo{pages}{217--273}.
\newblock \DOIprefix\doi{10.1016/0304-3975(92)90302-V}.
\bibitem[{Goguen et~al.(2000)Goguen, Winkler, Meseguer, Futatsugi and
  Jouannaud}]{goguen00iobj}
\bibinfo{author}{Goguen, J.A.}, \bibinfo{author}{Winkler, T.},
  \bibinfo{author}{Meseguer, J.}, \bibinfo{author}{Futatsugi, K.},
  \bibinfo{author}{Jouannaud, J.P.}, \bibinfo{year}{2000}.
\newblock \bibinfo{title}{Introducing {OBJ}}, in: \bibinfo{editor}{Goguen, J.},
  \bibinfo{editor}{Malcolm, G.} (Eds.), \bibinfo{booktitle}{Software
  Engineering with OBJ: algebraic specification in action}.
  \bibinfo{publisher}{Kluwer Academic Publishers}, pp. \bibinfo{pages}{3--167}.
\newblock \DOIprefix\doi{10.1007/978-1-4757-6541-0_1}.
\bibitem[{Guttag and et~al.(1993)}]{GuttagHGJMW93}
\bibinfo{author}{Guttag, J.V.}, \bibinfo{author}{et~al.}, \bibinfo{year}{1993}.
\newblock \bibinfo{title}{Larch: Languages and Tools for Formal Specification}.
\newblock Texts and Monographs in Computer Science,
  \bibinfo{publisher}{Springer}.
\newblock \DOIprefix\doi{10.1007/978-1-4612-2704-5}.
\bibitem[{Hrbacek and Jech(1999)}]{HrbaeekJech99}
\bibinfo{author}{Hrbacek, K.}, \bibinfo{author}{Jech, T.},
  \bibinfo{year}{1999}.
\newblock \bibinfo{title}{Introduction to Set Theory, 3rd ed}.
\newblock \bibinfo{publisher}{Marcel Dekker}.
\bibitem[{{Isabelle/HOL}(2022.09 accessed)}]{IsabelleHOL-wp}
\bibinfo{author}{{Isabelle/HOL}}, \bibinfo{year}{2022.09 accessed}.
\newblock \bibinfo{title}{Web page}.
\newblock \URLprefix \url{{https://isabelle.in.tum.de/dist/library/HOL/}}.
\bibitem[{Lucas et~al.(2005)Lucas, March{\'{e}} and Meseguer}]{LucasMM05}
\bibinfo{author}{Lucas, S.}, \bibinfo{author}{March{\'{e}}, C.},
  \bibinfo{author}{Meseguer, J.}, \bibinfo{year}{2005}.
\newblock \bibinfo{title}{Operational termination of conditional term rewriting
  systems}.
\newblock \bibinfo{journal}{Inf. Process. Lett.} \bibinfo{volume}{95},
  \bibinfo{pages}{446--453}.
\newblock \DOIprefix\doi{10.1016/j.ipl.2005.05.002}.
\bibitem[{{Maude}(2022.09 accessed)}]{maude-wp}
\bibinfo{author}{{Maude}}, \bibinfo{year}{2022.09 accessed}.
\newblock \bibinfo{title}{Web page}.
\newblock \URLprefix \url{{http://maude.cs.uiuc.edu/}}.
\bibitem[{Meseguer(1997)}]{meseguer97mel}
\bibinfo{author}{Meseguer, J.}, \bibinfo{year}{1997}.
\newblock \bibinfo{title}{Membership algebra as a logical framework for
  equational specification}, in: \bibinfo{booktitle}{Proc. WADT (LNCS-1376)},
  \bibinfo{publisher}{Springer}. pp. \bibinfo{pages}{18--61}.
\newblock \DOIprefix\doi{10.1007/3-540-64299-4\_26}.
\bibitem[{Meseguer(2012)}]{Meseguer12rwl20y}
\bibinfo{author}{Meseguer, J.}, \bibinfo{year}{2012}.
\newblock \bibinfo{title}{Twenty years of rewriting logic}.
\newblock \bibinfo{journal}{J. Log. Algebr. Program.} \bibinfo{volume}{81},
  \bibinfo{pages}{721--781}.
\newblock \DOIprefix\doi{10.1016/j.jlap.2012.06.003}.
\bibitem[{Meseguer(2017)}]{Meseguer17}
\bibinfo{author}{Meseguer, J.}, \bibinfo{year}{2017}.
\newblock \bibinfo{title}{Strict coherence of conditional rewriting modulo
  axioms}.
\newblock \bibinfo{journal}{Theor. Comput. Sci.} \bibinfo{volume}{672},
  \bibinfo{pages}{1--35}.
\newblock \DOIprefix\doi{10.1016/j.tcs.2016.12.026}.
\bibitem[{MiniSAT(2022.09 accessed)}]{minisat-wp}
\bibinfo{author}{MiniSAT}, \bibinfo{year}{2022.09 accessed}.
\newblock \bibinfo{title}{Web page}.
\newblock \URLprefix \url{http://minisat.se}.
\bibitem[{Nakamura et~al.(2022)Nakamura, Higashi, Sakakibara and
  Ogata}]{NakamuraHSO22}
\bibinfo{author}{Nakamura, M.}, \bibinfo{author}{Higashi, S.},
  \bibinfo{author}{Sakakibara, K.}, \bibinfo{author}{Ogata, K.},
  \bibinfo{year}{2022}.
\newblock \bibinfo{title}{Specification and verification of multitask real-time
  systems using the {OTS/CafeOBJ} method}.
\newblock \bibinfo{journal}{{IEICE} Trans. Fundam. Electron. Commun. Comput.
  Sci.} \bibinfo{volume}{105-A}, \bibinfo{pages}{823--832}.
\newblock \DOIprefix\doi{10.1587/transfun.2021map0007}.
\bibitem[{Nakamura et~al.(2014)Nakamura, Ogata and Futatsugi}]{NakamuraOF14}
\bibinfo{author}{Nakamura, M.}, \bibinfo{author}{Ogata, K.},
  \bibinfo{author}{Futatsugi, K.}, \bibinfo{year}{2014}.
\newblock \bibinfo{title}{Incremental proofs of termination, confluence and
  sufficient completeness of {OBJ} specifications}, in:
  \bibinfo{booktitle}{Specification, Algebra, and Software (LNCS-8373)},
  \bibinfo{publisher}{Springer}. pp. \bibinfo{pages}{92--109}.
\newblock \DOIprefix\doi{10.1007/978-3-642-54624-2\_5}.
\bibitem[{Nakamura et~al.(2021)Nakamura, Sakakibara, Okura and
  Ogata}]{NakamuraSOO21}
\bibinfo{author}{Nakamura, M.}, \bibinfo{author}{Sakakibara, K.},
  \bibinfo{author}{Okura, Y.}, \bibinfo{author}{Ogata, K.},
  \bibinfo{year}{2021}.
\newblock \bibinfo{title}{Formal verification of multitask hybrid systems by
  the {OTS/CafeOBJ} method}.
\newblock \bibinfo{journal}{Int. J. Softw. Eng. Knowl. Eng.}
  \bibinfo{volume}{31}, \bibinfo{pages}{1541--1559}.
\newblock \DOIprefix\doi{10.1142/S0218194021400118}.
\bibitem[{OBJ3(2022.09 accessed)}]{obj3-wp}
\bibinfo{author}{OBJ3}, \bibinfo{year}{2022.09 accessed}.
\newblock \bibinfo{title}{Web page}.
\newblock \URLprefix
  \url{https://www.kindsoftware.com/products/opensource/obj3/}.
\bibitem[{Ogata and Futatsugi(2003)}]{OgataF03}
\bibinfo{author}{Ogata, K.}, \bibinfo{author}{Futatsugi, K.},
  \bibinfo{year}{2003}.
\newblock \bibinfo{title}{Proof scores in the {OTS/CafeOBJ} method}, in:
  \bibinfo{booktitle}{Proc. 6th {IFIP} {WG} 6.1 {FMOODS} 2003 (LNCS-2884)},
  \bibinfo{publisher}{Springer}. pp. \bibinfo{pages}{170--184}.
\newblock \DOIprefix\doi{10.1007/978-3-540-39958-2\_12}.
\bibitem[{Ogata and Futatsugi(2008)}]{OgataF08}
\bibinfo{author}{Ogata, K.}, \bibinfo{author}{Futatsugi, K.},
  \bibinfo{year}{2008}.
\newblock \bibinfo{title}{Proof score approach to verification of liveness
  properties}.
\newblock \bibinfo{journal}{IEICE Trans.} \bibinfo{volume}{91-D},
  \bibinfo{pages}{2804--2817}.
\newblock \DOIprefix\doi{10.1093/ietisy/e91-d.12.2804}.
\bibitem[{Preining et~al.(2015)Preining, Ogata and Futatsugi}]{PreiningOF14}
\bibinfo{author}{Preining, N.}, \bibinfo{author}{Ogata, K.},
  \bibinfo{author}{Futatsugi, K.}, \bibinfo{year}{2015}.
\newblock \bibinfo{title}{Liveness properties in {CafeOBJ}}, in:
  \bibinfo{booktitle}{Proc. 24th {LOPSTR}, 2014 (LNCS-8981)},
  \bibinfo{publisher}{Springer}. pp. \bibinfo{pages}{182--198}.
\newblock \DOIprefix\doi{10.1007/978-3-319-17822-6\_11}.
\bibitem[{{PTcalc Manual}(2022.09 accessed)}]{ptcalc-wp}
\bibinfo{author}{{PTcalc Manual}}, \bibinfo{year}{2022.09 accessed}.
\newblock \bibinfo{title}{Web page}.
\newblock \URLprefix \url{{https://cafeobj.org/files/ptcalc.pdf}}.
\bibitem[{{PVS}(2022.09 accessed)}]{pvs-wp}
\bibinfo{author}{{PVS}}, \bibinfo{year}{2022.09 accessed}.
\newblock \bibinfo{title}{Web page}.
\newblock \URLprefix \url{{http://pvs.csl.sri.com/}}.
\bibitem[{Riesco and Ogata(2018)}]{RiescoO18}
\bibinfo{author}{Riesco, A.}, \bibinfo{author}{Ogata, K.},
  \bibinfo{year}{2018}.
\newblock \bibinfo{title}{Prove it! inferring formal proof scripts from
  {CafeOBJ} proof scores}.
\newblock \bibinfo{journal}{ACM Trans. Softw. Eng. Methodol.}
  \bibinfo{volume}{27}, \bibinfo{pages}{6:1--6:32}.
\newblock \DOIprefix\doi{10.1145/3208951}.
\bibitem[{Riesco and Ogata(2022)}]{RiescoO22}
\bibinfo{author}{Riesco, A.}, \bibinfo{author}{Ogata, K.},
  \bibinfo{year}{2022}.
\newblock \bibinfo{title}{An integrated tool set for verifying {CafeOBJ}
  specifications}.
\newblock \bibinfo{journal}{J. Syst. Softw.} \bibinfo{volume}{189},
  \bibinfo{pages}{111302}.
\newblock \DOIprefix\doi{10.1016/j.jss.2022.111302}.
\bibitem[{Riesco et~al.(2017)Riesco, Ogata and Futatsugi}]{RiescoOF17}
\bibinfo{author}{Riesco, A.}, \bibinfo{author}{Ogata, K.},
  \bibinfo{author}{Futatsugi, K.}, \bibinfo{year}{2017}.
\newblock \bibinfo{title}{A {Maude} environment for {CafeOBJ}}.
\newblock \bibinfo{journal}{Formal Asp. Comput.} \bibinfo{volume}{29},
  \bibinfo{pages}{309--334}.
\newblock \DOIprefix\doi{10.1007/s00165-016-0398-7}.
\bibitem[{Rodin(2022.09 accessed)}]{rodin-wp}
\bibinfo{author}{Rodin}, \bibinfo{year}{2022.09 accessed}.
\newblock \bibinfo{title}{Web page}.
\newblock \URLprefix \url{http://www.event-b.org}.
\bibitem[{Terese(2003)}]{terese03}
\bibinfo{author}{Terese}, \bibinfo{year}{2003}.
\newblock \bibinfo{title}{Term Rewriting Systems}.
\newblock \bibinfo{publisher}{Cambridge University Press}.
\bibitem[{VDM(2022.09 accessed)}]{vdm-wp}
\bibinfo{author}{VDM}, \bibinfo{year}{2022.09 accessed}.
\newblock \bibinfo{title}{Web page}.
\newblock \URLprefix \url{https://www.overturetool.org}.
\bibitem[{Wirsing(1986)}]{Wirsing86}
\bibinfo{author}{Wirsing, M.}, \bibinfo{year}{1986}.
\newblock \bibinfo{title}{Structured algebraic specifications: {A} kernel
  language}.
\newblock \bibinfo{journal}{Theor. Comput. Sci.} \bibinfo{volume}{42},
  \bibinfo{pages}{123--249}.
\newblock \DOIprefix\doi{10.1016/0304-3975(86)90051-4}.
\bibitem[{Yices(2022.09 accessed)}]{yices-wp}
\bibinfo{author}{Yices}, \bibinfo{year}{2022.09 accessed}.
\newblock \bibinfo{title}{Web page}.
\newblock \URLprefix \url{https://yices.csl.sri.com}.
\bibitem[{Yoshida et~al.(2015)Yoshida, Ogata and Futatsugi}]{YoshidaOF15}
\bibinfo{author}{Yoshida, H.}, \bibinfo{author}{Ogata, K.},
  \bibinfo{author}{Futatsugi, K.}, \bibinfo{year}{2015}.
\newblock \bibinfo{title}{Formalization and verification of declarative cloud
  orchestration}, in: \bibinfo{booktitle}{Proc. 17th {ICFEM} (LNCS-9407)},
  \bibinfo{publisher}{Springer}. pp. \bibinfo{pages}{33--49}.
\newblock \DOIprefix\doi{10.1007/978-3-319-25423-4\_3}.
\bibitem[{Z(2022.09 accessed)}]{z-wp}
\bibinfo{author}{Z}, \bibinfo{year}{2022.09 accessed}.
\newblock \bibinfo{title}{Web page}.
\newblock \URLprefix \url{https://formalmethods.fandom.com/wiki/Z_notation}.
\bibitem[{Z3(2022.09 accessed)}]{z3-wp}
\bibinfo{author}{Z3}, \bibinfo{year}{2022.09 accessed}.
\newblock \bibinfo{title}{Web page}.
\newblock \URLprefix \url{https://github.com/Z3Prover}.

\end{thebibliography}
